\documentclass[showpacs,twocolumn,superscriptaddress]{revtex4}
\usepackage{doi}
\usepackage{hyperref}
\hypersetup{
  colorlinks=true,        
  linkcolor=blue,         
  citecolor=cyan,         
}

\usepackage{graphicx}
\usepackage{dcolumn}
\usepackage{bm}
\usepackage{color}
\usepackage{enumitem}
\usepackage{amsmath}
\usepackage{amssymb}
\usepackage{orcidlink}
\usepackage{subcaption}
\usepackage{multirow}
\usepackage{graphicx} 
\begin{document}
\title{Gravitational wave signatures from periodic orbits around a Schwarzschild–Bertotti–Robinson black hole}

\author{Tursunali Xamidov}
\email{xamidovtursunali@gmail.com}
\affiliation{Institute of Fundamental and Applied Research, National Research University TIIAME, Kori Niyoziy 39, Tashkent 100000, Uzbekistan}
\affiliation{Institute for Theoretical Physics and Cosmology,
Zhejiang University of Technology, Hangzhou 310023, China}

\author{Sanjar Shaymatov}
\email{sanjar@astrin.uz}
\affiliation{Institute of Fundamental and Applied Research, National Research University TIIAME, Kori Niyoziy 39, Tashkent 100000, Uzbekistan}
\affiliation{Institute for Theoretical Physics and Cosmology,
Zhejiang University of Technology, Hangzhou 310023, China}
\affiliation{University of Tashkent for Applied Sciences, Str. Gavhar 1, Tashkent 100149, Uzbekistan}

\author{Qiang Wu}
\email{wuq@zjut.edu.cn}
\affiliation{Institute for Theoretical Physics \& Cosmology, Zhejiang University of Technology, Hangzhou 310023, China}
\affiliation{United Center for Gravitational Wave Physics (UCGWP), Zhejiang University of Technology, Hangzhou 310023, China}

\author{Tao Zhu}
\email{zhut05@zjut.edu.cn}
\affiliation{Institute for Theoretical Physics \& Cosmology, Zhejiang University of Technology, Hangzhou 310023, China}
\affiliation{United Center for Gravitational Wave Physics (UCGWP), Zhejiang University of Technology, Hangzhou 310023, China}

\date{\today}

\begin{abstract}

In this paper, we investigate periodic bound orbits and gravitational wave (GW) emission in the Schwarzschild–Bertotti–Robinson (Schwarzschild–BR) spacetime—an exact electrovacuum solution describing a static black hole (BH) immersed in a uniform magnetic field. We explore how the background magnetic field qualitatively alters the BH's gravitational dynamics, affecting timelike geodesics such as the marginally bound orbit (MBO) and the innermost stable circular orbit (ISCO). We then analyze periodic bound orbits using the frequency ratio ${\omega_{\varphi}}/{\omega_{r}}$, which characterizes the orbits by their azimuthal and radial motions. Based on the numerical kludge method we further compute the gravitational waveforms emitted from periodic orbits around a supermassive Schwarzschild–BR BH. We show that the background magnetic field significantly changes orbital frequencies, resonance conditions, zoom–whirl structures, and the resulting waveforms. Finally, we examine the frequency spectra in the mHz range and the detectability of these GW signals by computing the characteristic strain via a discrete Fourier transform on the time-domain waveforms, comparing the results with the sensitivity curves of space-based GW detectors such as LISA, Taiji, and TianQin. Our results show that intrinsically magnetic fields modify spacetime and leave observable imprints on extreme mass-ratio inspiral GWs, which may be tested by future observations.

\end{abstract}

\maketitle

\section{Introduction}

By demonstrating that gravity originates from spacetime curvature, General Relativity (GR) fundamentally changed our understanding of space and time \cite{Einstein1916}. It offers a robust description of black holes (BHs), which are both the end-state of massive stars and the most intriguing prediction of the theory. Recent detections of gravitational waves (GWs) \cite{Abbott16a,Abbott16b} and the imaging of supermassive BHs in M87* \cite{Akiyama19L1, Akiyama19L6} and Sgr A* \cite{Akiyama22L12, EHT2022L14} have further validated GR in both weak- and strong-field environments. Today, BHs act as cosmic laboratories for testing competing gravity theories and the limits of GR in astrophysics. Consequently, rigorous testing remains necessary to develop a fundamental understanding of phenomena such as GW physics, horizon structure, and the nature of gravity itself \cite{Will14LRR, Psaltis20PRL}.

Gravitational-wave analysis provides a powerful framework for probing the nature of gravity and the dynamics of astrophysical BHs. Current detectors, particularly space-based LISA \cite{Amaro-Seoane2017LISA} and Taiji \cite{10.1093/nsr/nwx116} detectors, are sensitive to a wide range of sources, including stellar-mass binaries and extreme mass-ratio inspirals (EMRIs) \cite{Hughes_2001,Amaro-Seoane18LRR,Babak17PRD}. EMRIs, consisting of a stellar-mass object orbiting a supermassive BH, radiate low-frequency GWs whose waveforms encode detailed information about particle dynamics and the underlying spacetime geometry, including deviations from circular motion. Furthermore, changes in any background spacetime modify orbital frequencies and resonance conditions, impacting the stability of circular orbits, the structure of the zoom–whirl dynamics, and the resulting EMRI GWs~\cite{Glampedakis02PRD, Ruangsri14PRD}. Analyzing these waveforms from EMRIs offers valuable insight into BH horizon structures and the nature of gravity. Hence, EMRIs serve as sensitive probes of gravitational theories, imprinting detectable signatures through their long-lived inspirals~\cite{Gair13CQG,Barack19CQG}. In this work, we study EMRI as a system of the Schwarzschild–Bertotti–Robinson spacetime, focusing on magnetic field effects on the gravitational waveforms from periodic orbits around BH.

In EMRI systems, a stellar-mass object orbiting a supermassive BH generates gravitational waveforms shaped by periodic orbits. These bound trajectories are fundamental to EMRI dynamics and provide an effective description of the long-lived inspiral phase through sequences of periodic orbits \cite{Levin_2008,Grossman_2009,Misra_2010}. These periodic orbits are modeled by requiring that a massive particle return to its initial position after a finite number of radial ${\omega_{r}}$ and angular ${\omega_{\varphi}}$ oscillations, exhibiting zoom–whirl motion around the central BH and satisfying a resonance configuration (\(r\)-\(\varphi\)). In addition, periodic orbits are uniquely characterized by the rational ratio of radial and azimuthal frequencies, ${\omega_{\varphi}}/{\omega_{r}}$. Each orbit is further classified by three integers: the zoom number $z$, the whirl number $w$, and the vertex number $v$ (see, e.g. \cite{Levin_2008,Levin_2009}). This formalism provides a systematic framework for modeling special periodic orbits and enables a detailed analysis of gravitational waveforms, zoom–whirl dynamics, and associated resonant and spectral features over complete inspiral phases \cite{Glampedakis02PRD}. It must be emphasized that, in Schwarzschild and Kerr spacetimes, such bound orbits have been extensively examined using three topological integers $(z,w,v)$ \cite{Levin_2008,Levin_2009,Bambhaniya20,Rana19}. Following similar approaches, there have been several studies on these lines \cite{Healy09PRL,Levin2010,Pugliese13,Babar17PRD,Liu18,Lin23,Yao23,Chan25,Lin22,Lin21,Deng20,Tu23,Deng20,Wei19,Zhang22,JIANG2024,Wang25,Wei25,Alloqulov26GW1,Sharipov25} exploring periodic orbits in a wide variety of contexts. Subsequently, a large amount of work has been devoted to investigating gravitational waveforms generated by periodic orbits in EMRIs for various BH spacetimes \cite[see, e.g.,][]{Barausse14PRD,Cardoso22PRD,Zhang25,Yang24JCAP,Shabbir25,Junior24,Yang24,Haroon25,Alloqulov25GW,Wang25JCAP,Lu25GW,Chen25,Li25,Ahmed25GW1,Ahmed25GW2}.

It should be noted that the proposed rotating Bertotti–Robinson (BR) spacetime, known as the Kerr–Bertotti–Robinson (KBR) solution, is an exact solution to the Einstein–Maxwell equations describing a rotating BH in a uniform electromagnetic field~\cite{Podolsky2025}. Featuring an AdS AdS\(_2 \times S^2\) geometry with conformal flatness and a constant field~\cite{Bertotti59, Robinson61}, it belongs to the Plebański–Demiański class~\cite{PLEBANSKI1976} and was first studied in detail by Carter~\cite{Brandon1968}. Unlike perturbative models, the KBR metric treats the electromagnetic field as an intrinsic part of the spacetime, offering a non-perturbative framework for studying magnetized rotating BHs. This framework is key for modeling environments where magnetic and gravitational energies are comparable, such as magnetically dominated accretion flows~\cite{Bocquet1995}. Its non-asymptotically flat geometry mimics a uniform background field, relevant in cosmology. Recent studies have shown that intrinsic magnetic fields significantly alter geodesics, BH shadows, and energy extraction mechanisms~\cite{Andersson2025,Wang2025vsx,Zeng:2025KRB,Vachher2025JCAP,Wang2025bjf,Siahaan2025KRB,Podolsky2025zlm,Mirkhaydarov2026MPP,Liu2025wwq}. In this work, we focus on the non-rotating case $a = 0$, where the solution reduces to the Schwarzschild–Bertotti–Robinson (Schwarzschild–BR) metric. This spacetime offers an ideal setting to explore periodic bound orbits and gravitational wave emission from EMRIs. Due to the magnetic field being an intrinsic part of the geometry, our analysis reveals novel features that set this system apart from the test-field Schwarzschild case. Specifically, intrinsic magnetic fields can qualitatively modify orbital frequencies and resonance conditions, impacting the stability of circular orbits, zoom–whirl structures, and the resulting gravitational waveforms. These results suggest that in astrophysical environments where magnetic fields play a dynamical role, controlled deviations from the Schwarzschild scenario arise and leave observable imprints on the gravitational radiation emitted by EMRIs.

The paper is organized as follows. In Sec.~\ref{sec2}, we introduce the Schwarzschild–Bertotti–Robinson (Schwarzschild–BR) spacetime and briefly review the formalism used to modeling test-particle motion, focusing on how the background magnetic field influences geodesic properties such as marginally bound orbits (MBO) and innermost stable circular orbits (ISCOs), which are key quantities for further exploring periodic bound orbits. In Sec.~\ref{sec3}, we analyze the periodic bound orbits around the Schwarzschild–BR BH using the rational number as the ratio of azimuthal and radial frequencies, ${\omega_{\varphi}}/{\omega_{r}}$, and examine how Schwarzschild–BR geometry affects the particle's energy, orbital angular momentum, and the allowed ($E-L$) parameter space for each periodic orbit. In Sec.~\ref{sec4}, we analyze the gravitational waveforms from periodic orbits in the EMRI system using the numerical kludge approach. We then examine the detectability of the emitted gravitational radiation from periodic EMRI orbits by computing the characteristic strain via a discrete Fourier transform of the time-domain signals. Finally, Sec.~\ref{summary} presents our conclusions and summarizes the main results. 

\section{Spacetime metric and timelike geodesics}\label{sec2}

\begin{figure*}[!htb]
    \centering
    \includegraphics[scale=0.5]{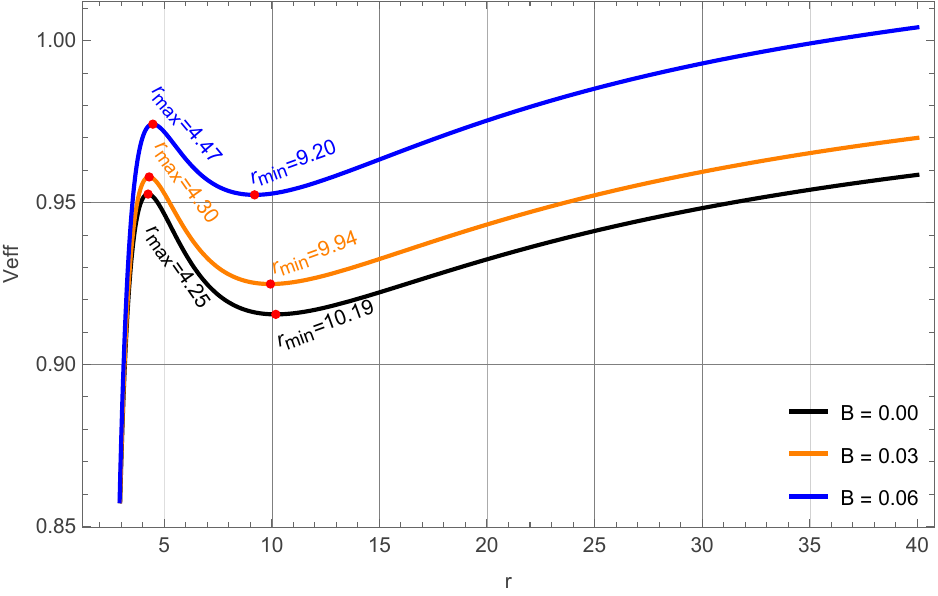}
    \includegraphics[scale=0.5]{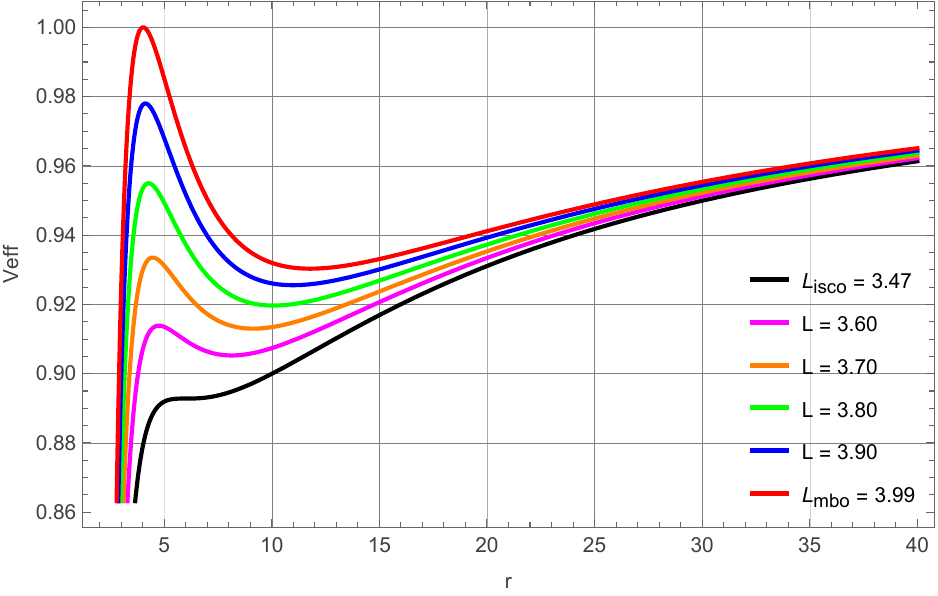}
    \caption{{The radial dependence of the effective potential for different values of the magnetic field $B$ and the orbital angular momentum $L$.} }
    \label{fig:effpotential}
\end{figure*}

A recent study~\cite{Podolsky2025} presented a new exact solution of the Einstein–Maxwell equations describing a Kerr BHimmersed in an asymptotically uniform external electromagnetic field, known as the Kerr–Bertotti–Robinson (KBR) spacetime. In the nonrotating limit, $a = 0$, this solution reduces to the Schwarzschild–Bertotti–Robinson (Schwarzschild–BR) metric and can be written as 
\begin{eqnarray}\label{Eq:metric} 
ds^{2} &=& -\frac{Q}{\Omega^{2}r^{2}} dt^2 + \frac{r^{2}}{\Omega^{2}Q} \, dr^{2} + \frac{r^{2}}{\Omega^{2}P} \, d\theta^{2} \nonumber\\&+& \frac{r^{2}P \sin^{2}\theta}{\Omega^{2}} d\varphi^{2}\, , 
\end{eqnarray}
where we use the following notation: 
\begin{align*}
P &= 1 + B^{2} M^{2} \cos^{2}\theta, \quad \\
Q &= (1 + B^{2}r^{2})\,\Delta, \quad \\
\Omega^{2} &= (1 + B^{2}r^{2}) - B^{2}\Delta\cos^{2}\theta\, , \quad \\
\Delta &= \left( 1 - B^{2}M^{2} \right)r^{2} - 2Mr\, . 
\end{align*}

Using the Lagrangian formalism, we further analyze the geodesic motion of a neutral test particle in the Schwarzschild--BR BH spacetime. The Lagrangian can be written as follows~\cite{1983mtbh.book.....C}:
\begin{equation}
\mathcal{L}=\frac{1}{2}m\,g_{\mu\nu}\,\frac{dx^{\mu}}{d\tau}\,\frac{dx^{\nu}}{d\tau},
\end{equation}
where $\tau$ denotes the proper time and $m$ is the mass of the test particle. For simplicity, we set $m=1$ and write the generalized momentum as
\begin{equation}
    p_{\mu}=\frac{\partial {\cal L}}{\partial \dot{x}^{\mu}}=g_{\mu \nu}\dot{x}^{\nu}\, .
\end{equation}
Since the spacetime metric \eqref{Eq:metric} does not explicitly depend on the coordinates $t$ and $\phi$, the particle’s energy $E$ and angular momentum $L$ are conserved. The conserved components of the four-momentum can be written as

\begin{eqnarray}\label{eq:conslaw}
p_{t} = -\frac{Q}{\Omega^2r^2}\dot{t}=-E \, ,\quad p_{\phi} = \frac{r^{2}P \sin^{2}\theta}{\Omega^{2}}\dot{\phi}=L \, , 
\end{eqnarray}
Using the conservation laws in Eq.~\eqref{eq:conslaw} and the normalization condition $g_{\mu\nu}p^{\mu}p^{\nu}=-1$, the timelike radial motion of a neutral particle in the equatorial plane ($\theta=\pi/2$) is governed by
\begin{eqnarray} \label{eq:radialeq}
\dot{r}^2 = \left(1+B^2 r^2\right)^2\left(E^2 - V_{eff}\right)\, ,
\end{eqnarray}
where
\begin{equation}
V_{eff}=\left(1-\frac{2M}{r}-B^{2} M^2\right)
\left(1+\frac{L^{2}}{r^{2}}\left(1+B^{2} r^{2}\right)\right)\, ,
\end{equation}

To provide further insight, we plot the radial dependence of the effective potential for different values of the magnetic field $B$ in Fig.~\ref{fig:effpotential} . As can be seen in Fig.~\ref{fig:effpotential} that, the increase in the magnetic field $B$ causes 
stable orbits to shift inward. Moreover, the right panel of Fig.~\ref{fig:effpotential} indicates that the effective potential barrier increases due to the particle’s orbital angular momentum $L$.

We then turn to analyze bound orbits around Schwarzschild-BR BH. Bound orbits exist only when the particle’s energy $E$ and orbital angular momentum $L$ satisfy the following conditions:
\begin{equation}
    L_{ISCO}\leq L \quad\mbox{and}\quad E_{ISCO}\leq E \leq E_{MBO}=1\, ,
\end{equation}
where $E_{ISCO}$ and $L_{ISCO}$ are the energy and orbital angular momentum of the particle on the innermost stable circular orbit (ISCO), respectively, while $E_{MBO}$ denotes the energy at the marginally bound orbit (MBO). The MBO is defined by
\begin{equation}
    V_{eff}=1 \quad\mbox{and}\quad \frac{d V_{eff}}{dr}=0\, .
\end{equation}

\begin{figure}[!htb]
    \centering
    \includegraphics[scale=0.5]{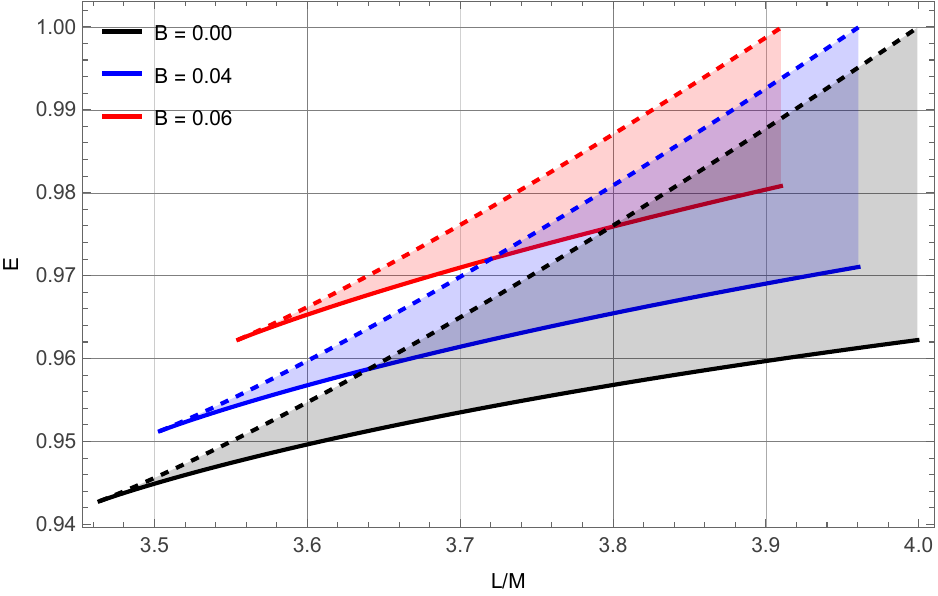}
    \caption{{The allowed parameter space of the energy and the angular momentum for the bound orbits around the Schwarzschild-BR BH with different values of the magnetic field $B$}}
    \label{fig:allowed}
\end{figure}

Here, it should be emphasized that since the analytical expressions for $r_{MBO}$ and $L_{MBO}$ are quite complicated, we explore these quantities numerically. The conditions for determining the ISCO are given by
\begin{equation}
    \dot{r}=0, \quad \frac{d V_{eff}}{dr}=0 \quad\mbox{and}\quad \frac{d^2 V_{eff}}{dr^2}=0\, .
\end{equation}
Solving these equations simultaneously permits us to obtain the ISCO parameters as follows:
\begin{eqnarray}
r_{ISCO}&=&\frac{6M}{1-B^2M^2}\, , \\
L_{ISCO}&=& \frac{2 \sqrt{3}M}{\sqrt{1-14 B^2M^2+B^4M^4}}\, , \\
E_{ISCO}&=& \frac{2\sqrt{2}}{3} \sqrt{\frac{\left(1-B^2M^2\right)^3}{1-14 B^2M^2+B^4M^4}}\, .
\end{eqnarray}
Table~\ref{tab:isco_mbo} lists the ISCO and MBO parameters for various values of the magnetic field. It can be seen from the table that, with increasing magnetic field $B$, the ISCO parameters and $r_{MBO}$ increase, whereas $L_{MBO}$ decreases. Fig.~\ref{fig:allowed} shows the allowed-parameter space in the $E$--$L$ plane for bound orbits. From Fig.~\ref{fig:allowed}, one can see that increasing the magnetic field $B$ shrinks and shifts the allowed region upward, indicating that larger energies are required for the particle to be on circular/bound orbit. 
\begin{table}[h!]
\centering
\begin{tabular}{c|cccccc}
\hline
$B$ & $R_{\mathrm{ISCO}}$ & $L_{\mathrm{ISCO}}$ & $E_{\mathrm{ISCO}}$ & $R_{\mathrm{MBO}}$ & $L_{\mathrm{MBO}}$ \\
\hline
0.00 & 6.00000 & 3.46410 & 0.94281 & 4.00000 & 4.00000 \\
0.02 & 6.00240 & 3.47384 & 0.94489 & 4.02606 & 3.99037 \\
0.04 & 6.00962 & 3.50356 & 0.95126 & 4.11043 & 3.96109 \\
0.06 & 6.02168 & 3.55482 & 0.96228 & 4.27777 & 3.91077 \\
0.08 & 6.03865 & 3.63049 & 0.97862 & 4.60983 & 3.83577 \\
\hline
\end{tabular}
\caption{The ISCO and MBO parameters for different values of the magnetic field $B$.}
\label{tab:isco_mbo}
\end{table}

\begin{figure*}[htbp]
\centering
 \includegraphics[scale=0.5]{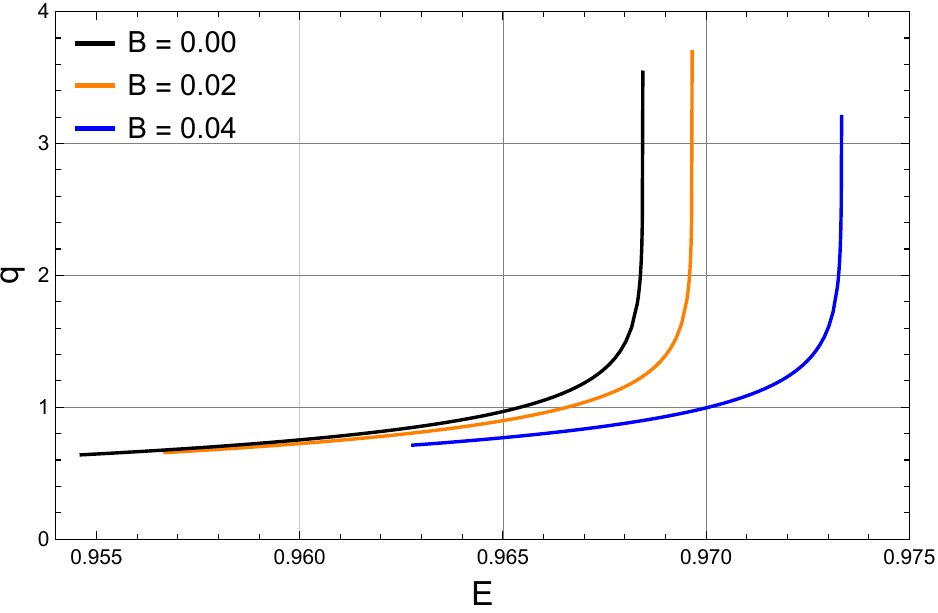}
 \includegraphics[scale=0.49]{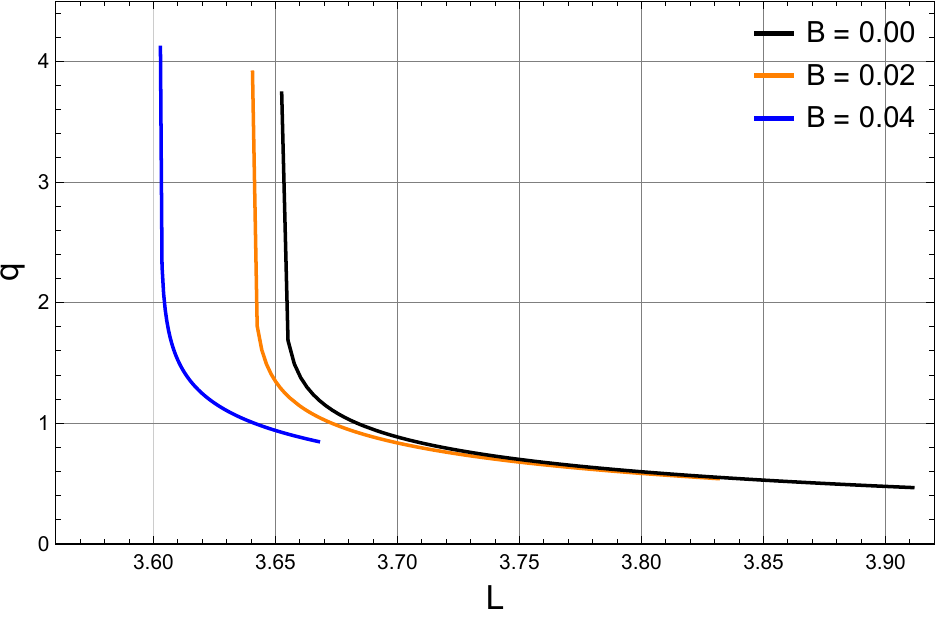}
 \caption{{Left panel: the dependence of the rational number $q$ on the energy of periodic orbits around the Schwarzschild-BR BH for different values of the magnetic field $B$. Here, the orbital angular momentum $L$ is fixed at $L=\frac{1}{2}(L_{MBO}+L_{ISCO})$. Right panel: the dependence of the rational number $q$ on the orbital angular momentum of periodic orbits around the Schwarzschild-BR BH for various combinations of the magnetic field $B$. Here, we set the energy as $E=0.96$.}}
 \label{fig:q}
\end{figure*}

\section{Periodic orbits around the Schwarzschild-BR black hole}\label{sec3}

In this section, we analyze the properties and structure of periodic orbits in the vicinity of the Schwarzschild–BR BH. Periodic orbits are particularly important among bound orbits, as their fundamental orbital frequencies form rational ratios. Each periodic orbit can be uniquely identified by three topological integers ($z,w,v$), which, as mentioned previously, correspond to the zoom, whirl, and vertex numbers. The rational number can be written in the form~\cite{Levin_2008}.
\begin{equation}
    q=\frac{\omega_{\phi}}{\omega_r}-1=w+\frac{v}{z}\, ,
\end{equation}
where $\omega_{\phi}$ and $\omega_r$ denote the angular and radial frequencies, respectively. From the equations of motion, one finds that the rational number can be written as ~\cite{2025JCAP...01..091Y,SHABBIR2025101816,JIANG2024}
\begin{equation}
    q=\frac{1}{\pi}\int^{r_2}_{r_1} \frac{\dot{\phi}}{\dot{r}}dr-1=\frac{1}{\pi} \int^{r_2}_{r_1}\frac{L}{r^2\sqrt{E^2-V_{eff}}}dr-1\, ,
\end{equation}
where $r_1$ and $r_2$ are the periapsis and apoapsis radii of the periodic orbits, respectively. Fig.~\ref{fig:q} depicts the dependence of the rational number $q$ on the particle's energy and orbital angular momentum for different values of the magnetic field $B$. The left panel shows that the rational number $q$ increases with energy $E$, rising rapidly as $E$ approaches its maximum value, while stronger magnetic fields shift $q$ toward larger energies. Meanwhile, the right panel illustrates the rational number $q$ as a function of the orbital angular momentum. As $L$ approaches its minimum, $q$ increases sharply and then decreases gradually for larger $L$, while stronger magnetic fields shift $q$ toward lower values of $L$.

\renewcommand{\arraystretch}{1.0}
\begin{table*}[]
\centering
\resizebox{1.0\textwidth}{!}{
\begin{tabular}{|c|c|c|c|c|c|c|c|c|c|}
\hline
$B$ & $L$ & $E_{(1,1,0)}$ & $E_{(1,2,0)}$ & $E_{(2,1,1)}$ & $E_{(2,2,1)}$ & $E_{(3,1,2)}$ & $E_{(3,2,2)}$ & $E_{(4,1,3)}$ & $E_{(4,2,3)}$ \\ \hline
0.00    & 3.73205 & 0.965425   & 0.968383   & 0.968026   & 0.968435   & 0.968225   & 0.968438   & 0.968285   & 0.968440   \\ \hline
0.02    & 3.73211 & 0.966600   & 0.969592   & 0.969224   & 0.969646   & 0.969428   & 0.969651   & 0.969490   & 0.969652   \\ \hline
0.04    & 3.73233 & 0.970057 & 0.973259 & 0.972851 & 0.973323 & 0.973075 & 0.973328 & 0.973144 & 0.973330   \\ \hline
0.06    & 3.73279 & 0.975942 & 0.979500 & 0.979010 & 0.979586 & 0.979273 & 0.979594 & 0.979357 & 0.979596   \\ \hline
\end{tabular}
}
\caption{The values of the energy $E$ are tabulated for different periodic orbits characterized by three integers $(z,w,v)$ and different values of the magnetic field $B$. Here, we set $L=\frac{1}{2}(L_{MBO}+L_{ISCO})$.}
\label{table1}
\end{table*}
\begin{table*}[]
\resizebox{1.0\textwidth}{!}{
\begin{tabular}{|c|c|c|c|c|c|c|c|c|}
\hline
$B$ & $L_{(1,1,0)}$ & $L_{(1,2,0)}$ & $L_{(2,1,1)}$ & $L_{(2,2,1)}$ & $L_{(3,1,2)}$ & $L_{(3,2,2)}$ & $L_{(4,1,3)}$ & $L_{(4,2,3)}$ \\ \hline
0.00    & 3.683588    & 3.653440    & 3.657596    & 3.652701    & 3.655335    & 3.652636    & 3.654621    & 3.652616    \\ \hline
0.01    & 3.681001 & 3.650471 & 3.654746 & 3.649742 & 3.652442 & 3.649675 & 3.651723 & 3.649653    \\ \hline
0.02    & 3.676378 & 3.641578 & 3.646132 & 3.640773 & 3.643699 & 3.640697 & 3.642922 & 3.640672    \\ \hline
0.03    & 3.660099    & 3.626450    & 3.631539    & 3.625485    & 3.628857    & 3.625485    & 3.627986    & 3.625355    \\ \hline
0.04    & 3.641478    & 3.604509    & 3.610550    & 3.603227    & 3.607439    & 3.603085    & 3.605901    & 3.603040    \\ \hline
\end{tabular}
}
\caption{The values of the orbital angular momentum $L$ are tabulated for different periodic orbits characterized by $(z,w,v)$ and various combinations of magnetic field $B$. Here, we set $E = 0.96$.}
\label{table2}
\end{table*}

\begin{figure*}[htbp]
    \centering
    \includegraphics[width=0.32\textwidth]{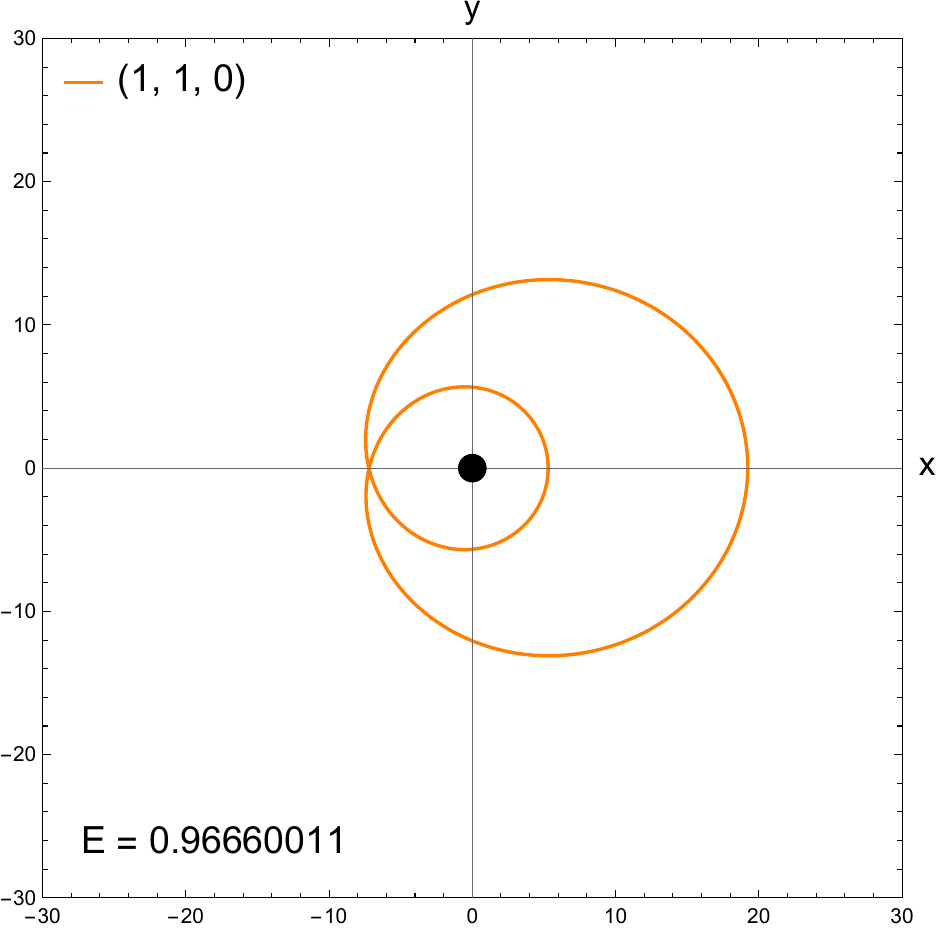} \hfill
    \includegraphics[width=0.32\textwidth]{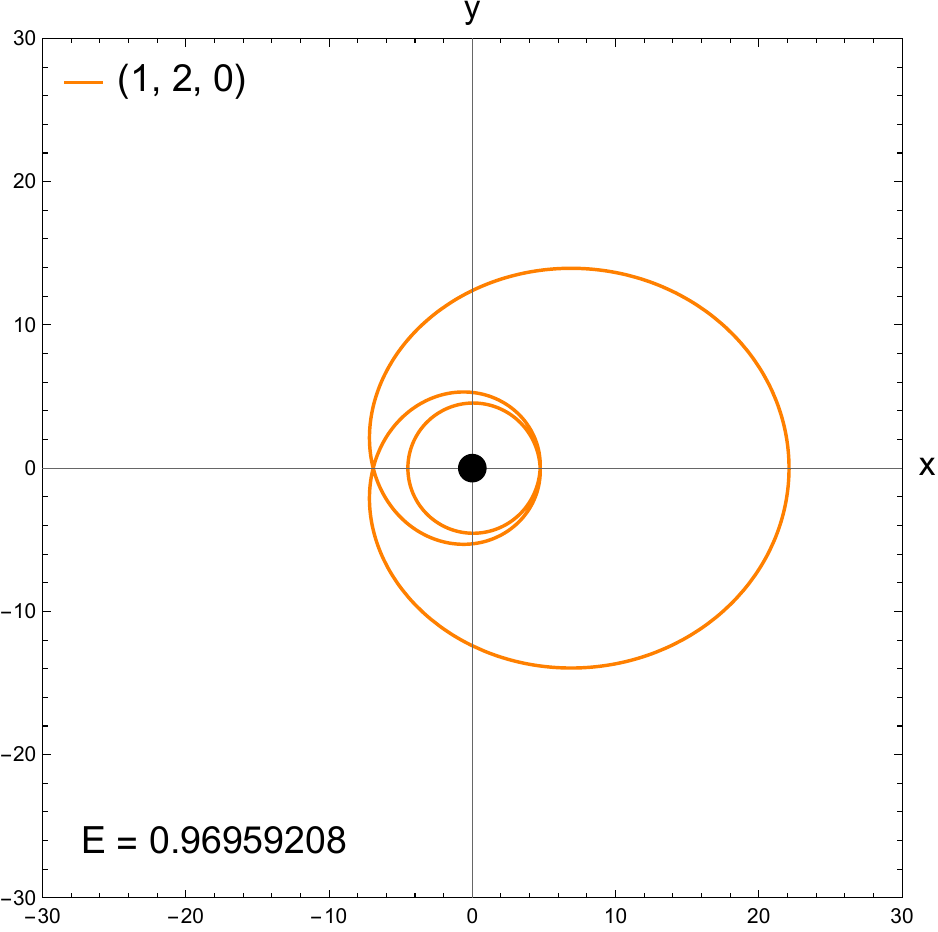} \hfill
    \includegraphics[width=0.32\textwidth]{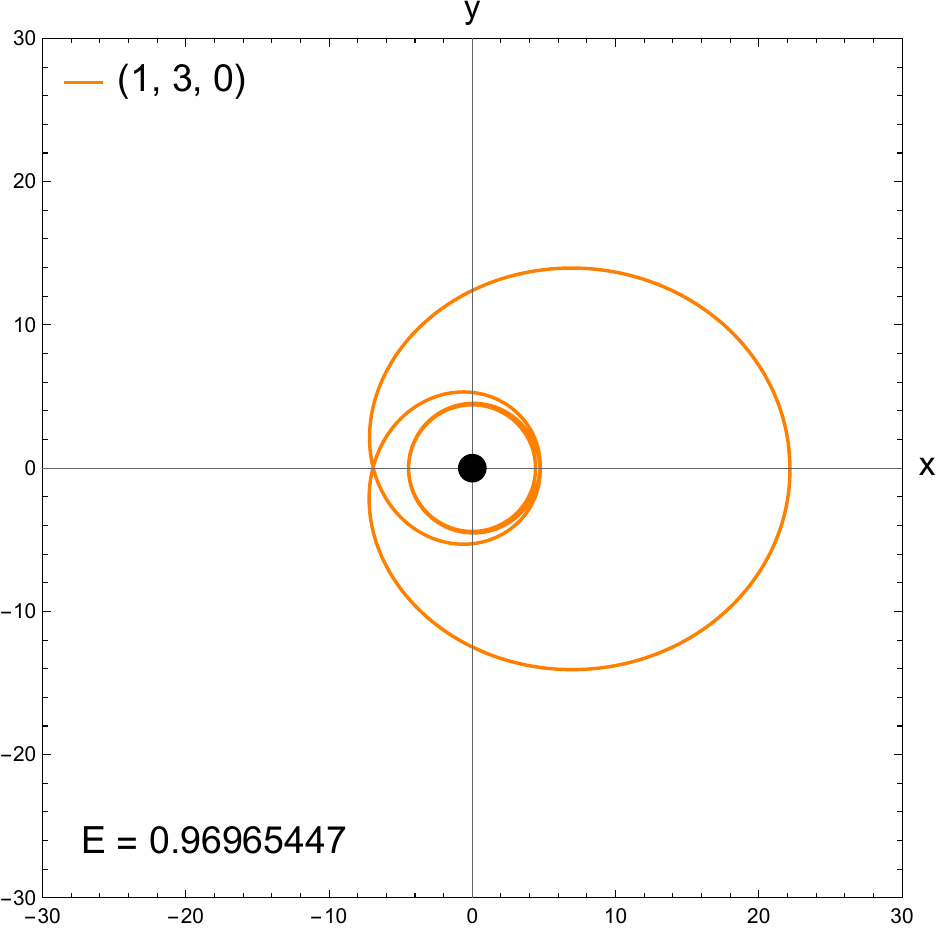} \\
    
    \vspace{0.2cm} 
    \includegraphics[width=0.32\textwidth]{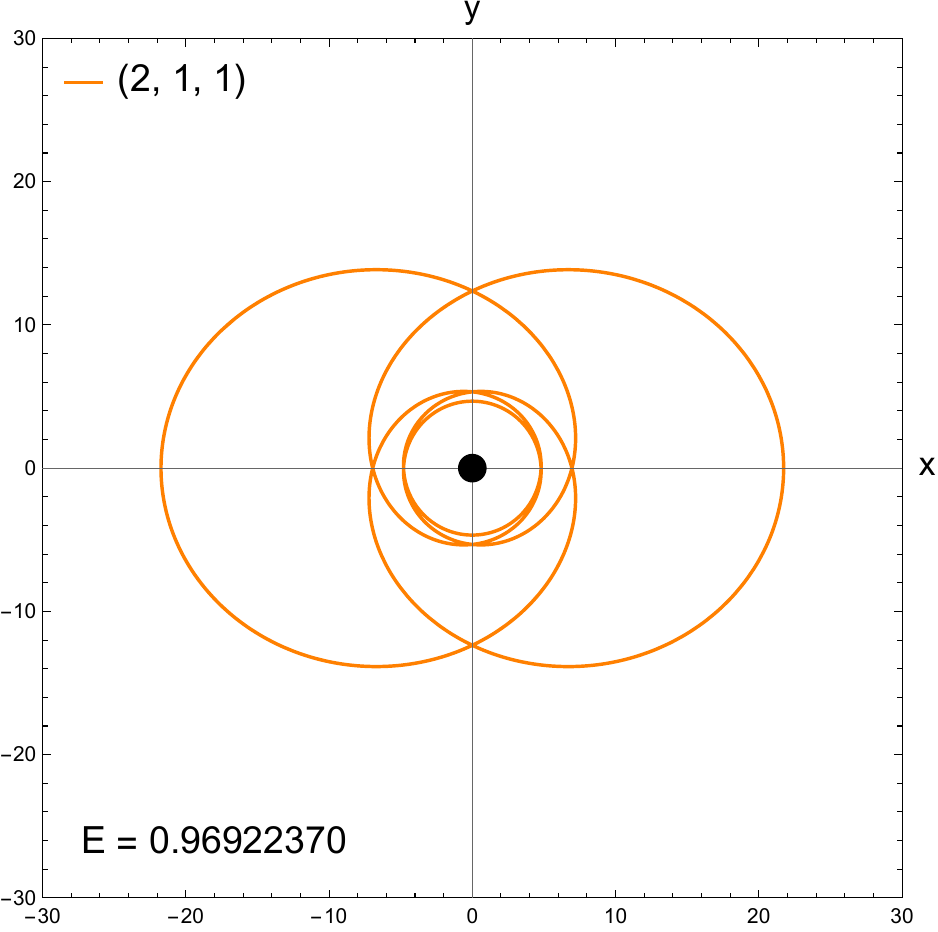} \hfill
    \includegraphics[width=0.32\textwidth]{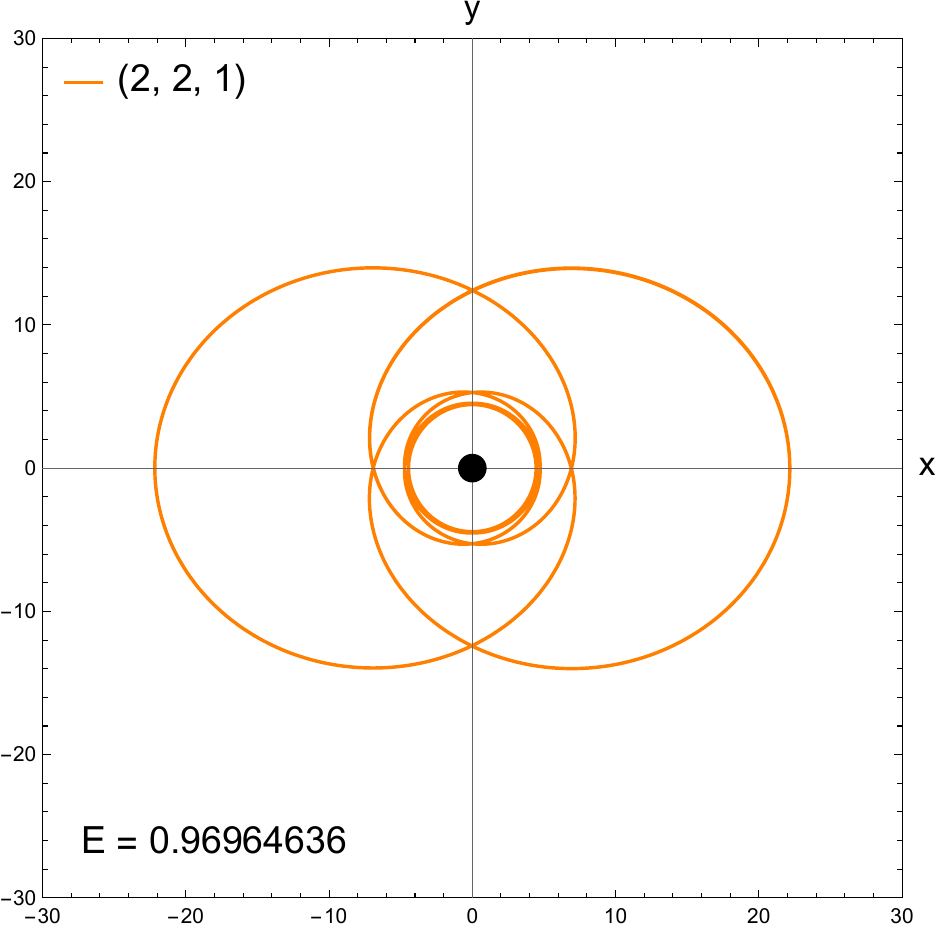} \hfill
    \includegraphics[width=0.32\textwidth]{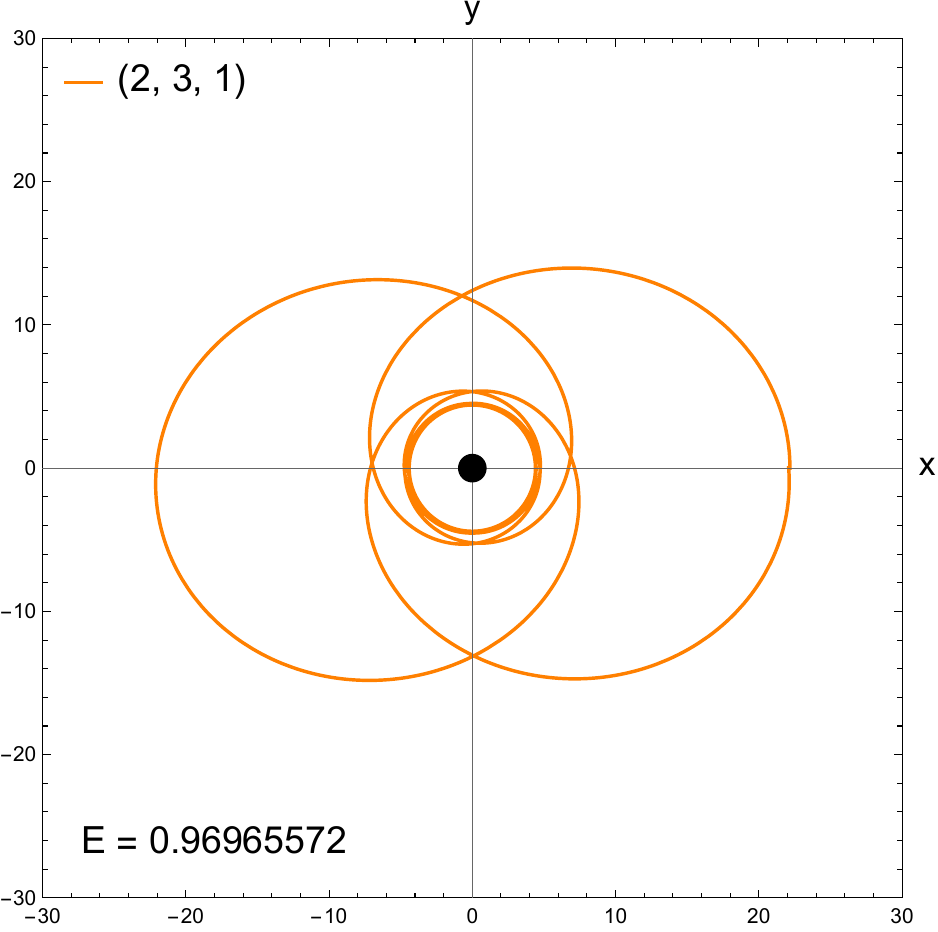} \\
    
    \vspace{0.2cm} 
    \includegraphics[width=0.32\textwidth]{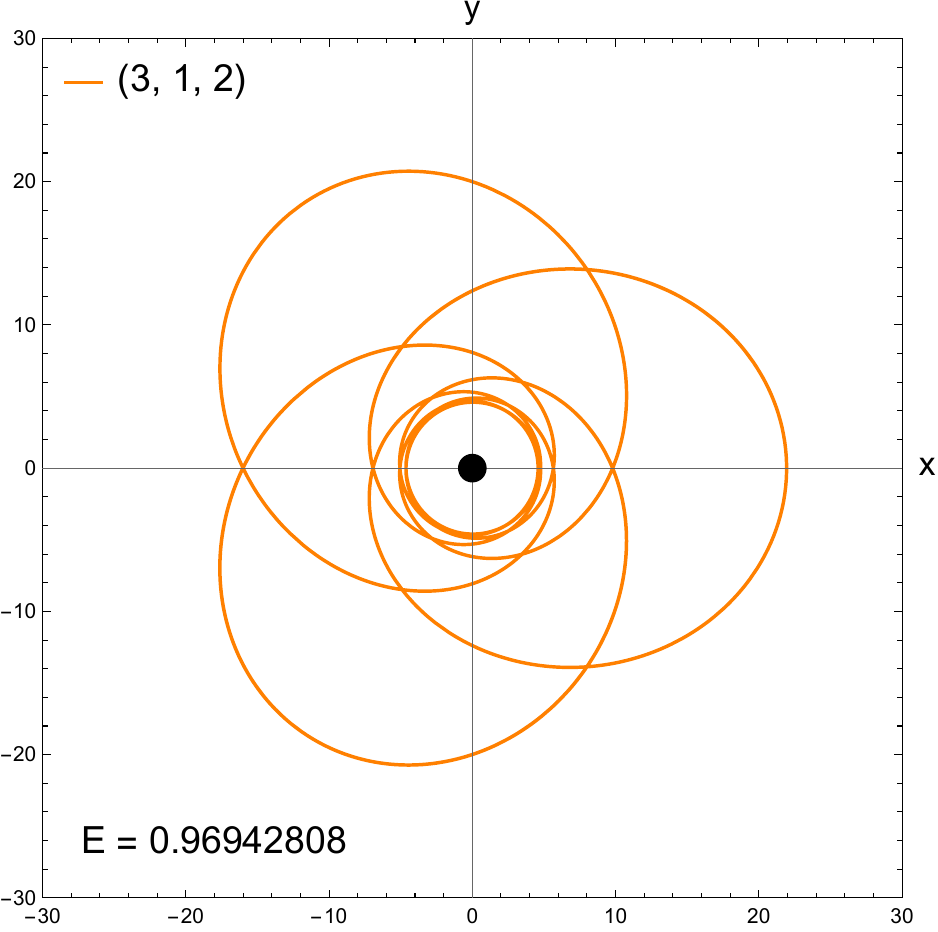} \hfill
    \includegraphics[width=0.32\textwidth]{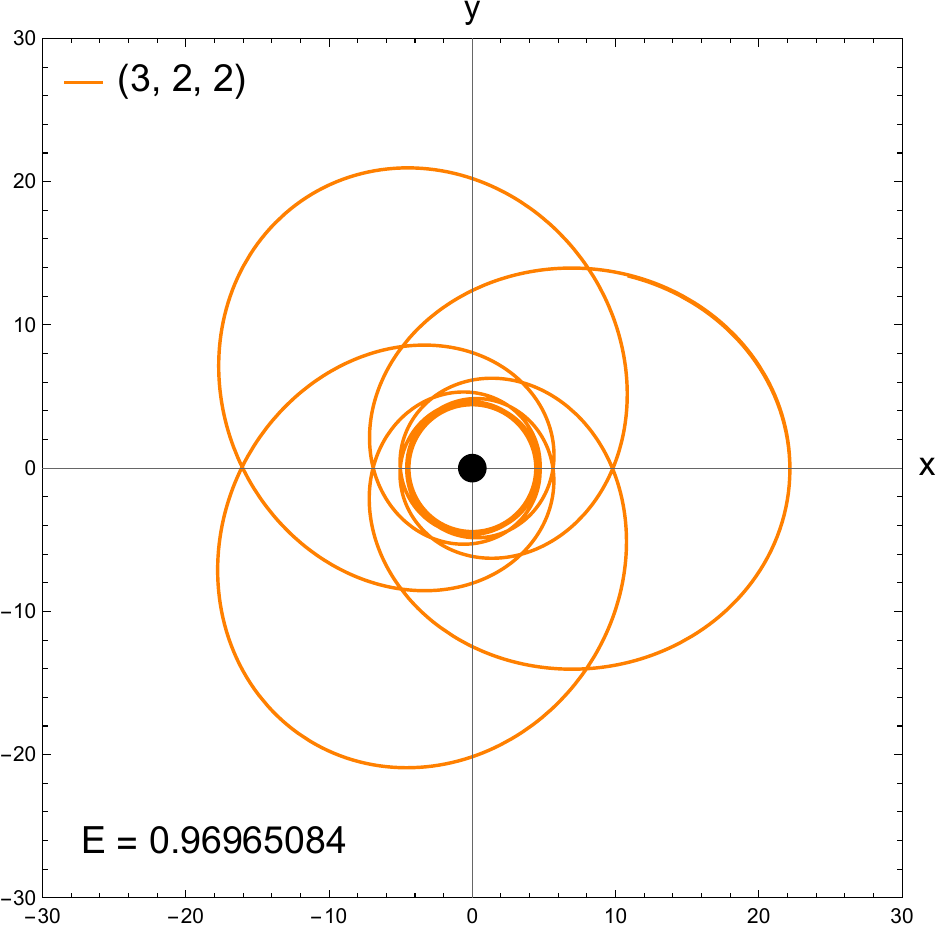} \hfill
    \includegraphics[width=0.32\textwidth]{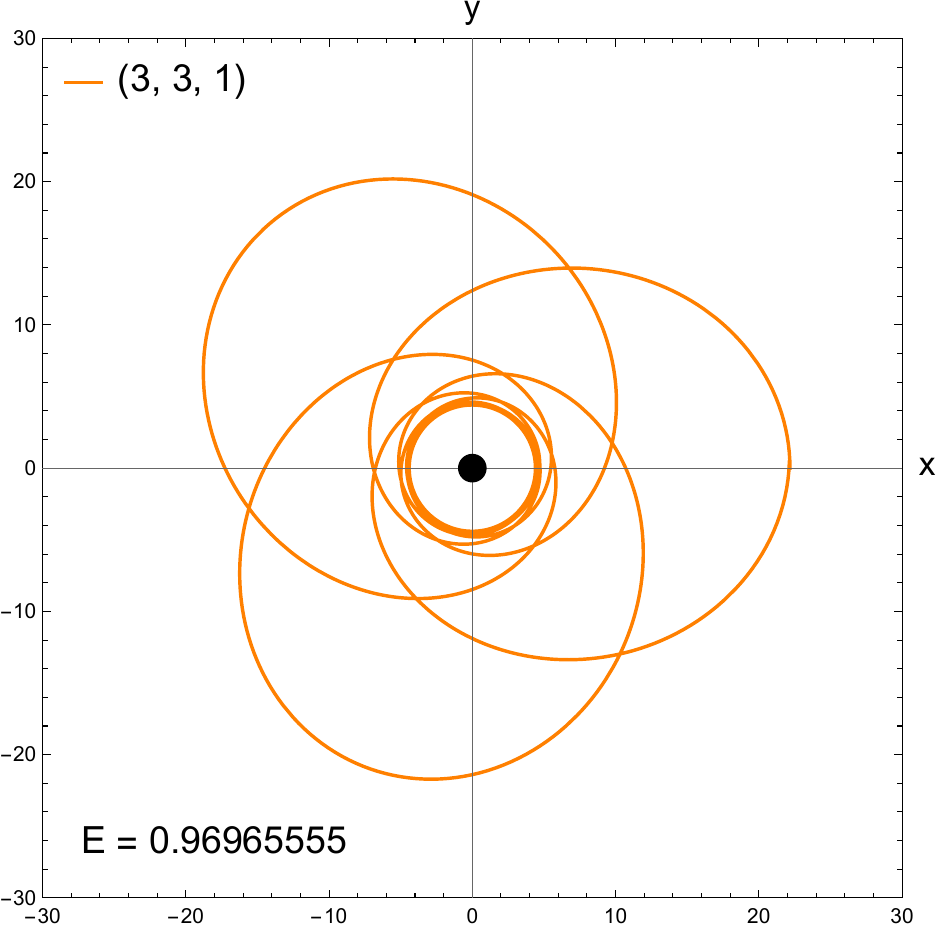} \\
    
    \vspace{0.2cm} 
    \includegraphics[width=0.32\textwidth]{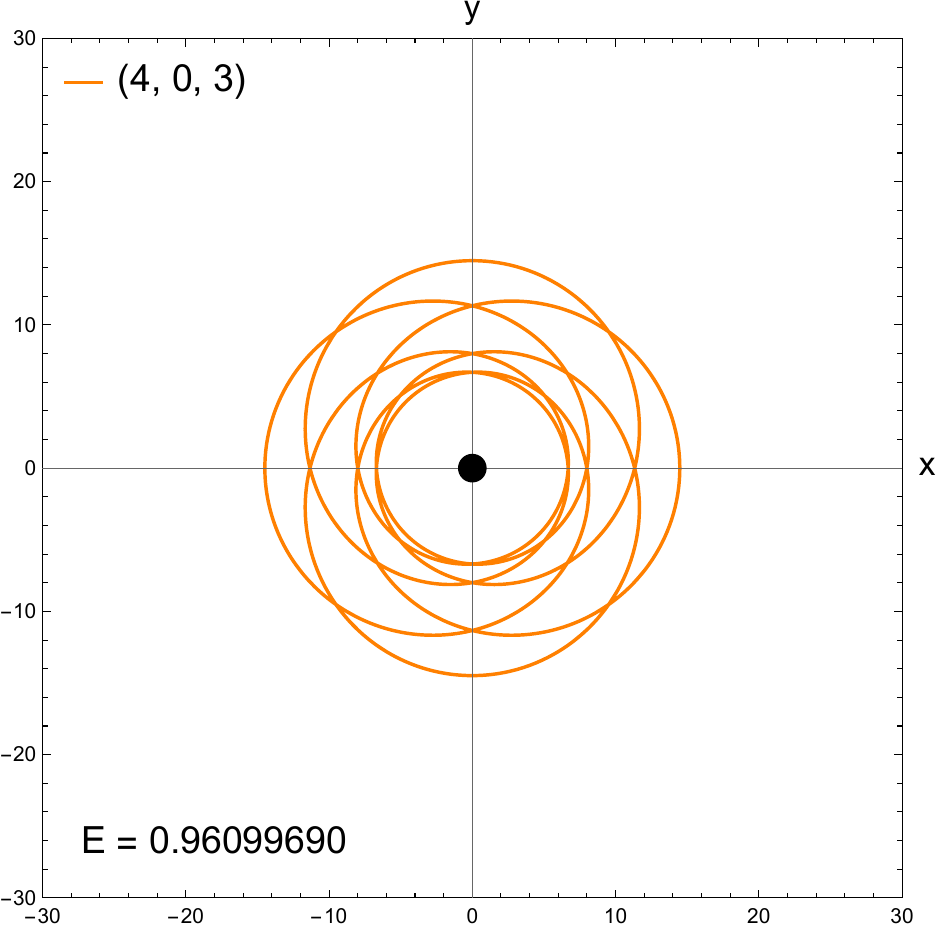} \hfill
    \includegraphics[width=0.32\textwidth]{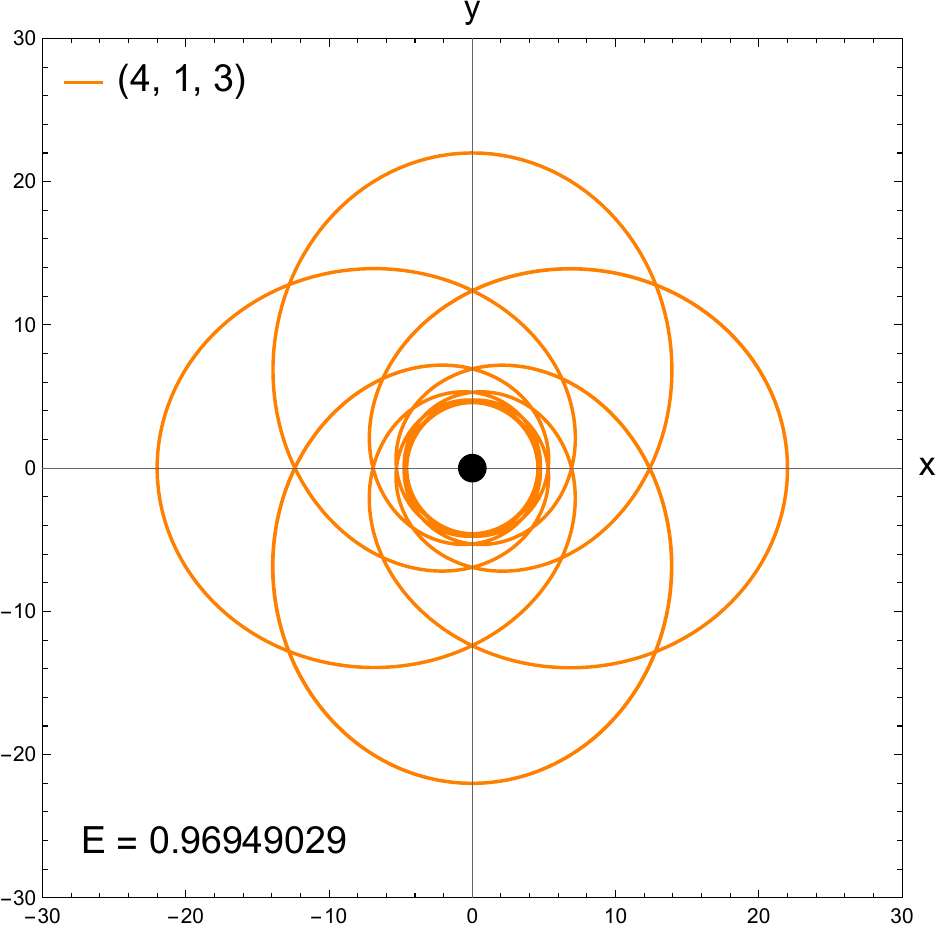} \hfill
    \includegraphics[width=0.32\textwidth]{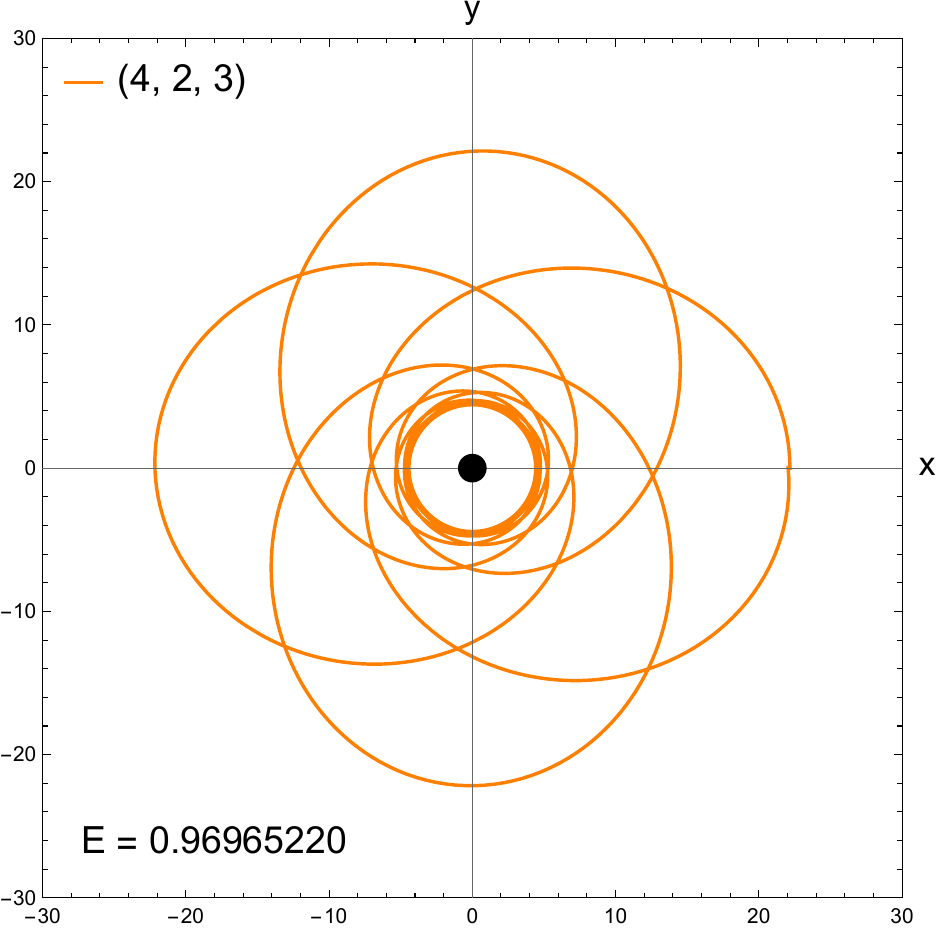}
    
    \caption{The figure demonstrates the periodic orbits for different $(z,w,v)$ around the Schwarzschild-BR BH in the case in which the magnetic field $B = 0.02$ and $L=\frac{1}{2}(L_{MBO}+L_{ISCO})$.}
    \label{fig:periodic}
\end{figure*}

\begin{figure*}[htbp]
    \centering
    \includegraphics[width=0.32\textwidth]{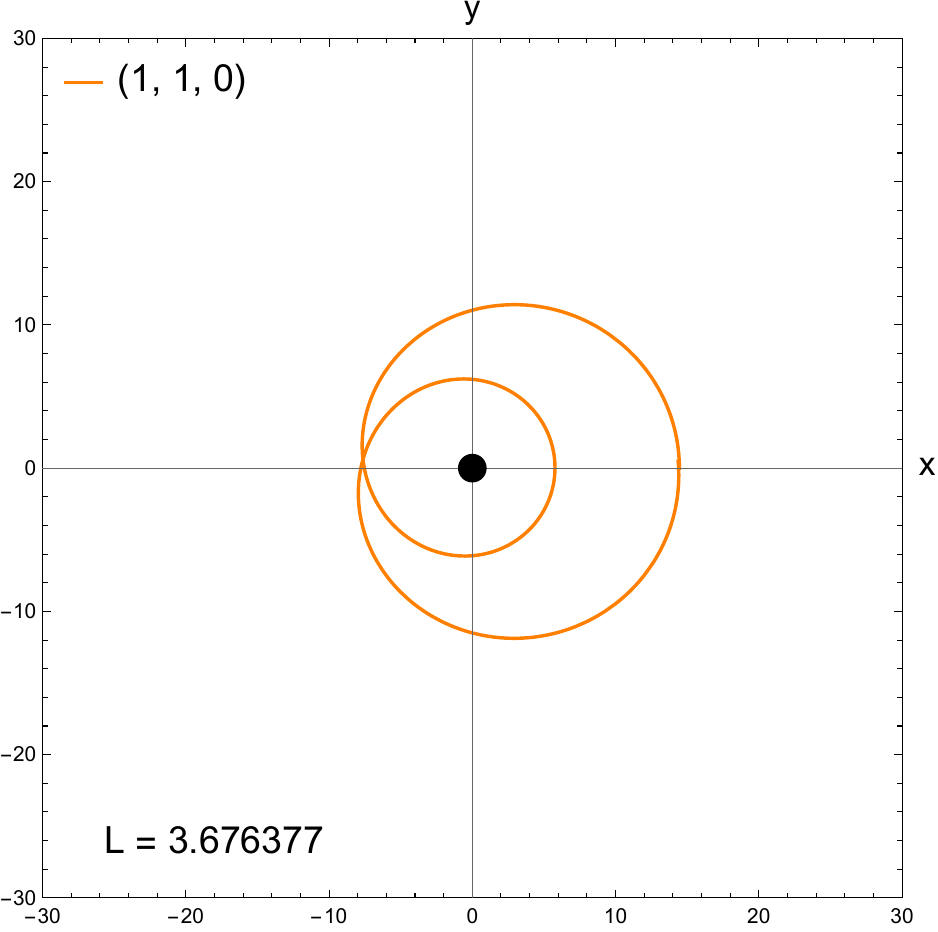} \hfill
    \includegraphics[width=0.32\textwidth]{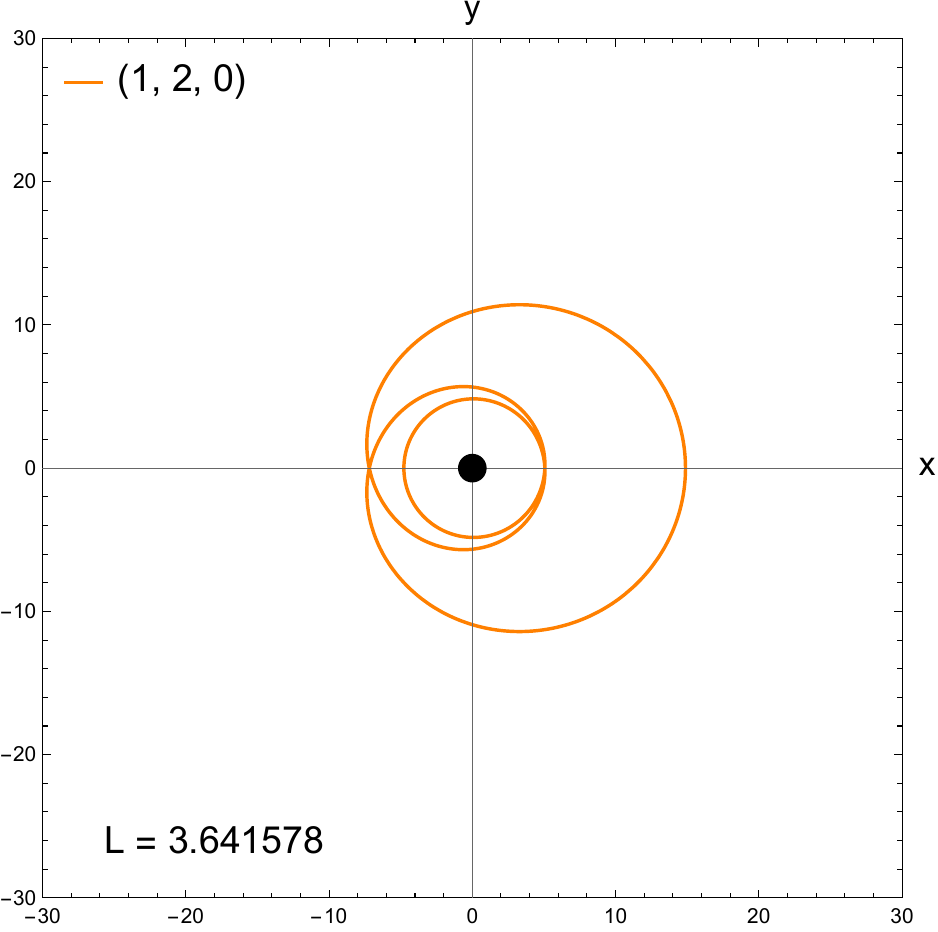} \hfill
    \includegraphics[width=0.32\textwidth]{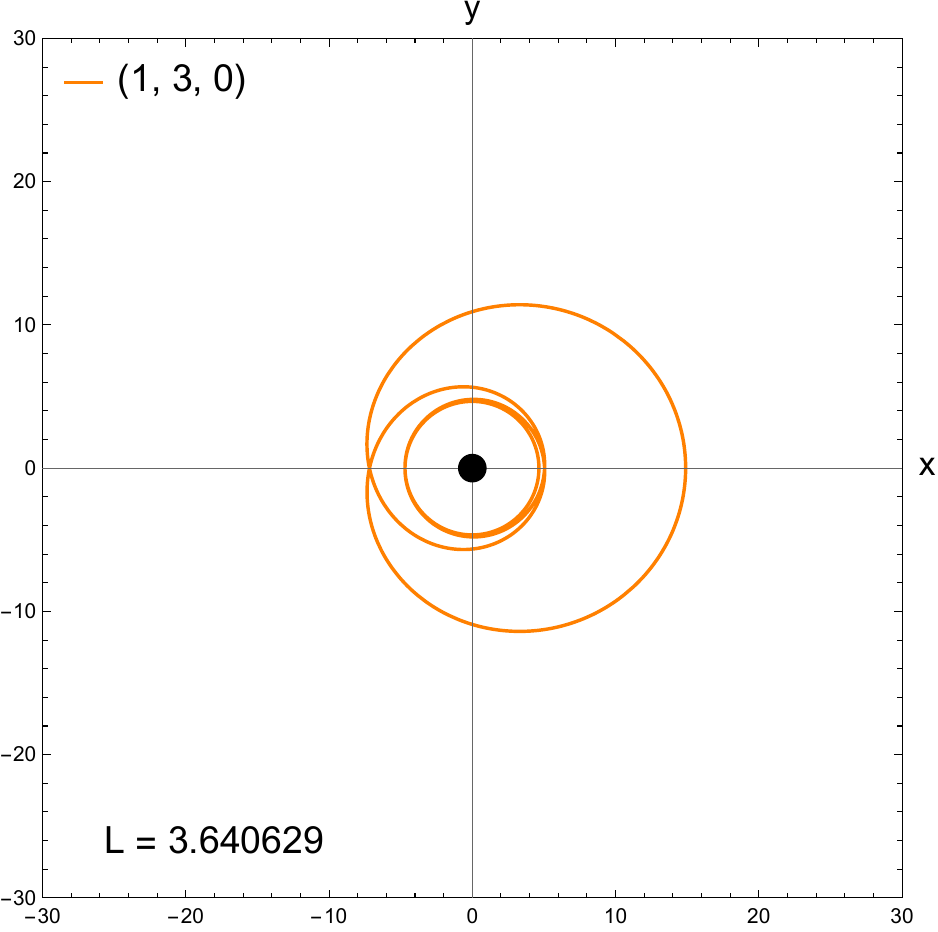} \\
    
    \vspace{0.2cm} 
    \includegraphics[width=0.32\textwidth]{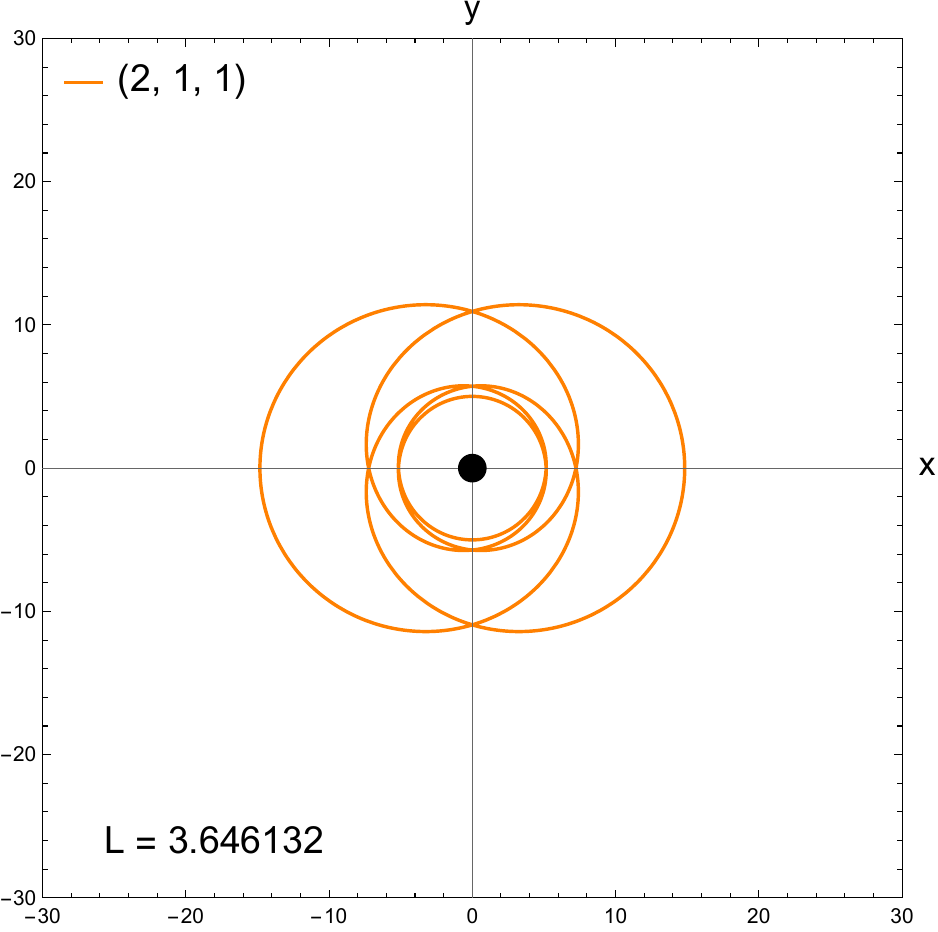} \hfill
    \includegraphics[width=0.32\textwidth]{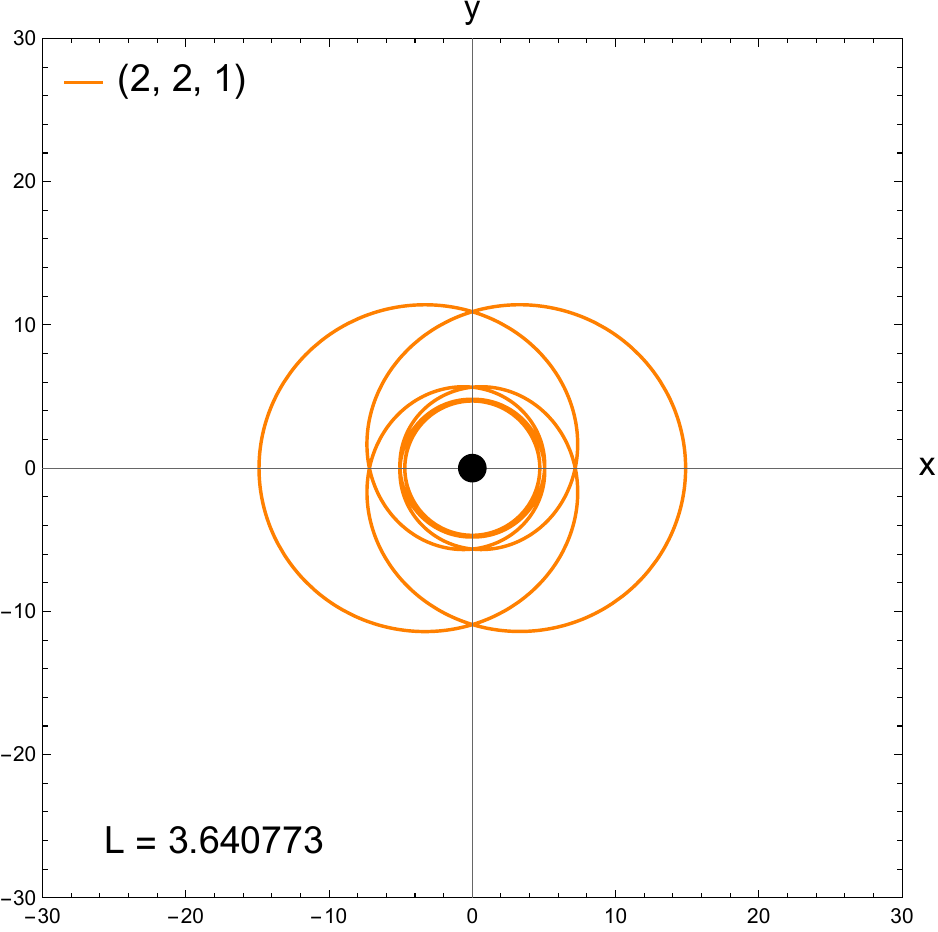} \hfill
    \includegraphics[width=0.32\textwidth]{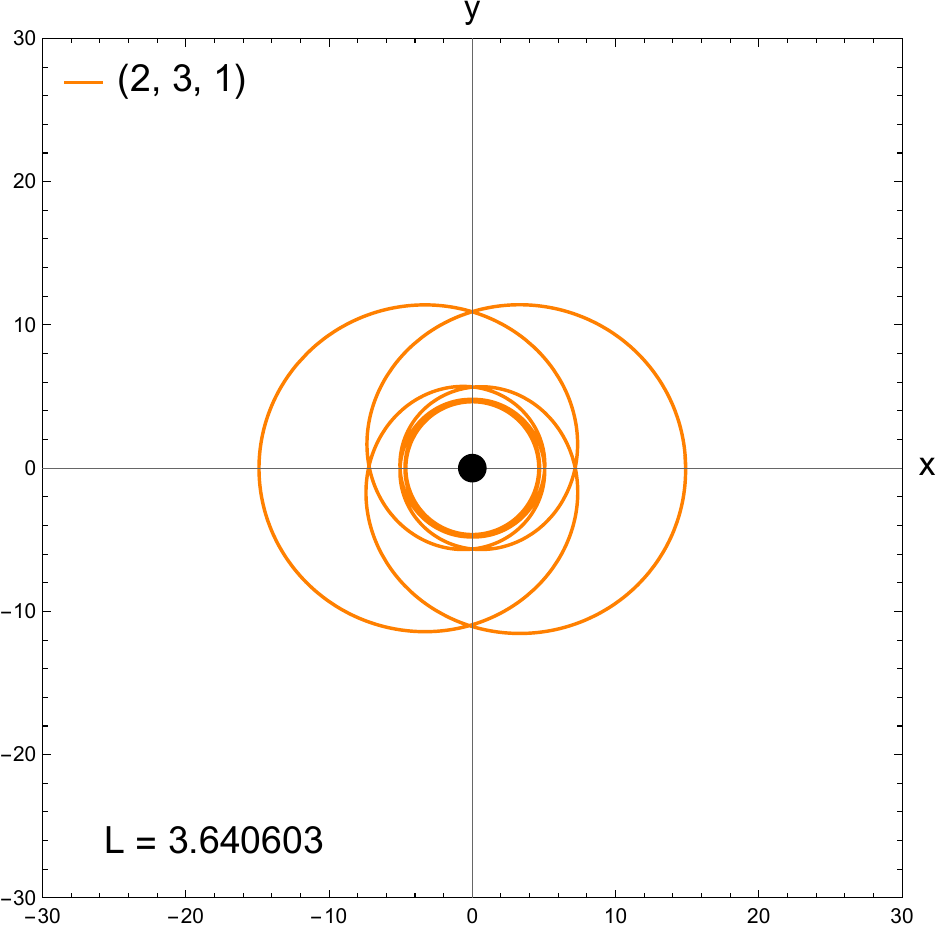} \\
    
    \vspace{0.2cm} 
    \includegraphics[width=0.32\textwidth]{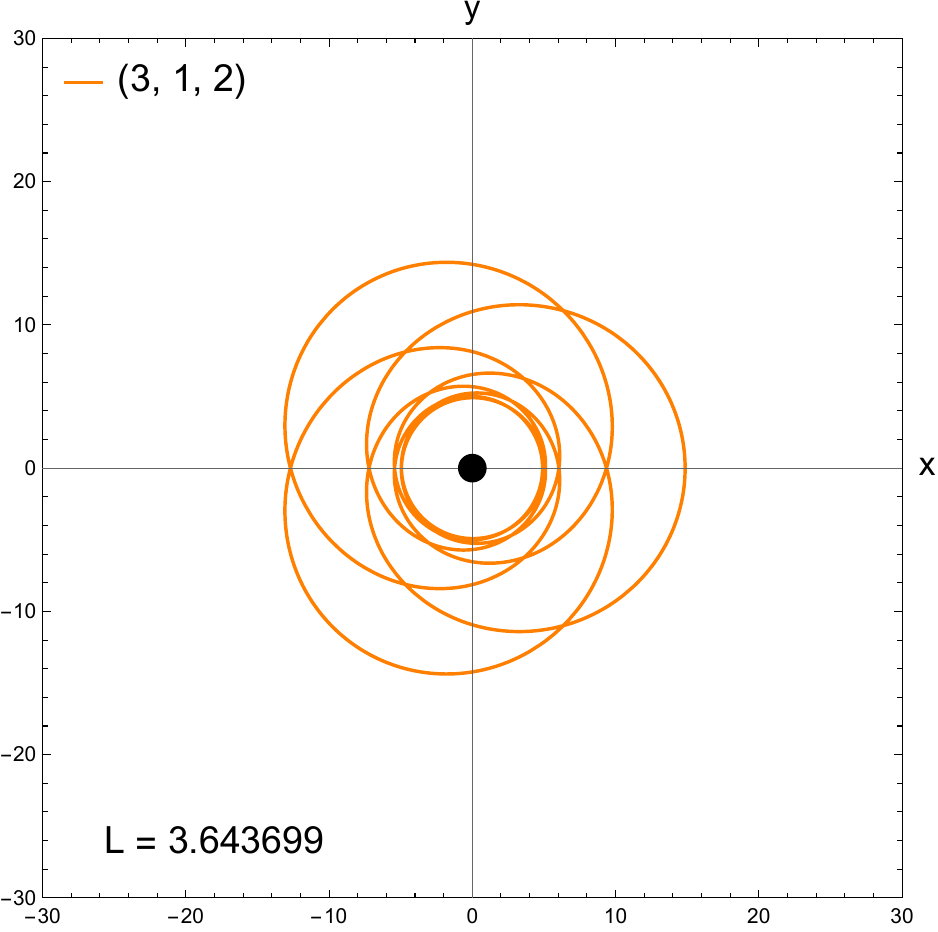} \hfill
    \includegraphics[width=0.32\textwidth]{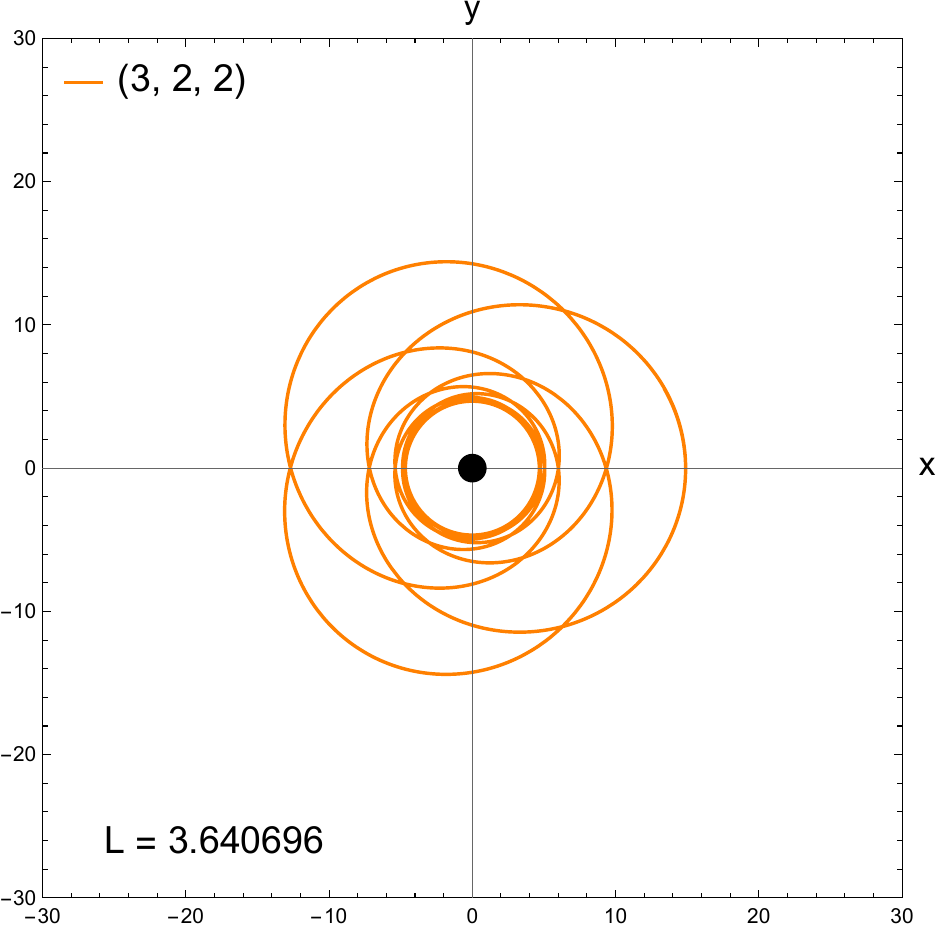} \hfill
    \includegraphics[width=0.32\textwidth]{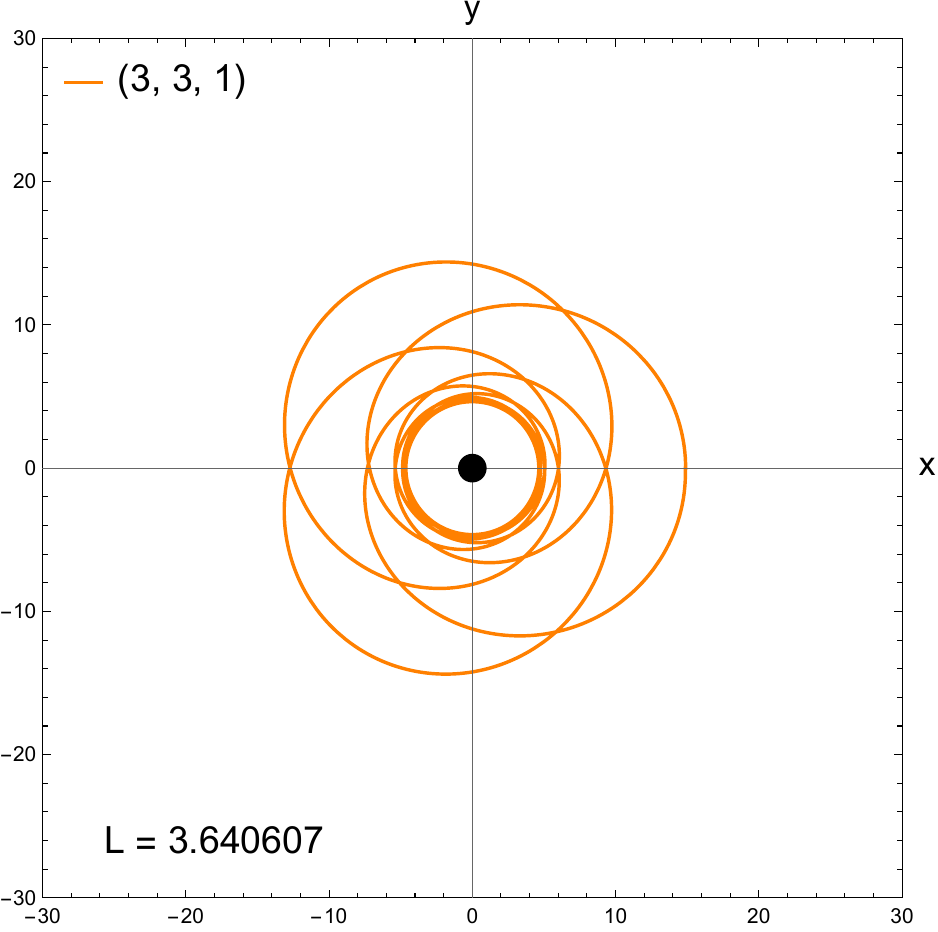} \\
    
    \vspace{0.2cm} 
    \includegraphics[width=0.32\textwidth]{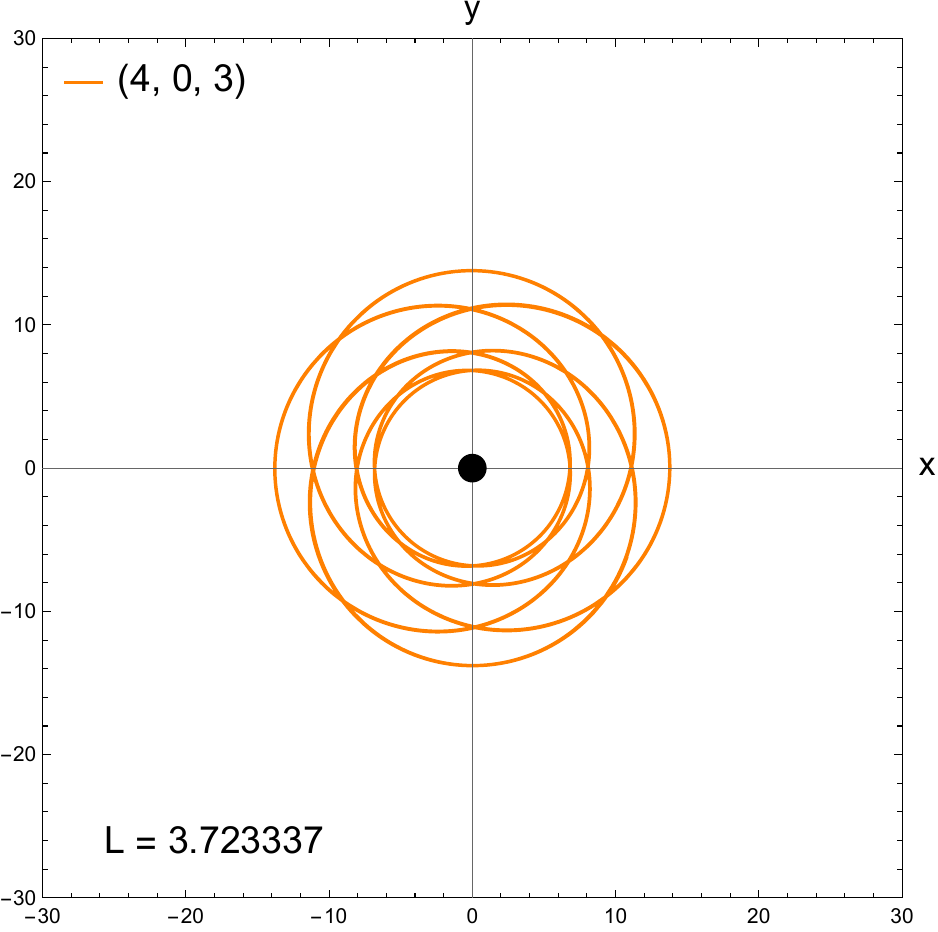} \hfill
    \includegraphics[width=0.32\textwidth]{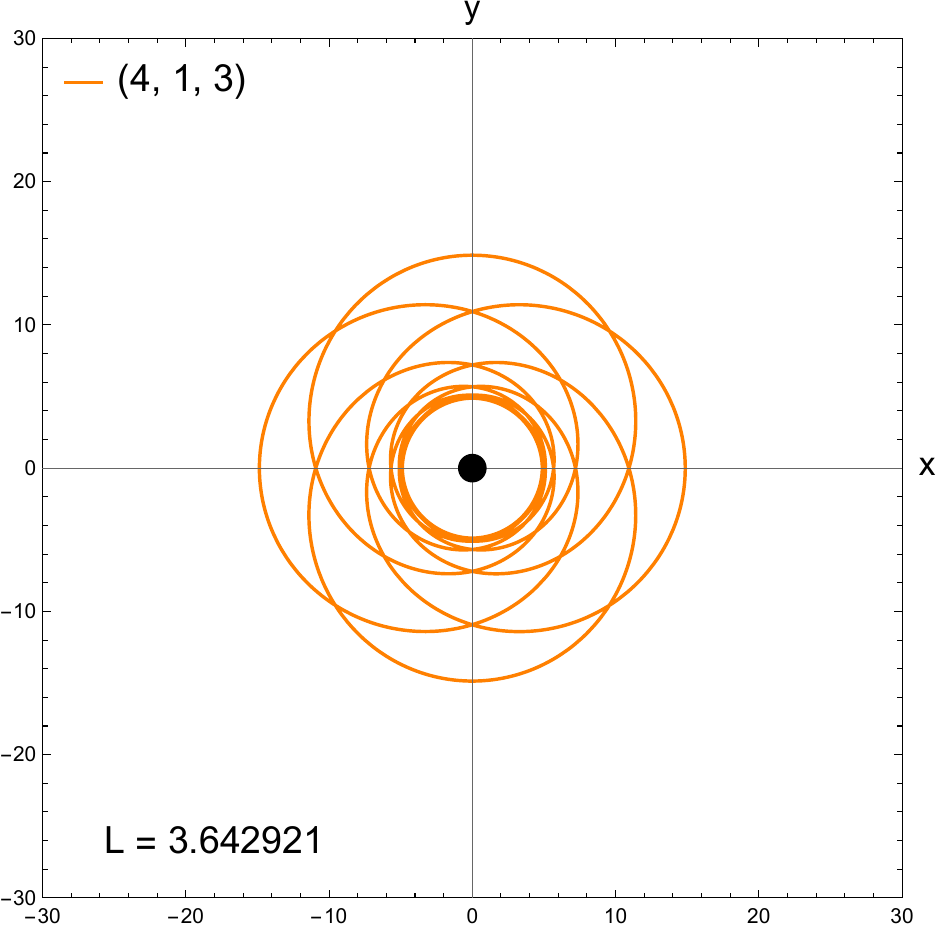} \hfill
    \includegraphics[width=0.32\textwidth]{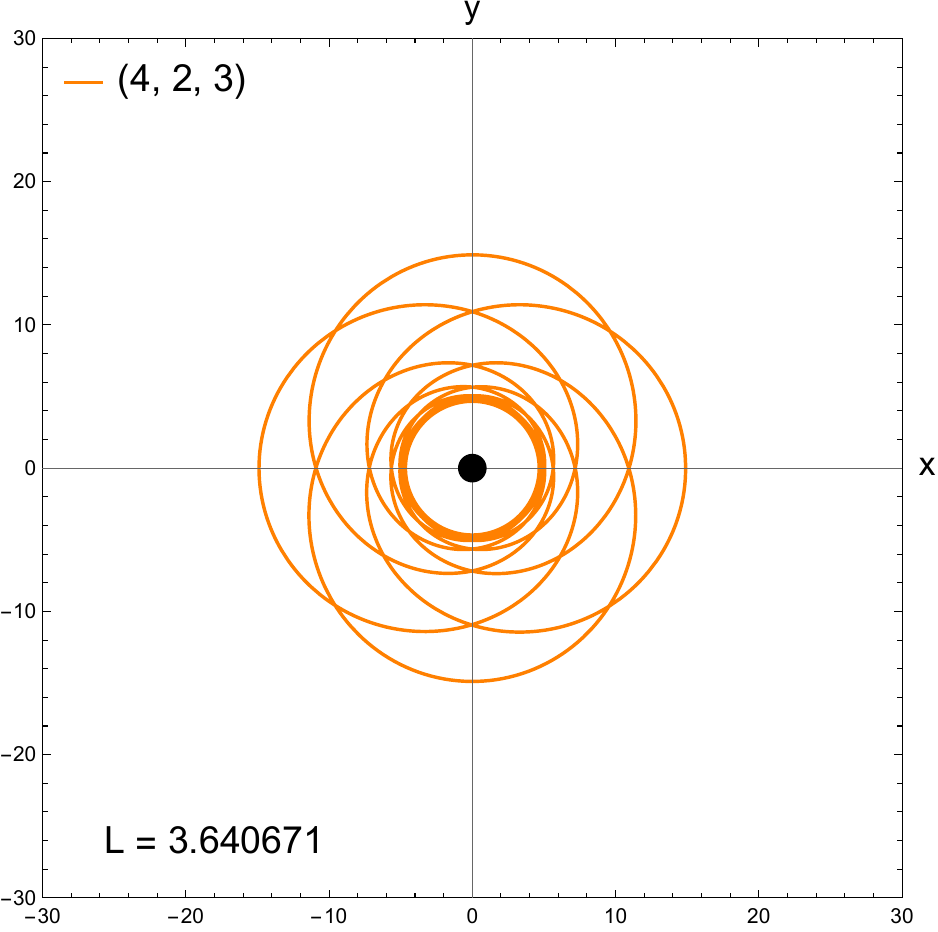}
    
    \caption{The figure demonstrates the periodic orbits for different $(z,w,v)$ around the Schwarzschild-BR BH in the case in which the magnetic field $B=0.02$ and $E=0.96$.}
    \label{fig:periodicL}
\end{figure*}

Furthermore, We numerically obtain the energies $E$ of periodic orbits with different $(z,w,v)$ at a fixed orbital angular momentum $L=\frac{1}{2}(L_{\rm MBO}+L_{\rm ISCO})$, as summarized in Table~\ref{table1}. Using these values, the corresponding periodic orbits around the Schwarzschild-BR BH are shown in Fig.~\ref{fig:periodic} for $B=0.02$. In addition, Table~\ref{table2} lists the orbital angular momentum $L$ for different periodic orbits at $E=0.96$ and $B=0.02$. These results are then used to construct the periodic orbits for various $(z,w,v)$, as presented in Fig.~\ref{fig:periodicL}. As seen from the figure, larger values of $z$ lead to more complex orbital structures, whereas larger $w$ corresponds to a greater number of revolutions between successive apoapses.
\begin{figure*}[htbp]
\begin{subfigure}[b]{0.35\textwidth}
\includegraphics[width=\textwidth]{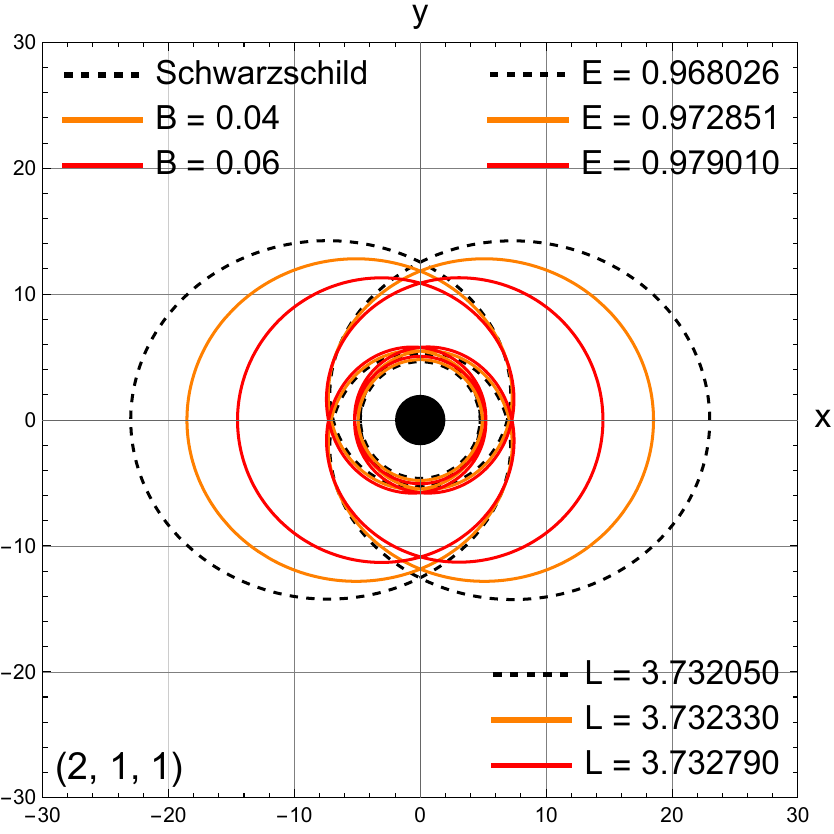}
\end{subfigure}
\begin{subfigure}[b]{0.64\textwidth}
\includegraphics[width=\textwidth]{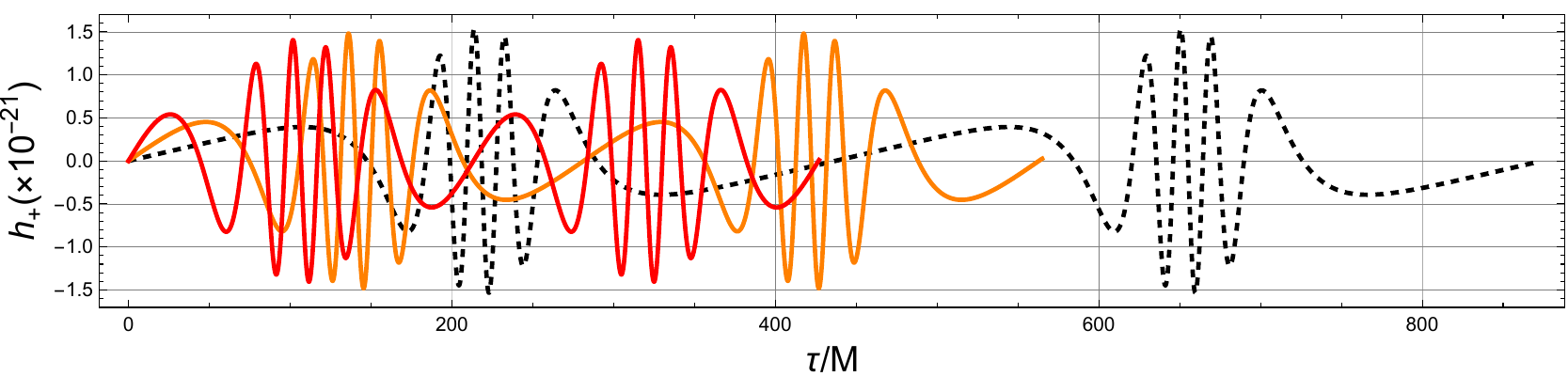}
\includegraphics[width=\textwidth]{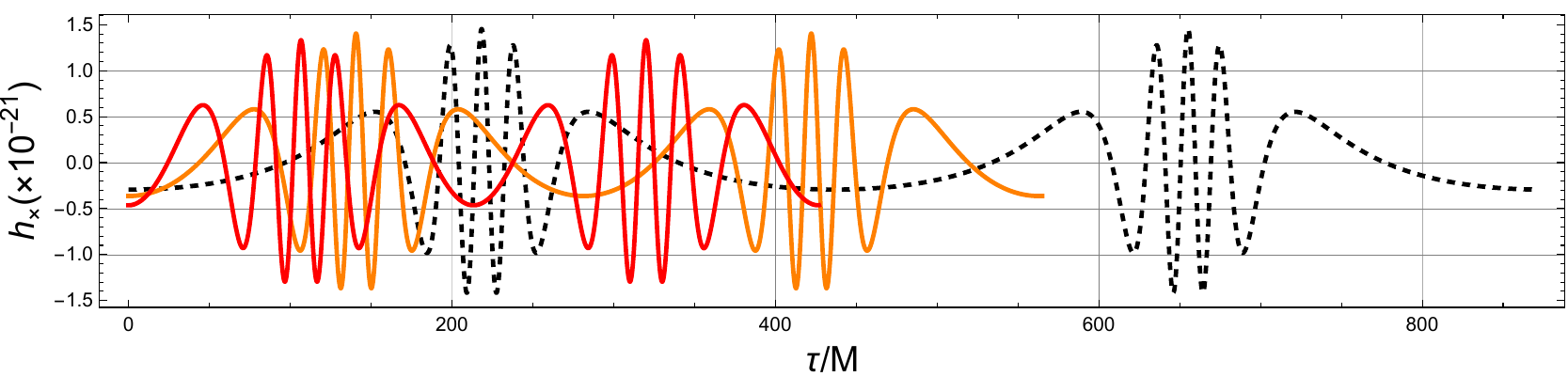}    
\end{subfigure} 
\begin{subfigure}[b]{0.35\textwidth}
\includegraphics[width=\textwidth]{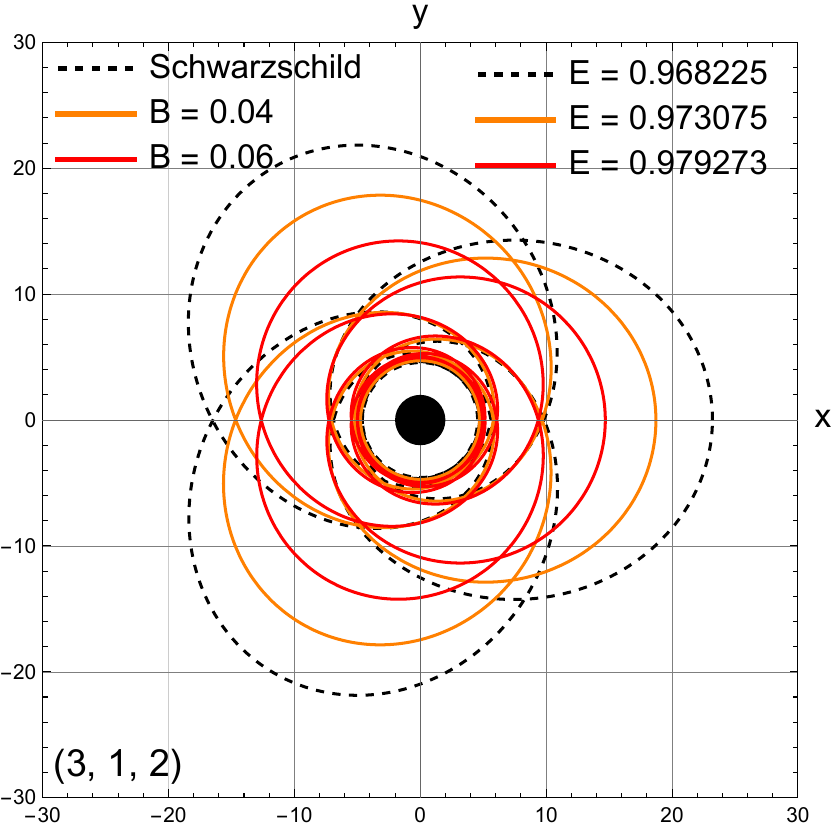}
\end{subfigure}
\begin{subfigure}[b]{0.64\textwidth}
\includegraphics[width=\textwidth]{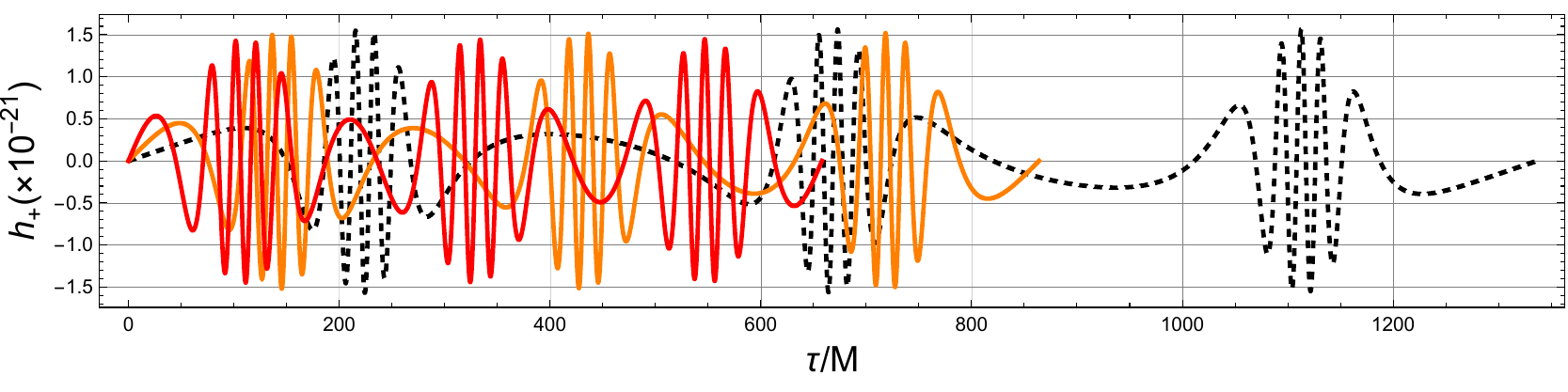}
\includegraphics[width=\textwidth]{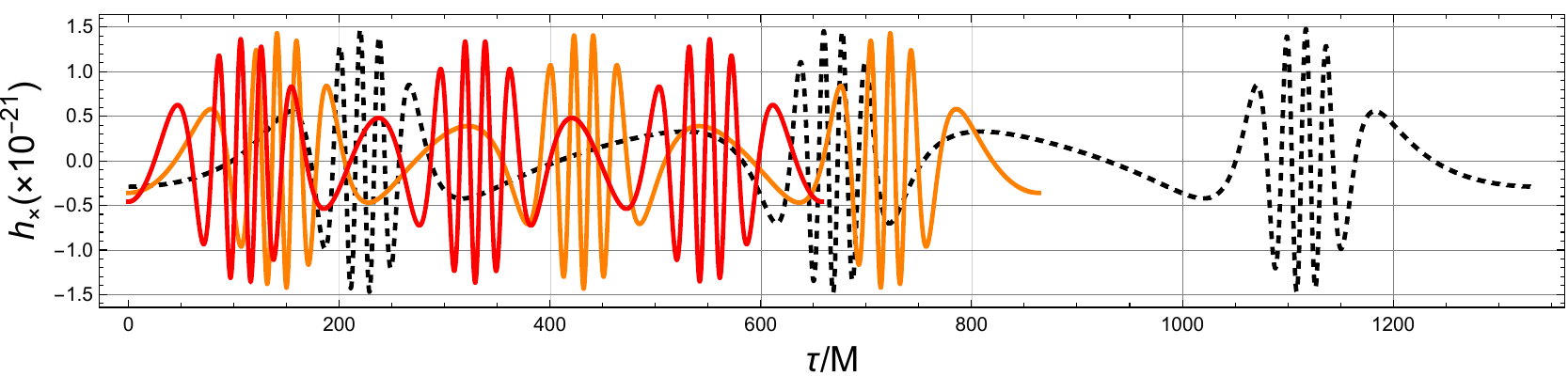}    
\end{subfigure} 
\begin{subfigure}[b]{0.35\textwidth}
\includegraphics[width=\textwidth]{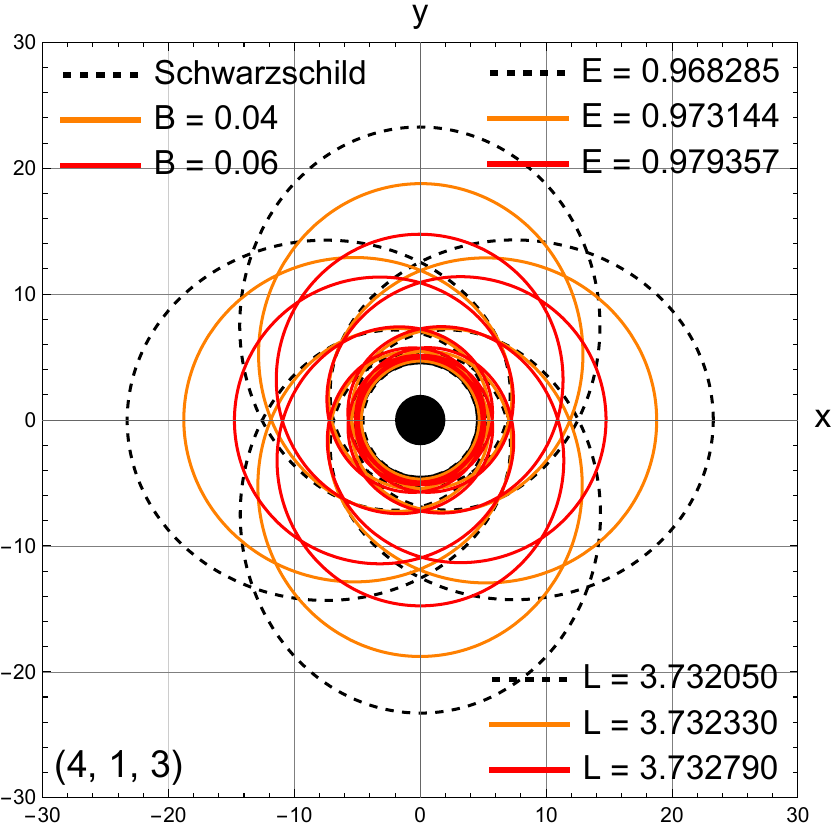}
\end{subfigure}
\begin{subfigure}[b]{0.64\textwidth}
\includegraphics[width=\textwidth]{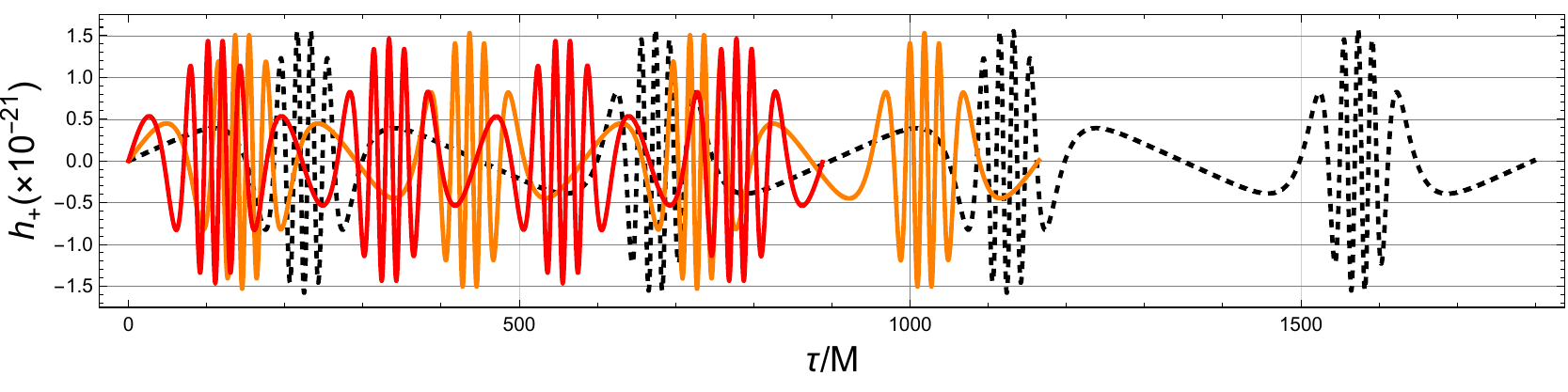}
\includegraphics[width=\textwidth]{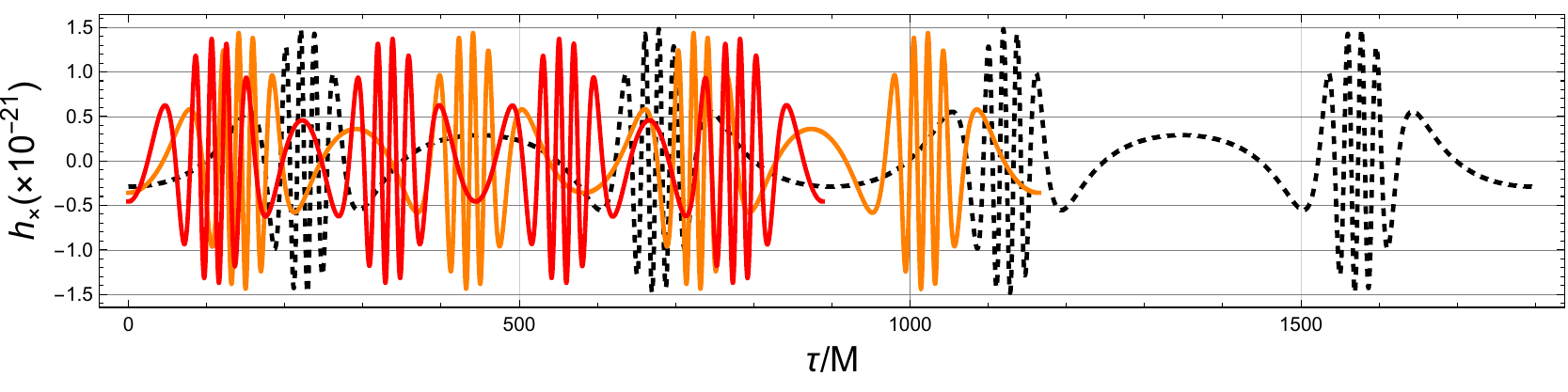}    
\end{subfigure} 
\caption{Periodic orbits and the corresponding gravitational waveforms for EMRI system consisting of a small object orbiting a supermassive Schwarzschild--BR BH for different values of the magnetic field $B$. The masses of the supermassive BH and the small object are of order $M \sim 10^{6} M_{\odot}$ and $m \sim 10 M_{\odot}$, respectively.}
\label{fig:waveforms}
\end{figure*}
\begin{figure*}[htbp]
\centering
 \includegraphics[scale=0.3]{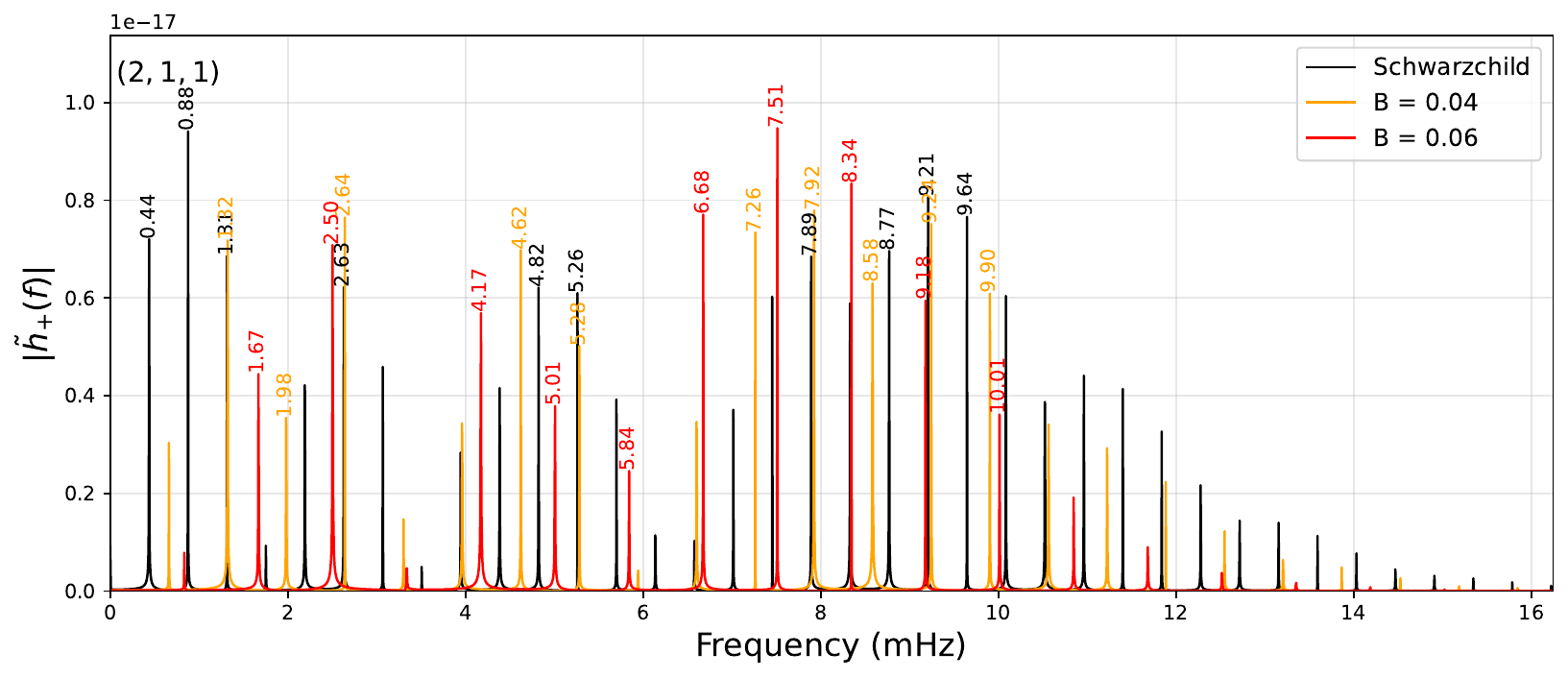}
 \includegraphics[scale=0.3]{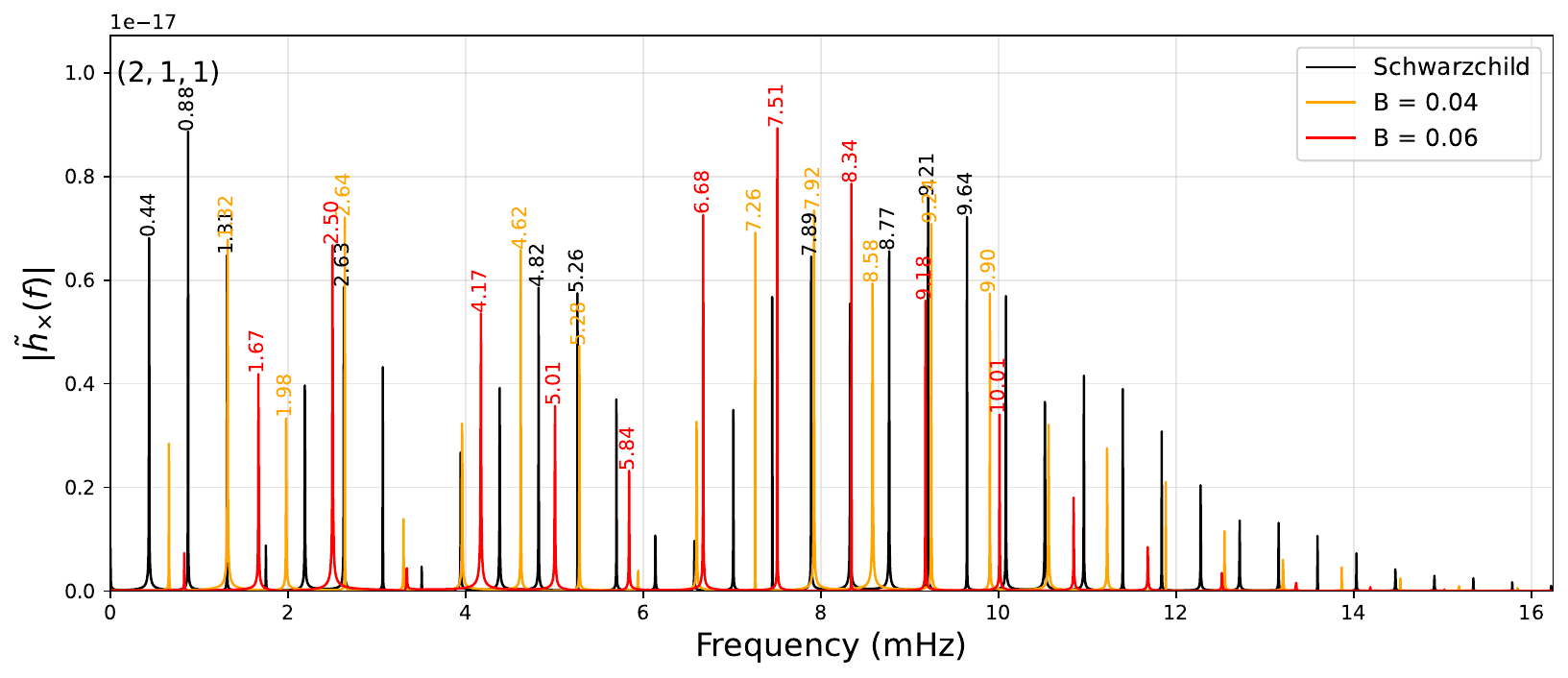}
 \includegraphics[scale=0.3]{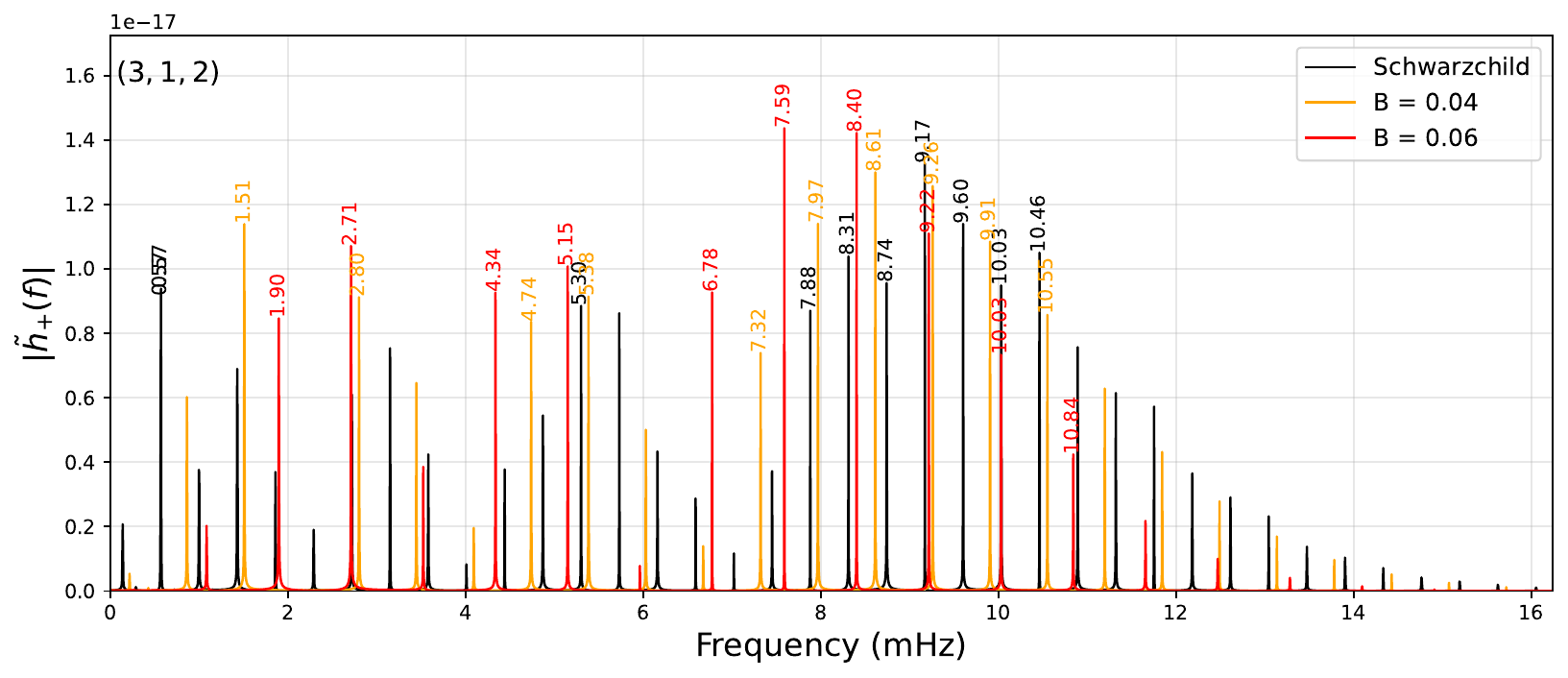}
 \includegraphics[scale=0.3]{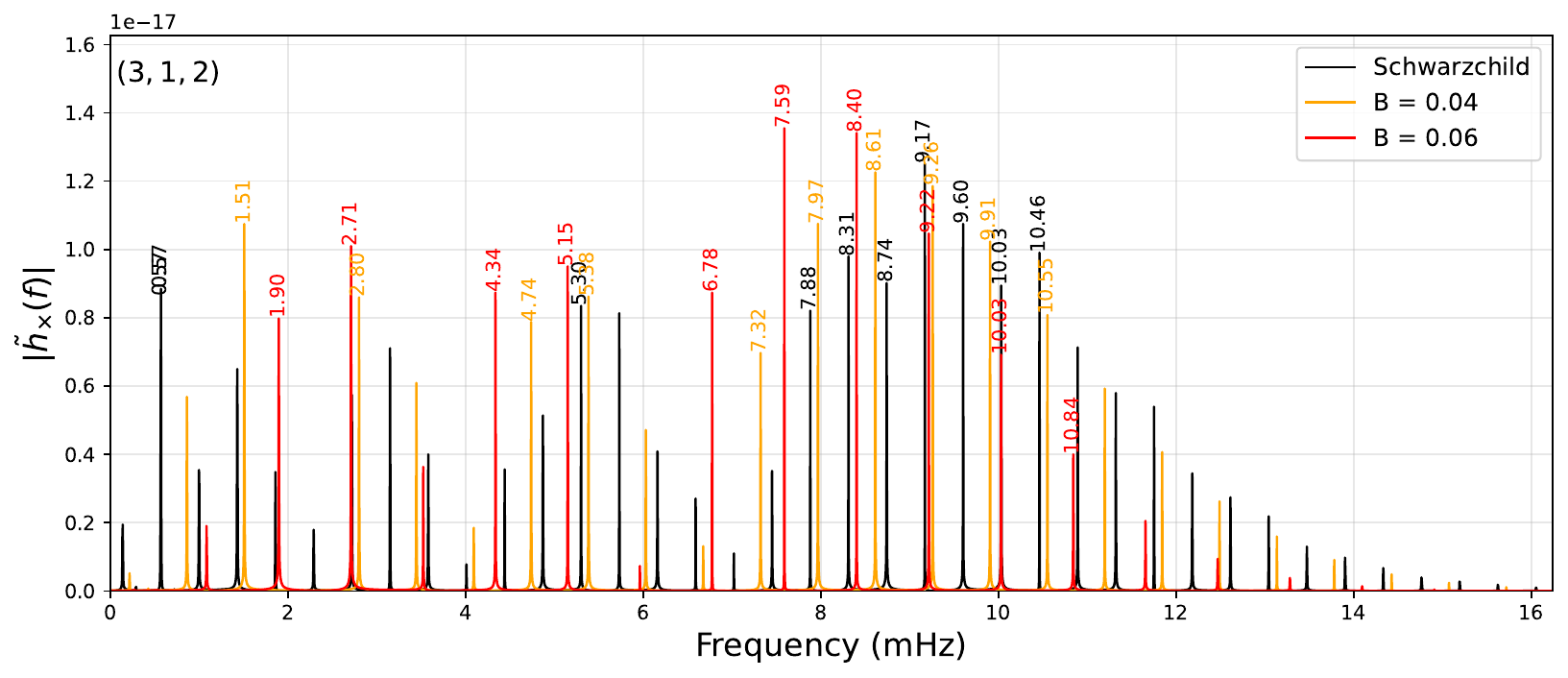}
  \includegraphics[scale=0.3]{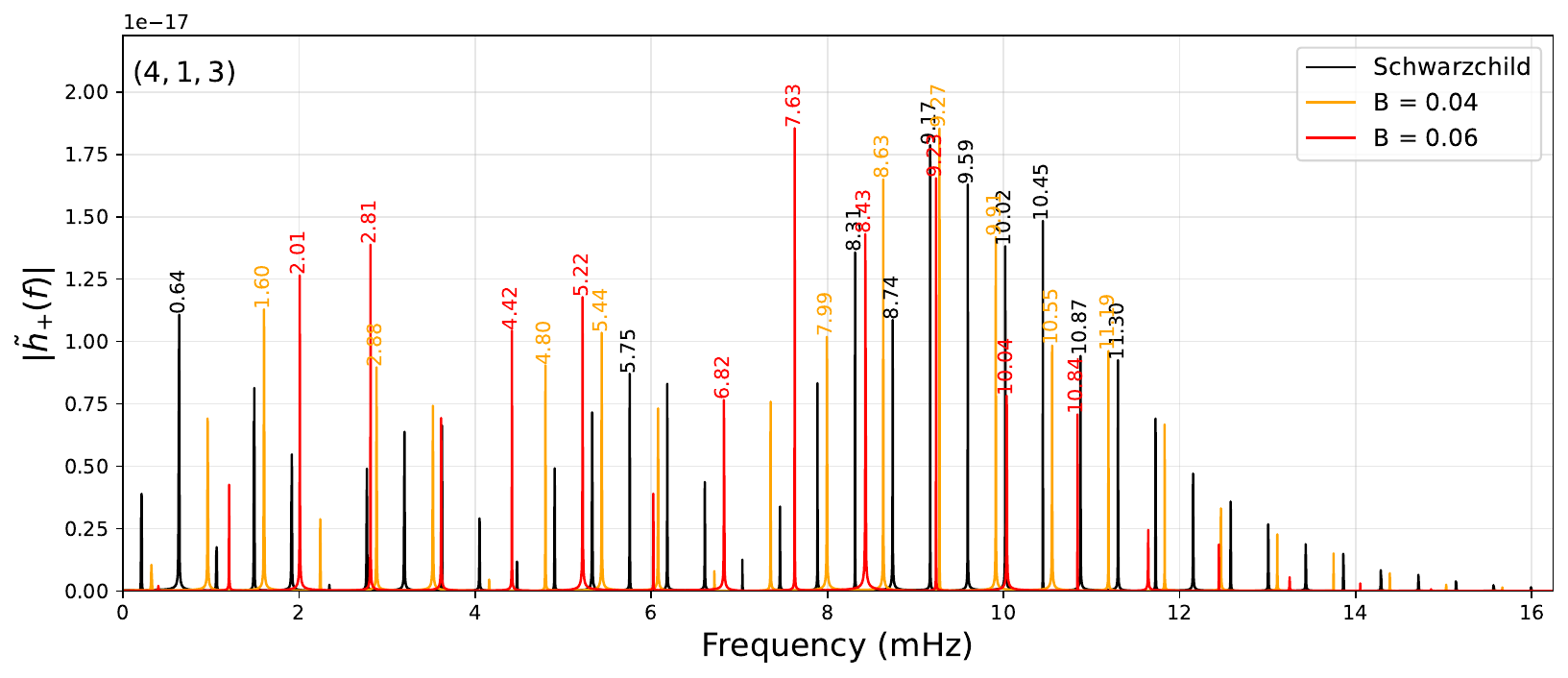}
 \includegraphics[scale=0.3]{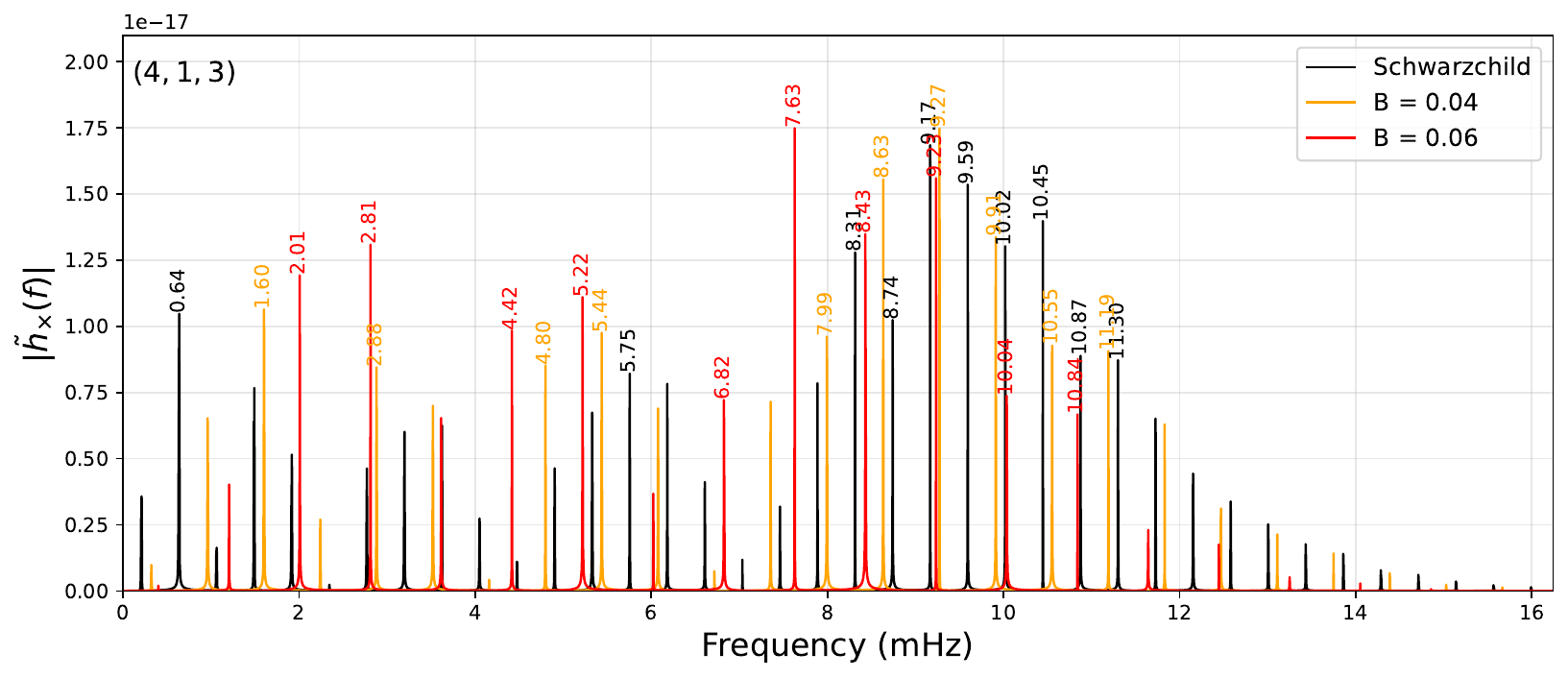}
 \caption{The frequency spectra of different waveforms corresponding to $(2, 1, 1)$, $(3, 1, 2)$ and $(4, 1, 3)$ orbits for different values of the magnetic field $B$.}
 \label{fig:spectrum}
\end{figure*}
\begin{figure*}[htbp]
\centering
 \includegraphics[scale=0.27]{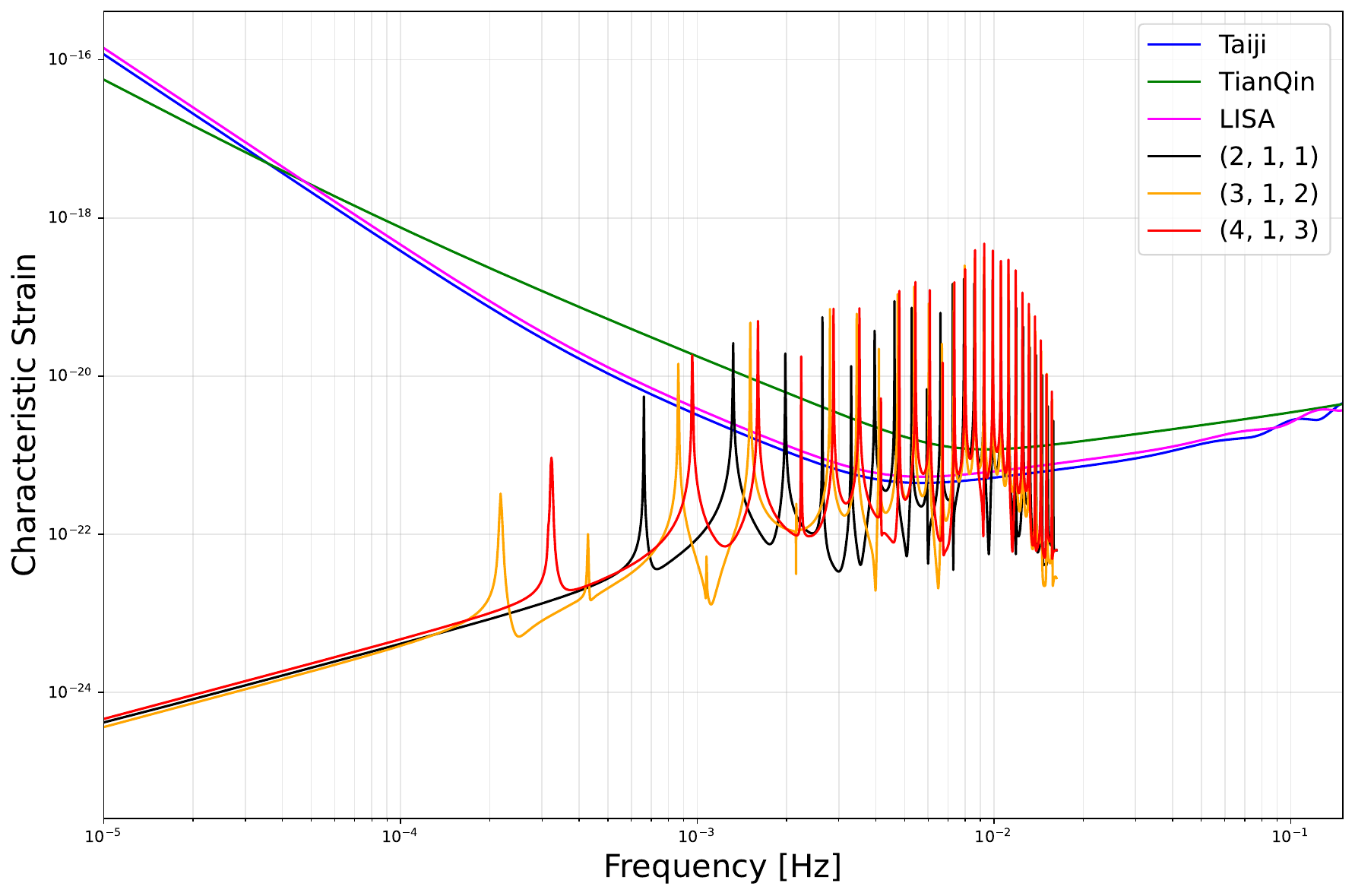}
 \includegraphics[scale=0.27]{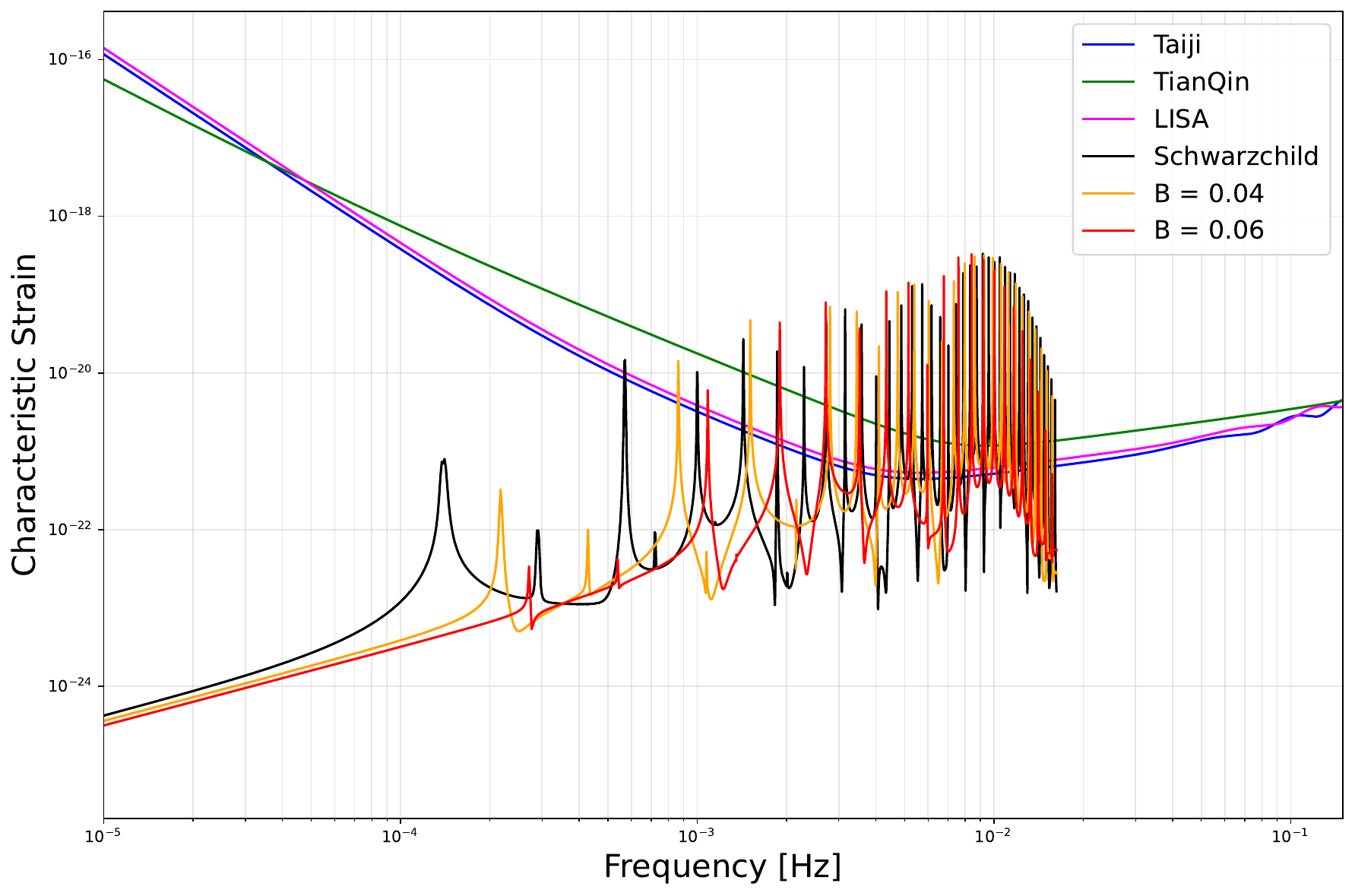}
 \caption{The characteristic strain of GWs compared with the sensitivity curves of space-based detectors, such as LISA, Taiji, and TianQin for different periodic orbits at the fixed $B=0.04$ (left panel) and for the $(3,1,2)$ orbit at different magnetic field $B$ strengths (right panel).}
 \label{fig:strain}
\end{figure*}

\section{NUMERICAL KLUDGE GRAVITATIONAL WAVEFORMS FROM PERIODIC ORBITS}\label{sec4}

In this section, we explore the gravitational waveforms from periodic orbits around a Schwarzschild–BR BH. We focus on an extreme mass-ratio inspiral (EMRI) system, where a stellar-mass compact object follows periodic trajectories around a supermassive BH. EMRIs are regarded as some of the most promising gravitational-wave sources for next-generation space-based detectors, including LISA, Taiji, and TianQin. The GWs emitted by these systems carry detailed information about the strong-field dynamical processes and the spacetime structure in the close vicinity of the central BH. For further analysis, we shall adopt the adiabatic approximation for EMRI waveform calculations, which is valid when the inspiral timescale is much longer than the orbital period and is widely used in EMRI waveform modeling. In this approximation, the loss of energy and angular momentum of the smaller object due to gravitational radiation occurs slowly and can be treated as negligible over many orbits, allowing the motion to be modeled as a geodesic in the spacetime of the supermassive BH. 

GWs emitted by periodic orbits around the central BH are calculated using a numerical “kludge” waveform model. In this approach, the orbital motion of the smaller object is first determined by numerically solving the geodesic equations of motion, which are given in Eqs.~\eqref{eq:conslaw} and \eqref{eq:radialeq} in the Schwarzschild-BR BH spacetime. The resulting trajectory is then used as input to the quadrupole formula for gravitational radiation to generate the corresponding waveform. This method provides a practical framework for obtaining approximate GW signals from EMRI systems and allows for an initial exploration of their dependence on the orbital dynamics and the underlying spacetime geometry. In the quadrupole approximation, the GW metric perturbation
$h_{ij}$ produced by a symmetric trace-free (STF) mass quadrupole moment $I_{ij}$
is given by (see \cite{2025JCAP...01..091Y})
\begin{equation}
h_{ij} = \frac{2G}{c^4D_L}\,\ddot{I}_{ij}\, .
\end{equation}
where $D_{L}$ is the luminosity distance to the source. For a point-like object of mass $m$, the symmetric trace-free (STF) mass quadrupole moment is given by
\begin{equation}
I^{ij}
=
\left[
\int d^{3}x \, x^{i} x^{j} \, T^{tt}(t,x^{i})
\right]^{\mathrm{STF}}\, ,
\end{equation}
where $T^{tt}$ denotes the $tt$ component of the stress-energy tensor for a point-like object moving along a trajectory $Z^{i}(t)$,
\begin{equation}
T^{tt}(t,x^{i}) = m \, \delta^{3}\!\left(x^{i} - Z^{i}(t)\right)\, .
\end{equation}
Here, the trajectory $Z^{i}(t)$ is obtained by numerically solving the geodesic
equations of motion in the background spacetime. We describe the motion of the point-like object orbiting the Schwarzschild-BR BH using Boyer-Lindquist coordinates $(r, \theta, \phi)$. The quadrupole formula for the EMRI system ($M \gg m$) in this coordinate system is given by~\cite{2025EPJC...85...36Z}
\begin{equation}\label{eq:perturbation}
h_{ij}=\frac{4\mu M}{D_{L}}\left(v_{i}v_{j}-\frac{M}{r}n_{i}n_{j}\right)\, .
\end{equation}
where $m$ and $M$ refer to the masses of the smaller object and the BH, respectively, while $\mu=Mm/(M+m)^{2}$ refers to the symmetric mass ratio of the system. Here, $v_i$ denotes the $i$ th component of the velocity vector of the smaller object with mass $m$, describing its orbital motion, whereas $n_i$ is the normalized unit radial vector directed from the BH toward the smaller object.
To simplify GW data analysis, we use a Cartesian coordinate system $(x,y,z)$ centered on the BH. The transformation from Boyer--Lindquist to Cartesian coordinates is given by
\begin{equation} 
x = r \sin\theta \cos\phi, \quad
y = r \sin\theta \sin\phi, \quad
z = r \cos\theta\,  .
\end{equation}
We then introduce a detector-adapted coordinate system $(X,Y,Z)$ whose origin coincides with that of the $(x,y,z)$ coordinate system, with both frames centered on the supermassive BH~\cite{Poisson_Will_2014,2025JCAP...01..091Y,2025EPJC...85...36Z}. The directions of the detector-adapted coordinate axes in the original $(x,y,z)$ coordinates are expressed as
\begin{equation}\label{xyz}
\begin{array}{l}
e_{X}=[\cos\zeta,-\sin\zeta,0]\, ,\\
e_{Y}=[\cos\iota\sin\zeta,\cos\iota\cos\zeta,-\sin\iota]\, ,\\
e_{Z}=[\sin\iota\sin\zeta,\sin\iota\cos\zeta,\cos\iota]\, .
\end{array}
\end{equation}
where, $\iota$ corresponds to the inclination angle between the orbital plane of the smaller object and the $X$-$Y$ plane, while $\zeta$ denotes the longitude of the pericenter defined in the orbital plane.  It is important to note that the vectors (i.e., $e_X$, $e_Y$, and $e_Z$) refer to the orthogonal coordinate basis of the detector frame, defined in Cartesian coordinates. These basis vectors are used to decompose the GW signal into its polarazition components, specifically "plus" and "cross" modes. This decomposition is essential because it reveals how the detector responds to incoming GWs from varying directions and polarizations. These GW polarizations $h_{+}$ and $h_{\times}$ are then written as follows~\cite{2025JCAP...01..091Y}:
\begin{eqnarray}
 h_{+}&=&\frac{1}{2}(e^i_X e^j_X-e^i_Ye^j_Y)h_{ij}, \\
 h_{\times}&=&\frac{1}{2}(e^i_Xe^j_Y+e^i_Ye^j_X)h_{ij}\, .
\end{eqnarray}
Using Eqs.~\eqref{eq:perturbation} and \eqref{xyz}, the corresponding above GW  polarizations can then be rewritten as follows~\cite{2025EPJC...85...36Z,2024arXiv241101858M}:
\begin{equation} \label{eq:h_plus}
    h_{+}=-\frac{2\mu M^{2}}{D_{L}r}\left(1+\cos^{2}\iota\right)\cos\left(2\phi+2\zeta\right)\, ,
\end{equation}
\begin{equation}\label{eq:h_cross}
    h_{\times}=-\frac{4\mu M^{2}}{D_{L}r}\cos\iota\sin\left(2\phi+2\zeta\right)\, ,
\end{equation}
where the coordinates $(r,\phi)$ are determined by numerically solving the equations of motion in Boyer--Lindquist coordinates. To visualize the GW radiations emitted by a small object moving along periodic orbits around the Schwarzschild-BR BH, we consider the EMRI system with $M \sim 10^{6} M_{\odot}$ and $m \sim 10 M_{\odot}$, choose $\zeta=\iota=\pi/4$, and set the luminosity distance to $D_{L}=200\,\text{Mpc}$.

In Fig.~\ref{fig:waveforms}, we show the gravitational waveforms ($h_{+}$ and $h_{\times}$) radiated over a single orbital period by the $(2,1,1)$, $(3,1,2)$, and $(4,1,3)$ periodic orbits for different values of magnetic field $B$. From Fig.~\ref{fig:waveforms}, it is obvious that distinct zoom and whirl stages are evident in the waveforms. The slowly varying, low-amplitude portions of the waveforms originate from the zoom phase, when the small object follows a highly elliptical trajectory far from the BH. Conversely, the rapidly oscillating segments are produced during the whirl phase, as the object approaches the BH and undergoes quasi-circular motion, leading to a sharp increase in the gravitational-wave frequency. From the trajectories shown in the left panel of Fig.~\ref{fig:waveforms}, we observe that for larger magnetic field strengths, the particle in the zoom phase moves closer to the BH compared to the pure Schwarzschild case, while the particle in the whirl phase moves a little farther away. Moreover, the orbital period decreases with increasing $B$, whereas periodic orbits with larger zoom numbers $z$ have longer periods and exhibit repeated waveform patterns.

The frequency-spectra representation of the gravitational waveforms is obtained by performing discrete Fourier transforms on the time-domain GW signals shown in Fig.~\ref{fig:waveforms}, converting them into the frequency domain. The absolute values of the frequency spectra $\tilde{h}_{+,\times}(f)$ are presented in Fig.~\ref{fig:spectrum}. From the spectra in Fig.~\ref{fig:spectrum}, we find that GWs emitted by EMRIs with different periodic orbits have characteristic frequencies in the mHz range, which makes them detectable by space-based gravitational-wave detectors. From Fig.~\ref{fig:spectrum}, we observe that the spectral component with the largest amplitude shifts toward lower frequencies as the magnetic field strength $B$ increases. However, according to Eqs.~\eqref{eq:h_plus} and \eqref{eq:h_cross}, the gravitational-wave amplitudes $h_{+}$ and $h_{\times}$ are enhanced when the emitting source follows a smaller-radius orbit, which corresponds to the whirl phase of the periodic motion. As can be seen in Fig.~\ref{fig:waveforms}, increasing the magnetic field $B$ causes the outer (zoom) orbits to shift inward, while the inner (whirl) orbits move outward. Consequently, the orbital radius during the whirl phase increases, leading to a longer orbital period and thus a lower characteristic GW frequency. This behavior explains the observed shift of the dominant spectral peak toward lower frequencies with increasing magnetic field $B$.
The characteristic GW strain is given by
\begin{equation}\label{eq:strain}
h_c(f) = 2f \left( \left| \tilde{h}_{+}(f) \right|^2 
+ \left| \tilde{h}_{\times}(f) \right|^2 \right)^{1/2} \, .
\end{equation}
Following Eq.~\ref{eq:strain}, in Fig.~\eqref{eq:strain}, we show the characteristic strain of GWs emitted by periodic EMRI orbits for different orbital configurations $(z, w,v)$ and magnetic field strengths, and compare the results with the sensitivity curves of space-based GW detectors, such as LISA, Taiji, and TianQin. For all orbital configurations and magnetic field values considered, portions of the characteristic strain exceed the detector sensitivity curves, indicating that these EMRI signals  are potentially detectable by future space-based GW observatories.

\section{Conclusions}\label{summary}

In this work, we presented a detailed analysis of timelike geodesic motion in the Schwarzschild-BR BH spacetime and investigated the influence of a magnetic field on bound orbits and gravitational-wave signals. Using the Lagrangian formalism, we derived the effective potential and analyzed its behavior to study the orbital dynamics. We found that stable orbits shift inward as the magnetic field $B$ increases. We further examined the ISCO and the MBO, showing that both the ISCO and MBO radii increase with increasing magnetic field strength (see Table~\ref{tab:isco_mbo}). In addition, an analysis of the allowed parametric space ($E$--$L$) reveals that the allowed region shrinks and shifts upward as the magnetic field increases, implying that larger energies are required to sustain bound orbits (see Fig.~\ref{fig:allowed}). 

We further investigated the influence of the magnetic field on the periodic orbits of a test particle around the Schwarzschild--BR BH. We showed that at higher magnetic field strengths, a particle requires larger energy to sustain the same periodic orbit, whereas for a fixed energy, a lower angular momentum is sufficient to maintain that orbit (see Fig.~\ref{fig:q}). We numerically computed the values of the energy $E$ and orbital angular momentum $L$ corresponding to various configurations $(z,w,v)$ of periodic orbits and summarized them in Tables~\ref{table1} and \ref{table2}. These values were then used to visualize the periodic orbits and to analyze the GWs emitted by a particle moving along such trajectories.

Subsequently, we studied the GW radiation emitted by periodic orbits in an extreme mass ratio inspiral (EMRI), where a stellar-mass object $m \sim 10M_{\odot}$ moves along periodic trajectories around a supermassive Schwarzschild–BR BH with mass $M \sim 10^{6}M_{\odot}$. Employing the numerical kludge method, we obtained the corresponding gravitational waveforms for the orbits $(2,1,1)$, $(3,1,2)$, and $(4,1,3)$, as illustrated in Fig.~\ref{fig:waveforms}. We also found that increasing the magnetic field shifts the zoom-phase orbit inward and the whirl-phase orbit outward relative to the pure Schwarzschild case, resulting in a reduction in the orbital period of periodic motion around the Schwarzschild--BR BH (see Fig.~\ref{fig:waveforms}).

Finally, we evaluated the detectability of GW radiation from periodic EMRI orbits by applying discrete Fourier transforms of the time-domain signals to obtain their frequency spectra. We found that the characteristic frequencies lie in the mHz range, making them accessible to space-based detectors. In addition, the dominant spectral peak shifts toward lower frequencies as the magnetic field strength $B$ increases (see Fig.~\ref{fig:spectrum}). We further computed the characteristic strain for various orbital configurations $(z,w,v)$ of periodic orbits and magnetic field strengths $B$ and compared the results with the sensitivity curves of space-based GW detectors, LISA, Taiji, and TianQin. For all configurations considered, portions of the characteristic strain exceed the detector noise curves, indicating that these EMRI GW signals are potentially detectable by future space-based observatories (see Fig.~\ref{fig:strain}).

Given the importance of EMRI gravitational waveforms as key observational signatures of supermassive BHs, our findings retain astrophysical significance even though they are based on a simplified model. This model provides insight into the role of intrinsically electromagnetic fields in modifying the background spacetime and influencing the GWs emitted by EMRI systems, thereby offering promising opportunities for future GW observations to test and constrain the effects of spacetimes where the magnetic field is intrinsic to the geometry.

\section*{Acknowledgements}

The authors wish to thank H. Lin for useful discussions. S.S. is supported by the National Natural Science Foundation of China under Grant No. W2433018. Q.W and T.Z. are supported by the National Natural Science Foundation of China under Grants No.~12275238, No.~12542053, and No.~11675143, the National Key Research and Development Program under Grant No. 2020YFC2201503, and the Zhejiang Provincial Natural Science Foundation of China under Grants No. LR21A050001 and No. LY20A050002, and the Fundamental Research Funds for the Provincial Universities of Zhejiang in China under Grant No. RF-A2019015.

\bibliographystyle{apsrev4-1}
\bibliography{KBR_GW_Ref,NewRef}

\begin{thebibliography}{79}%
\makeatletter
\providecommand \@ifxundefined [1]{%
 \@ifx{#1\undefined}
}%
\providecommand \@ifnum [1]{%
 \ifnum #1\expandafter \@firstoftwo
 \else \expandafter \@secondoftwo
 \fi
}%
\providecommand \@ifx [1]{%
 \ifx #1\expandafter \@firstoftwo
 \else \expandafter \@secondoftwo
 \fi
}%
\providecommand \natexlab [1]{#1}%
\providecommand \enquote  [1]{``#1''}%
\providecommand \bibnamefont  [1]{#1}%
\providecommand \bibfnamefont [1]{#1}%
\providecommand \citenamefont [1]{#1}%
\providecommand \href@noop [0]{\@secondoftwo}%
\providecommand \href [0]{\begingroup \@sanitize@url \@href}%
\providecommand \@href[1]{\@@startlink{#1}\@@href}%
\providecommand \@@href[1]{\endgroup#1\@@endlink}%
\providecommand \@sanitize@url [0]{\catcode `\\12\catcode `\$12\catcode
  `\&12\catcode `\#12\catcode `\^12\catcode `\_12\catcode `\%12\relax}%
\providecommand \@@startlink[1]{}%
\providecommand \@@endlink[0]{}%
\providecommand \url  [0]{\begingroup\@sanitize@url \@url }%
\providecommand \@url [1]{\endgroup\@href {#1}{\urlprefix }}%
\providecommand \urlprefix  [0]{URL }%
\providecommand \Eprint [0]{\href }%
\providecommand \doibase [0]{http://dx.doi.org/}%
\providecommand \selectlanguage [0]{\@gobble}%
\providecommand \bibinfo  [0]{\@secondoftwo}%
\providecommand \bibfield  [0]{\@secondoftwo}%
\providecommand \translation [1]{[#1]}%
\providecommand \BibitemOpen [0]{}%
\providecommand \bibitemStop [0]{}%
\providecommand \bibitemNoStop [0]{.\EOS\space}%
\providecommand \EOS [0]{\spacefactor3000\relax}%
\providecommand \BibitemShut  [1]{\csname bibitem#1\endcsname}%
\let\auto@bib@innerbib\@empty
\bibitem [{\citenamefont {{Einstein}}(1916)}]{Einstein1916}%
  \BibitemOpen
  \bibfield  {author} {\bibinfo {author} {\bibfnamefont {A.}~\bibnamefont
  {{Einstein}}},\ }\href {\doibase 10.1002/andp.19163540702} {\bibfield
  {journal} {\bibinfo  {journal} {Annalen der Physik}\ }\textbf {\bibinfo
  {volume} {354}},\ \bibinfo {pages} {769} (\bibinfo {year}
  {1916})}\BibitemShut {NoStop}%
\bibitem [{\citenamefont {{Abbott}}\ and\ \citenamefont {et~al. {(Virgo and
  LIGO Scientific Collaborations)}}(2016{\natexlab{a}})}]{Abbott16a}%
  \BibitemOpen
  \bibfield  {author} {\bibinfo {author} {\bibfnamefont {B.~P.}\ \bibnamefont
  {{Abbott}}}\ and\ \bibinfo {author} {\bibnamefont {et~al. {(Virgo and LIGO
  Scientific Collaborations)}}},\ }\href {\doibase
  10.1103/PhysRevLett.116.061102} {\bibfield  {journal} {\bibinfo  {journal}
  {Phys. Rev. Lett.}\ }\textbf {\bibinfo {volume} {116}},\ \bibinfo {eid}
  {061102} (\bibinfo {year} {2016}{\natexlab{a}})},\ \Eprint
  {http://arxiv.org/abs/1602.03837} {arXiv:1602.03837 [gr-qc]} \BibitemShut
  {NoStop}%
\bibitem [{\citenamefont {{Abbott}}\ and\ \citenamefont {et~al. {(Virgo and
  LIGO Scientific Collaborations)}}(2016{\natexlab{b}})}]{Abbott16b}%
  \BibitemOpen
  \bibfield  {author} {\bibinfo {author} {\bibfnamefont {B.~P.}\ \bibnamefont
  {{Abbott}}}\ and\ \bibinfo {author} {\bibnamefont {et~al. {(Virgo and LIGO
  Scientific Collaborations)}}},\ }\href {\doibase
  10.1103/PhysRevLett.116.241102} {\bibfield  {journal} {\bibinfo  {journal}
  {Phys. Rev. Lett.}\ }\textbf {\bibinfo {volume} {116}},\ \bibinfo {eid}
  {241102} (\bibinfo {year} {2016}{\natexlab{b}})},\ \Eprint
  {http://arxiv.org/abs/1602.03840} {arXiv:1602.03840 [gr-qc]} \BibitemShut
  {NoStop}%
\bibitem [{\citenamefont {{Akiyama}}\ and\ \citenamefont {et~al. {(Event
  Horizon Telescope Collaboration)}}(2019{\natexlab{a}})}]{Akiyama19L1}%
  \BibitemOpen
  \bibfield  {author} {\bibinfo {author} {\bibfnamefont {K.}~\bibnamefont
  {{Akiyama}}}\ and\ \bibinfo {author} {\bibnamefont {et~al. {(Event Horizon
  Telescope Collaboration)}}},\ }\href {\doibase 10.3847/2041-8213/ab0ec7}
  {\bibfield  {journal} {\bibinfo  {journal} {Astrophys. J.}\ }\textbf
  {\bibinfo {volume} {875}},\ \bibinfo {eid} {L1} (\bibinfo {year}
  {2019}{\natexlab{a}})},\ \Eprint {http://arxiv.org/abs/1906.11238}
  {arXiv:1906.11238 [astro-ph.GA]} \BibitemShut {NoStop}%
\bibitem [{\citenamefont {{Akiyama}}\ and\ \citenamefont {et~al. {(Event
  Horizon Telescope Collaboration)}}(2019{\natexlab{b}})}]{Akiyama19L6}%
  \BibitemOpen
  \bibfield  {author} {\bibinfo {author} {\bibfnamefont {K.}~\bibnamefont
  {{Akiyama}}}\ and\ \bibinfo {author} {\bibnamefont {et~al. {(Event Horizon
  Telescope Collaboration)}}},\ }\href {\doibase 10.3847/2041-8213/ab1141}
  {\bibfield  {journal} {\bibinfo  {journal} {Astrophys. J.}\ }\textbf
  {\bibinfo {volume} {875}},\ \bibinfo {eid} {L6} (\bibinfo {year}
  {2019}{\natexlab{b}})},\ \Eprint {http://arxiv.org/abs/1906.11243}
  {arXiv:1906.11243 [astro-ph.GA]} \BibitemShut {NoStop}%
\bibitem [{\citenamefont {{Akiyama}}\ and\ \citenamefont {et~al. {(Event
  Horizon Telescope Collaboration)}}(2022{\natexlab{a}})}]{Akiyama22L12}%
  \BibitemOpen
  \bibfield  {author} {\bibinfo {author} {\bibfnamefont {K.}~\bibnamefont
  {{Akiyama}}}\ and\ \bibinfo {author} {\bibnamefont {et~al. {(Event Horizon
  Telescope Collaboration)}}},\ }\href {\doibase 10.3847/2041-8213/ac6674}
  {\bibfield  {journal} {\bibinfo  {journal} {Astrophys. J. Lett.}\ }\textbf
  {\bibinfo {volume} {930}},\ \bibinfo {eid} {L12} (\bibinfo {year}
  {2022}{\natexlab{a}})}\BibitemShut {NoStop}%
\bibitem [{\citenamefont {{Akiyama}}\ and\ \citenamefont {et~al. {(Event
  Horizon Telescope Collaboration)}}(2022{\natexlab{b}})}]{EHT2022L14}%
  \BibitemOpen
  \bibfield  {author} {\bibinfo {author} {\bibfnamefont {K.}~\bibnamefont
  {{Akiyama}}}\ and\ \bibinfo {author} {\bibnamefont {et~al. {(Event Horizon
  Telescope Collaboration)}}},\ }\href {\doibase 10.3847/2041-8213/ac6429}
  {\bibfield  {journal} {\bibinfo  {journal} {Astrophys. J. Lett.}\ }\textbf
  {\bibinfo {volume} {930}},\ \bibinfo {eid} {L14} (\bibinfo {year}
  {2022}{\natexlab{b}})}\BibitemShut {NoStop}%
\bibitem [{\citenamefont {Will}(2014)}]{Will14LRR}%
  \BibitemOpen
  \bibfield  {author} {\bibinfo {author} {\bibfnamefont {C.~M.}\ \bibnamefont
  {Will}},\ }\href {\doibase 10.12942/lrr-2014-4} {\bibfield  {journal}
  {\bibinfo  {journal} {Living Rev. Rel.}\ }\textbf {\bibinfo {volume} {17}},\
  \bibinfo {pages} {4} (\bibinfo {year} {2014})},\ \Eprint
  {http://arxiv.org/abs/1403.7377} {arXiv:1403.7377 [gr-qc]} \BibitemShut
  {NoStop}%
\bibitem [{\citenamefont {Psaltis}\ \emph {et~al.}(2020)\citenamefont {Psaltis}
  \emph {et~al.}}]{Psaltis20PRL}%
  \BibitemOpen
  \bibfield  {author} {\bibinfo {author} {\bibfnamefont {D.}~\bibnamefont
  {Psaltis}} \emph {et~al.} (\bibinfo {collaboration} {Event Horizon
  Telescope}),\ }\href {\doibase 10.1103/PhysRevLett.125.141104} {\bibfield
  {journal} {\bibinfo  {journal} {Phys. Rev. Lett.}\ }\textbf {\bibinfo
  {volume} {125}},\ \bibinfo {pages} {141104} (\bibinfo {year} {2020})},\
  \Eprint {http://arxiv.org/abs/2010.01055} {arXiv:2010.01055 [gr-qc]}
  \BibitemShut {NoStop}%
\bibitem [{\citenamefont {{Amaro-Seoane}}\ and\ \citenamefont {et~al. {(Laser
  Interferometer Space Antenna)}}(2017)}]{Amaro-Seoane2017LISA}%
  \BibitemOpen
  \bibfield  {author} {\bibinfo {author} {\bibfnamefont {P.}~\bibnamefont
  {{Amaro-Seoane}}}\ and\ \bibinfo {author} {\bibnamefont {et~al. {(Laser
  Interferometer Space Antenna)}}},\ }\href {https://arxiv.org/abs/1702.00786}
  {\enquote {\bibinfo {title} {Laser interferometer space antenna},}\ }
  (\bibinfo {year} {2017}),\ \Eprint {http://arxiv.org/abs/1702.00786}
  {arXiv:1702.00786 [astro-ph.IM]} \BibitemShut {NoStop}%
\bibitem [{\citenamefont {Hu}\ and\ \citenamefont
  {Wu}(2017)}]{10.1093/nsr/nwx116}%
  \BibitemOpen
  \bibfield  {author} {\bibinfo {author} {\bibfnamefont {W.-R.}\ \bibnamefont
  {Hu}}\ and\ \bibinfo {author} {\bibfnamefont {Y.-L.}\ \bibnamefont {Wu}},\
  }\href {\doibase 10.1093/nsr/nwx116} {\bibfield  {journal} {\bibinfo
  {journal} {Natl. Sci. Rev.}\ }\textbf {\bibinfo {volume} {4}},\ \bibinfo
  {pages} {685} (\bibinfo {year} {2017})},\ \Eprint
  {http://arxiv.org/abs/https://academic.oup.com/nsr/article-pdf/4/5/685/31566708/nwx116.pdf}
  {https://academic.oup.com/nsr/article-pdf/4/5/685/31566708/nwx116.pdf}
  \BibitemShut {NoStop}%
\bibitem [{\citenamefont {Hughes}(2001)}]{Hughes_2001}%
  \BibitemOpen
  \bibfield  {author} {\bibinfo {author} {\bibfnamefont {S.~A.}\ \bibnamefont
  {Hughes}},\ }\href {\doibase 10.1088/0264-9381/18/19/314} {\bibfield
  {journal} {\bibinfo  {journal} {Class. Quantum Gravity}\ }\textbf {\bibinfo
  {volume} {18}},\ \bibinfo {pages} {4067–4073} (\bibinfo {year}
  {2001})}\BibitemShut {NoStop}%
\bibitem [{\citenamefont {{Amaro-Seoane}}(2018)}]{Amaro-Seoane18LRR}%
  \BibitemOpen
  \bibfield  {author} {\bibinfo {author} {\bibfnamefont {P.}~\bibnamefont
  {{Amaro-Seoane}}},\ }\href {\doibase 10.1007/s41114-018-0013-8} {\bibfield
  {journal} {\bibinfo  {journal} {Living Rev. Rel.}\ }\textbf {\bibinfo
  {volume} {21}},\ \bibinfo {eid} {4} (\bibinfo {year} {2018})},\ \Eprint
  {http://arxiv.org/abs/1205.5240} {arXiv:1205.5240 [astro-ph.CO]} \BibitemShut
  {NoStop}%
\bibitem [{\citenamefont {{Babak}}\ \emph {et~al.}(2017)\citenamefont
  {{Babak}}, \citenamefont {{Gair}}, \citenamefont {{Sesana}}, \citenamefont
  {{Barausse}}, \citenamefont {{Sopuerta}}, \citenamefont {{Berry}},
  \citenamefont {{Berti}}, \citenamefont {{Amaro-Seoane}}, \citenamefont
  {{Petiteau}},\ and\ \citenamefont {{Klein}}}]{Babak17PRD}%
  \BibitemOpen
  \bibfield  {author} {\bibinfo {author} {\bibfnamefont {S.}~\bibnamefont
  {{Babak}}}, \bibinfo {author} {\bibfnamefont {J.}~\bibnamefont {{Gair}}},
  \bibinfo {author} {\bibfnamefont {A.}~\bibnamefont {{Sesana}}}, \bibinfo
  {author} {\bibfnamefont {E.}~\bibnamefont {{Barausse}}}, \bibinfo {author}
  {\bibfnamefont {C.~F.}\ \bibnamefont {{Sopuerta}}}, \bibinfo {author}
  {\bibfnamefont {C.~P.~L.}\ \bibnamefont {{Berry}}}, \bibinfo {author}
  {\bibfnamefont {E.}~\bibnamefont {{Berti}}}, \bibinfo {author} {\bibfnamefont
  {P.}~\bibnamefont {{Amaro-Seoane}}}, \bibinfo {author} {\bibfnamefont
  {A.}~\bibnamefont {{Petiteau}}}, \ and\ \bibinfo {author} {\bibfnamefont
  {A.}~\bibnamefont {{Klein}}},\ }\href {\doibase 10.1103/PhysRevD.95.103012}
  {\bibfield  {journal} {\bibinfo  {journal} {Phys. Rev. D}\ }\textbf {\bibinfo
  {volume} {95}},\ \bibinfo {eid} {103012} (\bibinfo {year} {2017})},\ \Eprint
  {http://arxiv.org/abs/1703.09722} {arXiv:1703.09722 [gr-qc]} \BibitemShut
  {NoStop}%
\bibitem [{\citenamefont {Glampedakis}\ and\ \citenamefont
  {Kennefick}(2002)}]{Glampedakis02PRD}%
  \BibitemOpen
  \bibfield  {author} {\bibinfo {author} {\bibfnamefont {K.}~\bibnamefont
  {Glampedakis}}\ and\ \bibinfo {author} {\bibfnamefont {D.}~\bibnamefont
  {Kennefick}},\ }\href {\doibase 10.1103/PhysRevD.66.044002} {\bibfield
  {journal} {\bibinfo  {journal} {Phys. Rev. D}\ }\textbf {\bibinfo {volume}
  {66}},\ \bibinfo {pages} {044002} (\bibinfo {year} {2002})}\BibitemShut
  {NoStop}%
\bibitem [{\citenamefont {Ruangsri}\ and\ \citenamefont
  {Hughes}(2014)}]{Ruangsri14PRD}%
  \BibitemOpen
  \bibfield  {author} {\bibinfo {author} {\bibfnamefont {U.}~\bibnamefont
  {Ruangsri}}\ and\ \bibinfo {author} {\bibfnamefont {S.~A.}\ \bibnamefont
  {Hughes}},\ }\href {\doibase 10.1103/PhysRevD.89.084036} {\bibfield
  {journal} {\bibinfo  {journal} {Phys. Rev. D}\ }\textbf {\bibinfo {volume}
  {89}},\ \bibinfo {pages} {084036} (\bibinfo {year} {2014})}\BibitemShut
  {NoStop}%
\bibitem [{\citenamefont {Gair}\ \emph {et~al.}(2013)\citenamefont {Gair},
  \citenamefont {Vallisneri}, \citenamefont {Larson},\ and\ \citenamefont
  {Baker}}]{Gair13CQG}%
  \BibitemOpen
  \bibfield  {author} {\bibinfo {author} {\bibfnamefont {J.~R.}\ \bibnamefont
  {Gair}}, \bibinfo {author} {\bibfnamefont {M.}~\bibnamefont {Vallisneri}},
  \bibinfo {author} {\bibfnamefont {S.~L.}\ \bibnamefont {Larson}}, \ and\
  \bibinfo {author} {\bibfnamefont {J.~G.}\ \bibnamefont {Baker}},\ }\href
  {\doibase 10.1088/0264-9381/30/18/184001} {\bibfield  {journal} {\bibinfo
  {journal} {Class. Quantum Gravity}\ }\textbf {\bibinfo {volume} {30}},\
  \bibinfo {pages} {184001} (\bibinfo {year} {2013})}\BibitemShut {NoStop}%
\bibitem [{\citenamefont {Barack}\ \emph {et~al.}(2019)\citenamefont {Barack}
  \emph {et~al.}}]{Barack19CQG}%
  \BibitemOpen
  \bibfield  {author} {\bibinfo {author} {\bibfnamefont {L.}~\bibnamefont
  {Barack}} \emph {et~al.},\ }\href {\doibase 10.1088/1361-6382/ab0587}
  {\bibfield  {journal} {\bibinfo  {journal} {Class. Quantum Gravity}\ }\textbf
  {\bibinfo {volume} {36}},\ \bibinfo {pages} {143001} (\bibinfo {year}
  {2019})}\BibitemShut {NoStop}%
\bibitem [{\citenamefont {Levin}\ and\ \citenamefont
  {Perez-Giz}(2008)}]{Levin_2008}%
  \BibitemOpen
  \bibfield  {author} {\bibinfo {author} {\bibfnamefont {J.}~\bibnamefont
  {Levin}}\ and\ \bibinfo {author} {\bibfnamefont {G.}~\bibnamefont
  {Perez-Giz}},\ }\href {\doibase 10.1103/physrevd.77.103005} {\bibfield
  {journal} {\bibinfo  {journal} {Phys. Rev. D}\ }\textbf {\bibinfo {volume}
  {77}} (\bibinfo {year} {2008}),\ 10.1103/physrevd.77.103005}\BibitemShut
  {NoStop}%
\bibitem [{\citenamefont {Grossman}\ and\ \citenamefont
  {Levin}(2009)}]{Grossman_2009}%
  \BibitemOpen
  \bibfield  {author} {\bibinfo {author} {\bibfnamefont {R.}~\bibnamefont
  {Grossman}}\ and\ \bibinfo {author} {\bibfnamefont {J.}~\bibnamefont
  {Levin}},\ }\href {\doibase 10.1103/physrevd.79.043017} {\bibfield  {journal}
  {\bibinfo  {journal} {Phys. Rev. D}\ }\textbf {\bibinfo {volume} {79}}
  (\bibinfo {year} {2009}),\ 10.1103/physrevd.79.043017}\BibitemShut {NoStop}%
\bibitem [{\citenamefont {Misra}\ and\ \citenamefont
  {Levin}(2010{\natexlab{a}})}]{Misra_2010}%
  \BibitemOpen
  \bibfield  {author} {\bibinfo {author} {\bibfnamefont {V.}~\bibnamefont
  {Misra}}\ and\ \bibinfo {author} {\bibfnamefont {J.}~\bibnamefont {Levin}},\
  }\href {\doibase 10.1103/physrevd.82.083001} {\bibfield  {journal} {\bibinfo
  {journal} {Phys. Rev. D}\ }\textbf {\bibinfo {volume} {82}} (\bibinfo {year}
  {2010}{\natexlab{a}}),\ 10.1103/physrevd.82.083001}\BibitemShut {NoStop}%
\bibitem [{\citenamefont {Levin}(2009)}]{Levin_2009}%
  \BibitemOpen
  \bibfield  {author} {\bibinfo {author} {\bibfnamefont {J.}~\bibnamefont
  {Levin}},\ }\href {\doibase 10.1088/0264-9381/26/23/235010} {\bibfield
  {journal} {\bibinfo  {journal} {Class. Quantum Gravity}\ }\textbf {\bibinfo
  {volume} {26}},\ \bibinfo {pages} {235010} (\bibinfo {year}
  {2009})}\BibitemShut {NoStop}%
\bibitem [{\citenamefont {Bambhaniya}\ \emph {et~al.}(2021)\citenamefont
  {Bambhaniya}, \citenamefont {Solanki}, \citenamefont {Dey}, \citenamefont
  {Joshi}, \citenamefont {Joshi},\ and\ \citenamefont {Patel}}]{Bambhaniya20}%
  \BibitemOpen
  \bibfield  {author} {\bibinfo {author} {\bibfnamefont {P.}~\bibnamefont
  {Bambhaniya}}, \bibinfo {author} {\bibfnamefont {D.~N.}\ \bibnamefont
  {Solanki}}, \bibinfo {author} {\bibfnamefont {D.}~\bibnamefont {Dey}},
  \bibinfo {author} {\bibfnamefont {A.~B.}\ \bibnamefont {Joshi}}, \bibinfo
  {author} {\bibfnamefont {P.~S.}\ \bibnamefont {Joshi}}, \ and\ \bibinfo
  {author} {\bibfnamefont {V.}~\bibnamefont {Patel}},\ }\href {\doibase
  10.1140/epjc/s10052-021-08997-x} {\bibfield  {journal} {\bibinfo  {journal}
  {Eur. Phys. J. C}\ }\textbf {\bibinfo {volume} {81}},\ \bibinfo {pages} {205}
  (\bibinfo {year} {2021})},\ \Eprint {http://arxiv.org/abs/2007.12086}
  {arXiv:2007.12086 [gr-qc]} \BibitemShut {NoStop}%
\bibitem [{\citenamefont {Rana}\ and\ \citenamefont {Mangalam}(2019)}]{Rana19}%
  \BibitemOpen
  \bibfield  {author} {\bibinfo {author} {\bibfnamefont {P.}~\bibnamefont
  {Rana}}\ and\ \bibinfo {author} {\bibfnamefont {A.}~\bibnamefont
  {Mangalam}},\ }\href {\doibase 10.1088/1361-6382/ab004c} {\bibfield
  {journal} {\bibinfo  {journal} {Class. Quantum Gravity}\ }\textbf {\bibinfo
  {volume} {36}},\ \bibinfo {pages} {045009} (\bibinfo {year} {2019})},\
  \Eprint {http://arxiv.org/abs/1901.02730} {arXiv:1901.02730 [gr-qc]}
  \BibitemShut {NoStop}%
\bibitem [{\citenamefont {Healy}\ \emph {et~al.}(2009)\citenamefont {Healy},
  \citenamefont {Levin},\ and\ \citenamefont {Shoemaker}}]{Healy09PRL}%
  \BibitemOpen
  \bibfield  {author} {\bibinfo {author} {\bibfnamefont {J.}~\bibnamefont
  {Healy}}, \bibinfo {author} {\bibfnamefont {J.}~\bibnamefont {Levin}}, \ and\
  \bibinfo {author} {\bibfnamefont {D.}~\bibnamefont {Shoemaker}},\ }\href
  {\doibase 10.1103/PhysRevLett.103.131101} {\bibfield  {journal} {\bibinfo
  {journal} {Phys. Rev. Lett.}\ }\textbf {\bibinfo {volume} {103}},\ \bibinfo
  {pages} {131101} (\bibinfo {year} {2009})},\ \Eprint
  {http://arxiv.org/abs/0907.0671} {arXiv:0907.0671 [gr-qc]} \BibitemShut
  {NoStop}%
\bibitem [{\citenamefont {Misra}\ and\ \citenamefont
  {Levin}(2010{\natexlab{b}})}]{Levin2010}%
  \BibitemOpen
  \bibfield  {author} {\bibinfo {author} {\bibfnamefont {V.}~\bibnamefont
  {Misra}}\ and\ \bibinfo {author} {\bibfnamefont {J.}~\bibnamefont {Levin}},\
  }\href {\doibase 10.1103/PhysRevD.82.083001} {\bibfield  {journal} {\bibinfo
  {journal} {Phys. Rev. D}\ }\textbf {\bibinfo {volume} {82}},\ \bibinfo
  {pages} {083001} (\bibinfo {year} {2010}{\natexlab{b}})}\BibitemShut
  {NoStop}%
\bibitem [{\citenamefont {Pugliese}\ \emph {et~al.}(2017)\citenamefont
  {Pugliese}, \citenamefont {Quevedo},\ and\ \citenamefont
  {Ruffini}}]{Pugliese13}%
  \BibitemOpen
  \bibfield  {author} {\bibinfo {author} {\bibfnamefont {D.}~\bibnamefont
  {Pugliese}}, \bibinfo {author} {\bibfnamefont {H.}~\bibnamefont {Quevedo}}, \
  and\ \bibinfo {author} {\bibfnamefont {R.}~\bibnamefont {Ruffini}},\ }\href
  {\doibase 10.1140/epjc/s10052-017-4769-x} {\bibfield  {journal} {\bibinfo
  {journal} {Eur. Phys. J. C}\ }\textbf {\bibinfo {volume} {77}},\ \bibinfo
  {pages} {206} (\bibinfo {year} {2017})},\ \Eprint
  {http://arxiv.org/abs/1304.2940} {arXiv:1304.2940 [gr-qc]} \BibitemShut
  {NoStop}%
\bibitem [{\citenamefont {{Babar}}\ \emph {et~al.}(2017)\citenamefont
  {{Babar}}, \citenamefont {{Babar}},\ and\ \citenamefont
  {{Lim}}}]{Babar17PRD}%
  \BibitemOpen
  \bibfield  {author} {\bibinfo {author} {\bibfnamefont {G.~Z.}\ \bibnamefont
  {{Babar}}}, \bibinfo {author} {\bibfnamefont {A.~Z.}\ \bibnamefont
  {{Babar}}}, \ and\ \bibinfo {author} {\bibfnamefont {Y.-K.}\ \bibnamefont
  {{Lim}}},\ }\href {\doibase 10.1103/PhysRevD.96.084052} {\bibfield  {journal}
  {\bibinfo  {journal} {Phys. Rev. D}\ }\textbf {\bibinfo {volume} {96}},\
  \bibinfo {eid} {084052} (\bibinfo {year} {2017})},\ \Eprint
  {http://arxiv.org/abs/1710.09581} {arXiv:1710.09581 [gr-qc]} \BibitemShut
  {NoStop}%
\bibitem [{\citenamefont {Liu}\ \emph {et~al.}(2019)\citenamefont {Liu},
  \citenamefont {Ding},\ and\ \citenamefont {Jing}}]{Liu18}%
  \BibitemOpen
  \bibfield  {author} {\bibinfo {author} {\bibfnamefont {C.}~\bibnamefont
  {Liu}}, \bibinfo {author} {\bibfnamefont {C.}~\bibnamefont {Ding}}, \ and\
  \bibinfo {author} {\bibfnamefont {J.}~\bibnamefont {Jing}},\ }\href {\doibase
  10.1088/0253-6102/71/12/1461} {\bibfield  {journal} {\bibinfo  {journal}
  {Commun. Theor. Phys.}\ }\textbf {\bibinfo {volume} {71}},\ \bibinfo {pages}
  {1461} (\bibinfo {year} {2019})},\ \Eprint {http://arxiv.org/abs/1804.05883}
  {arXiv:1804.05883 [gr-qc]} \BibitemShut {NoStop}%
\bibitem [{\citenamefont {Lin}\ and\ \citenamefont {Deng}(2023)}]{Lin23}%
  \BibitemOpen
  \bibfield  {author} {\bibinfo {author} {\bibfnamefont {H.-Y.}\ \bibnamefont
  {Lin}}\ and\ \bibinfo {author} {\bibfnamefont {X.-M.}\ \bibnamefont {Deng}},\
  }\href {\doibase 10.1140/epjc/s10052-023-11487-x} {\bibfield  {journal}
  {\bibinfo  {journal} {Eur. Phys. J. C}\ }\textbf {\bibinfo {volume} {83}},\
  \bibinfo {pages} {311} (\bibinfo {year} {2023})}\BibitemShut {NoStop}%
\bibitem [{\citenamefont {Yao}\ and\ \citenamefont {Li}(2023)}]{Yao23}%
  \BibitemOpen
  \bibfield  {author} {\bibinfo {author} {\bibfnamefont {J.-T.}\ \bibnamefont
  {Yao}}\ and\ \bibinfo {author} {\bibfnamefont {X.}~\bibnamefont {Li}},\
  }\href {\doibase 10.1103/PhysRevD.108.084067} {\bibfield  {journal} {\bibinfo
   {journal} {Phys. Rev. D}\ }\textbf {\bibinfo {volume} {108}},\ \bibinfo
  {pages} {084067} (\bibinfo {year} {2023})}\BibitemShut {NoStop}%
\bibitem [{\citenamefont {Chan}\ and\ \citenamefont {Lim}(2025)}]{Chan25}%
  \BibitemOpen
  \bibfield  {author} {\bibinfo {author} {\bibfnamefont {Z.~C.~S.}\
  \bibnamefont {Chan}}\ and\ \bibinfo {author} {\bibfnamefont {Y.-K.}\
  \bibnamefont {Lim}},\ }\href {\doibase 10.1007/s10714-025-03368-3} {\bibfield
   {journal} {\bibinfo  {journal} {Gen. Rel. Grav.}\ }\textbf {\bibinfo
  {volume} {57}},\ \bibinfo {pages} {35} (\bibinfo {year} {2025})},\ \Eprint
  {http://arxiv.org/abs/2502.03082} {arXiv:2502.03082 [gr-qc]} \BibitemShut
  {NoStop}%
\bibitem [{\citenamefont {Lin}\ and\ \citenamefont {Deng}(2022)}]{Lin22}%
  \BibitemOpen
  \bibfield  {author} {\bibinfo {author} {\bibfnamefont {H.-Y.}\ \bibnamefont
  {Lin}}\ and\ \bibinfo {author} {\bibfnamefont {X.-M.}\ \bibnamefont {Deng}},\
  }\href {\doibase 10.1140/epjp/s13360-022-02391-6} {\bibfield  {journal}
  {\bibinfo  {journal} {Eur. Phys. J. Plus}\ }\textbf {\bibinfo {volume}
  {137}},\ \bibinfo {pages} {176} (\bibinfo {year} {2022})}\BibitemShut
  {NoStop}%
\bibitem [{\citenamefont {Lin}\ and\ \citenamefont {Deng}(2021)}]{Lin21}%
  \BibitemOpen
  \bibfield  {author} {\bibinfo {author} {\bibfnamefont {H.-Y.}\ \bibnamefont
  {Lin}}\ and\ \bibinfo {author} {\bibfnamefont {X.-M.}\ \bibnamefont {Deng}},\
  }\href {\doibase 10.1016/j.dark.2020.100745} {\bibfield  {journal} {\bibinfo
  {journal} {Phys. Dark Univ.}\ }\textbf {\bibinfo {volume} {31}},\ \bibinfo
  {pages} {100745} (\bibinfo {year} {2021})}\BibitemShut {NoStop}%
\bibitem [{\citenamefont {Deng}(2020)}]{Deng20}%
  \BibitemOpen
  \bibfield  {author} {\bibinfo {author} {\bibfnamefont {X.-M.}\ \bibnamefont
  {Deng}},\ }\href {\doibase 10.1140/epjc/s10052-020-8067-7} {\bibfield
  {journal} {\bibinfo  {journal} {Eur. Phys. J. C}\ }\textbf {\bibinfo {volume}
  {80}},\ \bibinfo {pages} {489} (\bibinfo {year} {2020})}\BibitemShut
  {NoStop}%
\bibitem [{\citenamefont {Tu}\ \emph {et~al.}(2023)\citenamefont {Tu},
  \citenamefont {Zhu},\ and\ \citenamefont {Wang}}]{Tu23}%
  \BibitemOpen
  \bibfield  {author} {\bibinfo {author} {\bibfnamefont {Z.-Y.}\ \bibnamefont
  {Tu}}, \bibinfo {author} {\bibfnamefont {T.}~\bibnamefont {Zhu}}, \ and\
  \bibinfo {author} {\bibfnamefont {A.}~\bibnamefont {Wang}},\ }\href {\doibase
  10.1103/PhysRevD.108.024035} {\bibfield  {journal} {\bibinfo  {journal}
  {Phys. Rev. D}\ }\textbf {\bibinfo {volume} {108}},\ \bibinfo {pages}
  {024035} (\bibinfo {year} {2023})},\ \Eprint
  {http://arxiv.org/abs/2304.14160} {arXiv:2304.14160 [gr-qc]} \BibitemShut
  {NoStop}%
\bibitem [{\citenamefont {Wei}\ \emph {et~al.}(2019)\citenamefont {Wei},
  \citenamefont {Yang},\ and\ \citenamefont {Liu}}]{Wei19}%
  \BibitemOpen
  \bibfield  {author} {\bibinfo {author} {\bibfnamefont {S.-W.}\ \bibnamefont
  {Wei}}, \bibinfo {author} {\bibfnamefont {J.}~\bibnamefont {Yang}}, \ and\
  \bibinfo {author} {\bibfnamefont {Y.-X.}\ \bibnamefont {Liu}},\ }\href
  {\doibase 10.1103/PhysRevD.99.104016} {\bibfield  {journal} {\bibinfo
  {journal} {Phys. Rev. D}\ }\textbf {\bibinfo {volume} {99}},\ \bibinfo
  {pages} {104016} (\bibinfo {year} {2019})},\ \Eprint
  {http://arxiv.org/abs/1904.03129} {arXiv:1904.03129 [gr-qc]} \BibitemShut
  {NoStop}%
\bibitem [{\citenamefont {Zhang}\ and\ \citenamefont {Xie}(2022)}]{Zhang22}%
  \BibitemOpen
  \bibfield  {author} {\bibinfo {author} {\bibfnamefont {J.}~\bibnamefont
  {Zhang}}\ and\ \bibinfo {author} {\bibfnamefont {Y.}~\bibnamefont {Xie}},\
  }\href {\doibase 10.1140/epjc/s10052-022-10846-4} {\bibfield  {journal}
  {\bibinfo  {journal} {Eur. Phys. J. C}\ }\textbf {\bibinfo {volume} {82}},\
  \bibinfo {pages} {854} (\bibinfo {year} {2022})}\BibitemShut {NoStop}%
\bibitem [{\citenamefont {Jiang}\ \emph {et~al.}(2024)\citenamefont {Jiang},
  \citenamefont {Alloqulov}, \citenamefont {Wu}, \citenamefont {Shaymatov},\
  and\ \citenamefont {Zhu}}]{JIANG2024}%
  \BibitemOpen
  \bibfield  {author} {\bibinfo {author} {\bibfnamefont {H.}~\bibnamefont
  {Jiang}}, \bibinfo {author} {\bibfnamefont {M.}~\bibnamefont {Alloqulov}},
  \bibinfo {author} {\bibfnamefont {Q.}~\bibnamefont {Wu}}, \bibinfo {author}
  {\bibfnamefont {S.}~\bibnamefont {Shaymatov}}, \ and\ \bibinfo {author}
  {\bibfnamefont {T.}~\bibnamefont {Zhu}},\ }\href {\doibase
  https://doi.org/10.1016/j.dark.2024.101627} {\bibfield  {journal} {\bibinfo
  {journal} {Phys. Dark Universe}\ }\textbf {\bibinfo {volume} {46}},\ \bibinfo
  {pages} {101627} (\bibinfo {year} {2024})}\BibitemShut {NoStop}%
\bibitem [{\citenamefont {Wang}\ \emph
  {et~al.}(2025{\natexlab{a}})\citenamefont {Wang}, \citenamefont {Zhang},
  \citenamefont {Zhu},\ and\ \citenamefont {Wei}}]{Wang25}%
  \BibitemOpen
  \bibfield  {author} {\bibinfo {author} {\bibfnamefont {C.-H.}\ \bibnamefont
  {Wang}}, \bibinfo {author} {\bibfnamefont {Y.-P.}\ \bibnamefont {Zhang}},
  \bibinfo {author} {\bibfnamefont {T.}~\bibnamefont {Zhu}}, \ and\ \bibinfo
  {author} {\bibfnamefont {S.-W.}\ \bibnamefont {Wei}},\ }\href@noop {} {\
  (\bibinfo {year} {2025}{\natexlab{a}})},\ \Eprint
  {http://arxiv.org/abs/2508.20558} {arXiv:2508.20558 [gr-qc]} \BibitemShut
  {NoStop}%
\bibitem [{\citenamefont {Wei}\ \emph {et~al.}(2025)\citenamefont {Wei},
  \citenamefont {Zhang}, \citenamefont {Xie},\ and\ \citenamefont
  {Yin}}]{Wei25}%
  \BibitemOpen
  \bibfield  {author} {\bibinfo {author} {\bibfnamefont {Z.-L.}\ \bibnamefont
  {Wei}}, \bibinfo {author} {\bibfnamefont {J.}~\bibnamefont {Zhang}}, \bibinfo
  {author} {\bibfnamefont {Y.}~\bibnamefont {Xie}}, \ and\ \bibinfo {author}
  {\bibfnamefont {P.-L.}\ \bibnamefont {Yin}},\ }\href {\doibase
  10.1140/epjc/s10052-025-14437-x} {\bibfield  {journal} {\bibinfo  {journal}
  {Eur. Phys. J. C}\ }\textbf {\bibinfo {volume} {85}},\ \bibinfo {pages} {698}
  (\bibinfo {year} {2025})}\BibitemShut {NoStop}%
\bibitem [{\citenamefont {Alloqulov}\ \emph {et~al.}(2026)\citenamefont
  {Alloqulov}, \citenamefont {Shaymatov}, \citenamefont {Ahmedov},\ and\
  \citenamefont {Zhu}}]{Alloqulov26GW1}%
  \BibitemOpen
  \bibfield  {author} {\bibinfo {author} {\bibfnamefont {M.}~\bibnamefont
  {Alloqulov}}, \bibinfo {author} {\bibfnamefont {S.}~\bibnamefont
  {Shaymatov}}, \bibinfo {author} {\bibfnamefont {B.}~\bibnamefont {Ahmedov}},
  \ and\ \bibinfo {author} {\bibfnamefont {T.}~\bibnamefont {Zhu}},\ }\href
  {\doibase 10.1140/epjc/s10052-025-15251-1} {\bibfield  {journal} {\bibinfo
  {journal} {Eur. Phys. J. C}\ }\textbf {\bibinfo {volume} {86}},\ \bibinfo
  {pages} {117} (\bibinfo {year} {2026})}\BibitemShut {NoStop}%
\bibitem [{\citenamefont {Sharipov}\ \emph {et~al.}(2025)\citenamefont
  {Sharipov}, \citenamefont {Xamidov}, \citenamefont {Wu}, \citenamefont
  {Shaymatov},\ and\ \citenamefont {Zhu}}]{Sharipov25}%
  \BibitemOpen
  \bibfield  {author} {\bibinfo {author} {\bibfnamefont {J.}~\bibnamefont
  {Sharipov}}, \bibinfo {author} {\bibfnamefont {T.}~\bibnamefont {Xamidov}},
  \bibinfo {author} {\bibfnamefont {Q.}~\bibnamefont {Wu}}, \bibinfo {author}
  {\bibfnamefont {S.}~\bibnamefont {Shaymatov}}, \ and\ \bibinfo {author}
  {\bibfnamefont {T.}~\bibnamefont {Zhu}},\ }\href@noop {} {\  (\bibinfo {year}
  {2025})},\ \Eprint {http://arxiv.org/abs/2511.10043} {arXiv:2511.10043
  [gr-qc]} \BibitemShut {NoStop}%
\bibitem [{\citenamefont {{Barausse}}\ \emph {et~al.}(2014)\citenamefont
  {{Barausse}}, \citenamefont {{Cardoso}},\ and\ \citenamefont
  {{Pani}}}]{Barausse14PRD}%
  \BibitemOpen
  \bibfield  {author} {\bibinfo {author} {\bibfnamefont {E.}~\bibnamefont
  {{Barausse}}}, \bibinfo {author} {\bibfnamefont {V.}~\bibnamefont
  {{Cardoso}}}, \ and\ \bibinfo {author} {\bibfnamefont {P.}~\bibnamefont
  {{Pani}}},\ }\href {\doibase 10.1103/PhysRevD.89.104059} {\bibfield
  {journal} {\bibinfo  {journal} {Phys. Rev. D}\ }\textbf {\bibinfo {volume}
  {89}},\ \bibinfo {eid} {104059} (\bibinfo {year} {2014})},\ \Eprint
  {http://arxiv.org/abs/1404.7149} {arXiv:1404.7149 [gr-qc]} \BibitemShut
  {NoStop}%
\bibitem [{\citenamefont {{Cardoso}}\ \emph {et~al.}(2022)\citenamefont
  {{Cardoso}}, \citenamefont {{Destounis}}, \citenamefont {{Duque}},
  \citenamefont {{Macedo}},\ and\ \citenamefont {{Maselli}}}]{Cardoso22PRD}%
  \BibitemOpen
  \bibfield  {author} {\bibinfo {author} {\bibfnamefont {V.}~\bibnamefont
  {{Cardoso}}}, \bibinfo {author} {\bibfnamefont {K.}~\bibnamefont
  {{Destounis}}}, \bibinfo {author} {\bibfnamefont {F.}~\bibnamefont
  {{Duque}}}, \bibinfo {author} {\bibfnamefont {R.~P.}\ \bibnamefont
  {{Macedo}}}, \ and\ \bibinfo {author} {\bibfnamefont {A.}~\bibnamefont
  {{Maselli}}},\ }\href {\doibase 10.1103/PhysRevD.105.L061501} {\bibfield
  {journal} {\bibinfo  {journal} {Phys. Rev. D}\ }\textbf {\bibinfo {volume}
  {105}},\ \bibinfo {eid} {L061501} (\bibinfo {year} {2022})},\ \Eprint
  {http://arxiv.org/abs/2109.00005} {arXiv:2109.00005 [gr-qc]} \BibitemShut
  {NoStop}%
\bibitem [{\citenamefont {Zhang}\ and\ \citenamefont {Zhu}(2025)}]{Zhang25}%
  \BibitemOpen
  \bibfield  {author} {\bibinfo {author} {\bibfnamefont {C.}~\bibnamefont
  {Zhang}}\ and\ \bibinfo {author} {\bibfnamefont {T.}~\bibnamefont {Zhu}},\
  }\href {https://arxiv.org/abs/2511.14080} {\enquote {\bibinfo {title}
  {Periodic orbits and their gravitational wave radiations in
  $\gamma$-metric},}\ } (\bibinfo {year} {2025}),\ \Eprint
  {http://arxiv.org/abs/2511.14080} {arXiv:2511.14080 [gr-qc]} \BibitemShut
  {NoStop}%
\bibitem [{\citenamefont {Yang}\ \emph
  {et~al.}(2025{\natexlab{a}})\citenamefont {Yang}, \citenamefont {Zhang},
  \citenamefont {Zhu}, \citenamefont {Zhao},\ and\ \citenamefont
  {Liu}}]{Yang24JCAP}%
  \BibitemOpen
  \bibfield  {author} {\bibinfo {author} {\bibfnamefont {S.}~\bibnamefont
  {Yang}}, \bibinfo {author} {\bibfnamefont {Y.-P.}\ \bibnamefont {Zhang}},
  \bibinfo {author} {\bibfnamefont {T.}~\bibnamefont {Zhu}}, \bibinfo {author}
  {\bibfnamefont {L.}~\bibnamefont {Zhao}}, \ and\ \bibinfo {author}
  {\bibfnamefont {Y.-X.}\ \bibnamefont {Liu}},\ }\href {\doibase
  10.1088/1475-7516/2025/01/091} {\bibfield  {journal} {\bibinfo  {journal}
  {JCAP}\ }\textbf {\bibinfo {volume} {01}},\ \bibinfo {pages} {091} (\bibinfo
  {year} {2025}{\natexlab{a}})},\ \Eprint {http://arxiv.org/abs/2407.00283}
  {arXiv:2407.00283 [gr-qc]} \BibitemShut {NoStop}%
\bibitem [{\citenamefont {Shabbir}\ \emph
  {et~al.}(2025{\natexlab{a}})\citenamefont {Shabbir}, \citenamefont {Jamil},\
  and\ \citenamefont {Azreg-A{\"\i}nou}}]{Shabbir25}%
  \BibitemOpen
  \bibfield  {author} {\bibinfo {author} {\bibfnamefont {O.}~\bibnamefont
  {Shabbir}}, \bibinfo {author} {\bibfnamefont {M.}~\bibnamefont {Jamil}}, \
  and\ \bibinfo {author} {\bibfnamefont {M.}~\bibnamefont {Azreg-A{\"\i}nou}},\
  }\href {\doibase 10.1016/j.dark.2025.101816} {\bibfield  {journal} {\bibinfo
  {journal} {Phys. Dark Univ.}\ }\textbf {\bibinfo {volume} {47}},\ \bibinfo
  {pages} {101816} (\bibinfo {year} {2025}{\natexlab{a}})},\ \Eprint
  {http://arxiv.org/abs/2501.04367} {arXiv:2501.04367 [gr-qc]} \BibitemShut
  {NoStop}%
\bibitem [{\citenamefont {Junior}\ \emph {et~al.}(2025)\citenamefont {Junior},
  \citenamefont {Junior}, \citenamefont {Lobo}, \citenamefont {Rodrigues},
  \citenamefont {Rubiera-Garcia}, \citenamefont {da~Silva},\ and\ \citenamefont
  {Vieira}}]{Junior24}%
  \BibitemOpen
  \bibfield  {author} {\bibinfo {author} {\bibfnamefont {E.~L.~B.}\
  \bibnamefont {Junior}}, \bibinfo {author} {\bibfnamefont {J.~T. S.~S.}\
  \bibnamefont {Junior}}, \bibinfo {author} {\bibfnamefont {F.~S.~N.}\
  \bibnamefont {Lobo}}, \bibinfo {author} {\bibfnamefont {M.~E.}\ \bibnamefont
  {Rodrigues}}, \bibinfo {author} {\bibfnamefont {D.}~\bibnamefont
  {Rubiera-Garcia}}, \bibinfo {author} {\bibfnamefont {L.~F.~D.}\ \bibnamefont
  {da~Silva}}, \ and\ \bibinfo {author} {\bibfnamefont {H.~A.}\ \bibnamefont
  {Vieira}},\ }\href {\doibase 10.1140/epjc/s10052-025-14299-3} {\bibfield
  {journal} {\bibinfo  {journal} {Eur. Phys. J. C}\ }\textbf {\bibinfo {volume}
  {85}},\ \bibinfo {pages} {557} (\bibinfo {year} {2025})},\ \Eprint
  {http://arxiv.org/abs/2412.00769} {arXiv:2412.00769 [gr-qc]} \BibitemShut
  {NoStop}%
\bibitem [{\citenamefont {Yang}\ \emph
  {et~al.}(2025{\natexlab{b}})\citenamefont {Yang}, \citenamefont {Zhang},
  \citenamefont {Zhu}, \citenamefont {Zhao},\ and\ \citenamefont
  {Liu}}]{Yang24}%
  \BibitemOpen
  \bibfield  {author} {\bibinfo {author} {\bibfnamefont {S.}~\bibnamefont
  {Yang}}, \bibinfo {author} {\bibfnamefont {Y.-P.}\ \bibnamefont {Zhang}},
  \bibinfo {author} {\bibfnamefont {T.}~\bibnamefont {Zhu}}, \bibinfo {author}
  {\bibfnamefont {L.}~\bibnamefont {Zhao}}, \ and\ \bibinfo {author}
  {\bibfnamefont {Y.-X.}\ \bibnamefont {Liu}},\ }\href {\doibase
  10.1088/1674-1137/adef1a} {\bibfield  {journal} {\bibinfo  {journal} {Chin.
  Phys.}\ }\textbf {\bibinfo {volume} {49}},\ \bibinfo {pages} {115107}
  (\bibinfo {year} {2025}{\natexlab{b}})},\ \Eprint
  {http://arxiv.org/abs/2412.04302} {arXiv:2412.04302 [gr-qc]} \BibitemShut
  {NoStop}%
\bibitem [{\citenamefont {Haroon}\ and\ \citenamefont {Zhu}(2025)}]{Haroon25}%
  \BibitemOpen
  \bibfield  {author} {\bibinfo {author} {\bibfnamefont {S.}~\bibnamefont
  {Haroon}}\ and\ \bibinfo {author} {\bibfnamefont {T.}~\bibnamefont {Zhu}},\
  }\href {\doibase 10.1103/ckdt-wtsl} {\bibfield  {journal} {\bibinfo
  {journal} {Phys. Rev. D}\ }\textbf {\bibinfo {volume} {112}},\ \bibinfo
  {pages} {044046} (\bibinfo {year} {2025})},\ \Eprint
  {http://arxiv.org/abs/2502.09171} {arXiv:2502.09171 [gr-qc]} \BibitemShut
  {NoStop}%
\bibitem [{\citenamefont {Alloqulov}\ \emph {et~al.}(2025)\citenamefont
  {Alloqulov}, \citenamefont {Xamidov}, \citenamefont {Shaymatov},\ and\
  \citenamefont {Ahmedov}}]{Alloqulov25GW}%
  \BibitemOpen
  \bibfield  {author} {\bibinfo {author} {\bibfnamefont {M.}~\bibnamefont
  {Alloqulov}}, \bibinfo {author} {\bibfnamefont {T.}~\bibnamefont {Xamidov}},
  \bibinfo {author} {\bibfnamefont {S.}~\bibnamefont {Shaymatov}}, \ and\
  \bibinfo {author} {\bibfnamefont {B.}~\bibnamefont {Ahmedov}},\ }\href
  {\doibase 10.1140/epjc/s10052-025-14529-8} {\bibfield  {journal} {\bibinfo
  {journal} {Eur. Phys. J. C}\ }\textbf {\bibinfo {volume} {85}},\ \bibinfo
  {pages} {798} (\bibinfo {year} {2025})},\ \Eprint
  {http://arxiv.org/abs/2504.05236} {arXiv:2504.05236 [gr-qc]} \BibitemShut
  {NoStop}%
\bibitem [{\citenamefont {Wang}\ \emph
  {et~al.}(2025{\natexlab{b}})\citenamefont {Wang}, \citenamefont {Meng},
  \citenamefont {Zhang}, \citenamefont {Zhu},\ and\ \citenamefont
  {Wei}}]{Wang25JCAP}%
  \BibitemOpen
  \bibfield  {author} {\bibinfo {author} {\bibfnamefont {C.-H.}\ \bibnamefont
  {Wang}}, \bibinfo {author} {\bibfnamefont {X.-C.}\ \bibnamefont {Meng}},
  \bibinfo {author} {\bibfnamefont {Y.-P.}\ \bibnamefont {Zhang}}, \bibinfo
  {author} {\bibfnamefont {T.}~\bibnamefont {Zhu}}, \ and\ \bibinfo {author}
  {\bibfnamefont {S.-W.}\ \bibnamefont {Wei}},\ }\href {\doibase
  10.1088/1475-7516/2025/07/021} {\bibfield  {journal} {\bibinfo  {journal} {J.
  Cosmol. Astropart. Phys.}\ }\textbf {\bibinfo {volume} {2025}},\ \bibinfo
  {pages} {021} (\bibinfo {year} {2025}{\natexlab{b}})}\BibitemShut {NoStop}%
\bibitem [{\citenamefont {Lu}\ and\ \citenamefont {Zhu}(2025)}]{Lu25GW}%
  \BibitemOpen
  \bibfield  {author} {\bibinfo {author} {\bibfnamefont {S.}~\bibnamefont
  {Lu}}\ and\ \bibinfo {author} {\bibfnamefont {T.}~\bibnamefont {Zhu}},\
  }\href {\doibase 10.1016/j.dark.2025.102141} {\bibfield  {journal} {\bibinfo
  {journal} {Phys. Dark Universe}\ }\textbf {\bibinfo {volume} {50}},\ \bibinfo
  {pages} {102141} (\bibinfo {year} {2025})},\ \Eprint
  {http://arxiv.org/abs/2505.00294} {arXiv:2505.00294 [gr-qc]} \BibitemShut
  {NoStop}%
\bibitem [{\citenamefont {Chen}\ and\ \citenamefont {Yang}(2025)}]{Chen25}%
  \BibitemOpen
  \bibfield  {author} {\bibinfo {author} {\bibfnamefont {J.}~\bibnamefont
  {Chen}}\ and\ \bibinfo {author} {\bibfnamefont {J.}~\bibnamefont {Yang}},\
  }\href {\doibase 10.1140/epjc/s10052-025-14457-7} {\bibfield  {journal}
  {\bibinfo  {journal} {Eur. Phys. J. C}\ }\textbf {\bibinfo {volume} {85}},\
  \bibinfo {pages} {726} (\bibinfo {year} {2025})},\ \Eprint
  {http://arxiv.org/abs/2505.02660} {arXiv:2505.02660 [gr-qc]} \BibitemShut
  {NoStop}%
\bibitem [{\citenamefont {Li}\ \emph {et~al.}(2025)\citenamefont {Li},
  \citenamefont {Qiao},\ and\ \citenamefont {Tao}}]{Li25}%
  \BibitemOpen
  \bibfield  {author} {\bibinfo {author} {\bibfnamefont {G.-H.}\ \bibnamefont
  {Li}}, \bibinfo {author} {\bibfnamefont {C.-K.}\ \bibnamefont {Qiao}}, \ and\
  \bibinfo {author} {\bibfnamefont {J.}~\bibnamefont {Tao}},\ }\href@noop {} {\
   (\bibinfo {year} {2025})},\ \Eprint {http://arxiv.org/abs/2510.24989}
  {arXiv:2510.24989 [gr-qc]} \BibitemShut {NoStop}%
\bibitem [{\citenamefont {Ahmed}\ \emph
  {et~al.}(2025{\natexlab{a}})\citenamefont {Ahmed}, \citenamefont {Wu},
  \citenamefont {Ghosh},\ and\ \citenamefont {Zhu}}]{Ahmed25GW1}%
  \BibitemOpen
  \bibfield  {author} {\bibinfo {author} {\bibfnamefont {F.}~\bibnamefont
  {Ahmed}}, \bibinfo {author} {\bibfnamefont {Q.}~\bibnamefont {Wu}}, \bibinfo
  {author} {\bibfnamefont {S.~G.}\ \bibnamefont {Ghosh}}, \ and\ \bibinfo
  {author} {\bibfnamefont {T.}~\bibnamefont {Zhu}},\ }\href@noop {} {\
  (\bibinfo {year} {2025}{\natexlab{a}})},\ \Eprint
  {http://arxiv.org/abs/2511.08456} {arXiv:2511.08456 [gr-qc]} \BibitemShut
  {NoStop}%
\bibitem [{\citenamefont {Ahmed}\ \emph
  {et~al.}(2025{\natexlab{b}})\citenamefont {Ahmed}, \citenamefont {Wu},
  \citenamefont {Ghosh},\ and\ \citenamefont {Zhu}}]{Ahmed25GW2}%
  \BibitemOpen
  \bibfield  {author} {\bibinfo {author} {\bibfnamefont {F.}~\bibnamefont
  {Ahmed}}, \bibinfo {author} {\bibfnamefont {Q.}~\bibnamefont {Wu}}, \bibinfo
  {author} {\bibfnamefont {S.~G.}\ \bibnamefont {Ghosh}}, \ and\ \bibinfo
  {author} {\bibfnamefont {T.}~\bibnamefont {Zhu}},\ }\href
  {https://arxiv.org/abs/2512.24036} {\enquote {\bibinfo {title} {Signatures of
  quantum-corrected black holes in gravitational waves from periodic orbits},}\
  } (\bibinfo {year} {2025}{\natexlab{b}}),\ \Eprint
  {http://arxiv.org/abs/2512.24036} {arXiv:2512.24036 [gr-qc]} \BibitemShut
  {NoStop}%
\bibitem [{\citenamefont {{Podolsk{\'y}}}\ and\ \citenamefont
  {{Ovcharenko}}(2025)}]{Podolsky2025}%
  \BibitemOpen
  \bibfield  {author} {\bibinfo {author} {\bibfnamefont {J.}~\bibnamefont
  {{Podolsk{\'y}}}}\ and\ \bibinfo {author} {\bibfnamefont {H.}~\bibnamefont
  {{Ovcharenko}}},\ }\href {\doibase 10.1103/rfgv-ybz5} {\bibfield  {journal}
  {\bibinfo  {journal} {Phys. Rev. Lett.}\ }\textbf {\bibinfo {volume} {135}},\
  \bibinfo {eid} {181401} (\bibinfo {year} {2025})},\ \Eprint
  {http://arxiv.org/abs/2507.05199} {arXiv:2507.05199 [gr-qc]} \BibitemShut
  {NoStop}%
\bibitem [{\citenamefont {Bertotti}(1959)}]{Bertotti59}%
  \BibitemOpen
  \bibfield  {author} {\bibinfo {author} {\bibfnamefont {B.}~\bibnamefont
  {Bertotti}},\ }\href {\doibase 10.1103/PhysRev.116.1331} {\bibfield
  {journal} {\bibinfo  {journal} {Phys. Rev.}\ }\textbf {\bibinfo {volume}
  {116}},\ \bibinfo {pages} {1331} (\bibinfo {year} {1959})}\BibitemShut
  {NoStop}%
\bibitem [{\citenamefont {Robinson}(1961)}]{Robinson61}%
  \BibitemOpen
  \bibfield  {author} {\bibinfo {author} {\bibfnamefont {I.}~\bibnamefont
  {Robinson}},\ }\href {\doibase 10.1063/1.1703712} {\bibfield  {journal}
  {\bibinfo  {journal} {J. Math. Phys.}\ }\textbf {\bibinfo {volume} {2}},\
  \bibinfo {pages} {290} (\bibinfo {year} {1961})}\BibitemShut {NoStop}%
\bibitem [{\citenamefont {Plebanski}\ and\ \citenamefont
  {Demianski}(1976)}]{PLEBANSKI1976}%
  \BibitemOpen
  \bibfield  {author} {\bibinfo {author} {\bibfnamefont {J.}~\bibnamefont
  {Plebanski}}\ and\ \bibinfo {author} {\bibfnamefont {M.}~\bibnamefont
  {Demianski}},\ }\href {\doibase https://doi.org/10.1016/0003-4916(76)90240-2}
  {\bibfield  {journal} {\bibinfo  {journal} {Ann. Phys.}\ }\textbf {\bibinfo
  {volume} {98}},\ \bibinfo {pages} {98} (\bibinfo {year} {1976})}\BibitemShut
  {NoStop}%
\bibitem [{\citenamefont {Carter}(1968)}]{Brandon1968}%
  \BibitemOpen
  \bibfield  {author} {\bibinfo {author} {\bibfnamefont {B.}~\bibnamefont
  {Carter}},\ }\href {\doibase 10.1103/PhysRev.174.1559} {\bibfield  {journal}
  {\bibinfo  {journal} {Phys. Rev.}\ }\textbf {\bibinfo {volume} {174}},\
  \bibinfo {pages} {1559} (\bibinfo {year} {1968})}\BibitemShut {NoStop}%
\bibitem [{\citenamefont {Bocquet}\ \emph {et~al.}(1995)\citenamefont
  {Bocquet}, \citenamefont {Bonazzola}, \citenamefont {Gourgoulhon},\ and\
  \citenamefont {Novak}}]{Bocquet1995}%
  \BibitemOpen
  \bibfield  {author} {\bibinfo {author} {\bibfnamefont {M.}~\bibnamefont
  {Bocquet}}, \bibinfo {author} {\bibfnamefont {S.}~\bibnamefont {Bonazzola}},
  \bibinfo {author} {\bibfnamefont {E.}~\bibnamefont {Gourgoulhon}}, \ and\
  \bibinfo {author} {\bibfnamefont {J.}~\bibnamefont {Novak}},\ }\href@noop {}
  {\bibfield  {journal} {\bibinfo  {journal} {Astron. Astrophys.}\ }\textbf
  {\bibinfo {volume} {301}},\ \bibinfo {pages} {757} (\bibinfo {year}
  {1995})},\ \Eprint {http://arxiv.org/abs/gr-qc/9503044} {arXiv:gr-qc/9503044}
  \BibitemShut {NoStop}%
\bibitem [{\citenamefont {Andersson}\ \emph {et~al.}(2025)\citenamefont
  {Andersson}, \citenamefont {Gray},\ and\ \citenamefont
  {Oancea}}]{Andersson2025}%
  \BibitemOpen
  \bibfield  {author} {\bibinfo {author} {\bibfnamefont {L.}~\bibnamefont
  {Andersson}}, \bibinfo {author} {\bibfnamefont {F.}~\bibnamefont {Gray}}, \
  and\ \bibinfo {author} {\bibfnamefont {M.~A.}\ \bibnamefont {Oancea}},\
  }\href@noop {} {\  (\bibinfo {year} {2025})},\ \Eprint
  {http://arxiv.org/abs/2512.07677} {arXiv:2512.07677 [gr-qc]} \BibitemShut
  {NoStop}%
\bibitem [{\citenamefont {Wang}\ \emph
  {et~al.}(2025{\natexlab{c}})\citenamefont {Wang}, \citenamefont {Hou},
  \citenamefont {Wan}, \citenamefont {Guo},\ and\ \citenamefont
  {Chen}}]{Wang2025vsx}%
  \BibitemOpen
  \bibfield  {author} {\bibinfo {author} {\bibfnamefont {X.}~\bibnamefont
  {Wang}}, \bibinfo {author} {\bibfnamefont {Y.}~\bibnamefont {Hou}}, \bibinfo
  {author} {\bibfnamefont {X.}~\bibnamefont {Wan}}, \bibinfo {author}
  {\bibfnamefont {M.}~\bibnamefont {Guo}}, \ and\ \bibinfo {author}
  {\bibfnamefont {B.}~\bibnamefont {Chen}},\ }\href@noop {} {\  (\bibinfo
  {year} {2025}{\natexlab{c}})},\ \Eprint {http://arxiv.org/abs/2507.22494}
  {arXiv:2507.22494 [gr-qc]} \BibitemShut {NoStop}%
\bibitem [{\citenamefont {Zeng}\ and\ \citenamefont
  {Wang}(2025)}]{Zeng:2025KRB}%
  \BibitemOpen
  \bibfield  {author} {\bibinfo {author} {\bibfnamefont {X.-X.}\ \bibnamefont
  {Zeng}}\ and\ \bibinfo {author} {\bibfnamefont {K.}~\bibnamefont {Wang}},\
  }\href {\doibase 10.1103/vc96-snjm} {\bibfield  {journal} {\bibinfo
  {journal} {Phys. Rev. D}\ }\textbf {\bibinfo {volume} {112}},\ \bibinfo
  {pages} {064032} (\bibinfo {year} {2025})},\ \Eprint
  {http://arxiv.org/abs/2507.21777} {arXiv:2507.21777 [gr-qc]} \BibitemShut
  {NoStop}%
\bibitem [{\citenamefont {Vachher}\ \emph {et~al.}(2025)\citenamefont
  {Vachher}, \citenamefont {Kumar},\ and\ \citenamefont
  {Ghosh}}]{Vachher2025JCAP}%
  \BibitemOpen
  \bibfield  {author} {\bibinfo {author} {\bibfnamefont {A.}~\bibnamefont
  {Vachher}}, \bibinfo {author} {\bibfnamefont {A.}~\bibnamefont {Kumar}}, \
  and\ \bibinfo {author} {\bibfnamefont {S.~G.}\ \bibnamefont {Ghosh}},\ }\href
  {\doibase 10.1088/1475-7516/2025/11/021} {\bibfield  {journal} {\bibinfo
  {journal} {J. Cosmol. Astropart. Phys.}\ }\textbf {\bibinfo {volume}
  {2025}},\ \bibinfo {pages} {021} (\bibinfo {year} {2025})}\BibitemShut
  {NoStop}%
\bibitem [{\citenamefont {Wang}(2025)}]{Wang2025bjf}%
  \BibitemOpen
  \bibfield  {author} {\bibinfo {author} {\bibfnamefont {T.}~\bibnamefont
  {Wang}},\ }\href@noop {} {\  (\bibinfo {year} {2025})},\ \Eprint
  {http://arxiv.org/abs/2508.04684} {arXiv:2508.04684 [gr-qc]} \BibitemShut
  {NoStop}%
\bibitem [{\citenamefont {Siahaan}(2025)}]{Siahaan2025KRB}%
  \BibitemOpen
  \bibfield  {author} {\bibinfo {author} {\bibfnamefont {H.~M.}\ \bibnamefont
  {Siahaan}},\ }\href@noop {} {\  (\bibinfo {year} {2025})},\ \Eprint
  {http://arxiv.org/abs/2512.12533} {arXiv:2512.12533 [gr-qc]} \BibitemShut
  {NoStop}%
\bibitem [{\citenamefont {Podolsky}(2025)}]{Podolsky2025zlm}%
  \BibitemOpen
  \bibfield  {author} {\bibinfo {author} {\bibfnamefont {J.}~\bibnamefont
  {Podolsky}},\ }in\ \href@noop {} {\emph {\bibinfo {booktitle} {{24th
  International Conference on General Relativity and Gravitation (GR24) and
  16th Edoardo Amaldi Conference on Gravitational (Amaldi16) Waves}}}}\
  (\bibinfo {year} {2025})\ \Eprint {http://arxiv.org/abs/2511.01029}
  {arXiv:2511.01029 [gr-qc]} \BibitemShut {NoStop}%
\bibitem [{\citenamefont {Mirkhaydarov}\ \emph {et~al.}(2026)\citenamefont
  {Mirkhaydarov}, \citenamefont {Xamidov}, \citenamefont {Sheoran},
  \citenamefont {Shaymatov},\ and\ \citenamefont
  {Nandan}}]{Mirkhaydarov2026MPP}%
  \BibitemOpen
  \bibfield  {author} {\bibinfo {author} {\bibfnamefont {M.}~\bibnamefont
  {Mirkhaydarov}}, \bibinfo {author} {\bibfnamefont {T.}~\bibnamefont
  {Xamidov}}, \bibinfo {author} {\bibfnamefont {P.}~\bibnamefont {Sheoran}},
  \bibinfo {author} {\bibfnamefont {S.}~\bibnamefont {Shaymatov}}, \ and\
  \bibinfo {author} {\bibfnamefont {H.}~\bibnamefont {Nandan}},\ }\href
  {https://arxiv.org/abs/2601.09919} {\enquote {\bibinfo {title} {Non-monotonic
  enhancement of the magnetic penrose process in kerr-bertotti-robinson
  spacetime and its implication for electron acceleration},}\ } (\bibinfo
  {year} {2026}),\ \Eprint {http://arxiv.org/abs/2601.09919} {arXiv:2601.09919
  [gr-qc]} \BibitemShut {NoStop}%
\bibitem [{\citenamefont {Liu}\ \emph {et~al.}(2025)\citenamefont {Liu},
  \citenamefont {Liu}, \citenamefont {Wu},\ and\ \citenamefont
  {Liu}}]{Liu2025wwq}%
  \BibitemOpen
  \bibfield  {author} {\bibinfo {author} {\bibfnamefont {W.}~\bibnamefont
  {Liu}}, \bibinfo {author} {\bibfnamefont {Y.}~\bibnamefont {Liu}}, \bibinfo
  {author} {\bibfnamefont {D.}~\bibnamefont {Wu}}, \ and\ \bibinfo {author}
  {\bibfnamefont {Y.-X.}\ \bibnamefont {Liu}},\ }\href@noop {} {\  (\bibinfo
  {year} {2025})},\ \Eprint {http://arxiv.org/abs/2511.06017} {arXiv:2511.06017
  [gr-qc]} \BibitemShut {NoStop}%
\bibitem [{\citenamefont {Chandrasekhar}(1984)}]{1983mtbh.book.....C}%
  \BibitemOpen
  \bibfield  {author} {\bibinfo {author} {\bibfnamefont {S.}~\bibnamefont
  {Chandrasekhar}},\ }\href {https://doi.org/10.1007/978-94-009-6469-3_2}
  {\emph {\bibinfo {title} {General Relativity and Gravitation: Invited Papers
  and Discussion Reports of the 10th International Conference on General
  Relativity and Gravitation, Padua, July 3--8, 1983}}},\ edited by\ \bibinfo
  {editor} {\bibfnamefont {B.}~\bibnamefont {Bertotti}}, \bibinfo {editor}
  {\bibfnamefont {F.}~\bibnamefont {de~Felice}}, \ and\ \bibinfo {editor}
  {\bibfnamefont {A.}~\bibnamefont {Pascolini}}\ (\bibinfo  {publisher}
  {Springer Netherlands},\ \bibinfo {address} {Dordrecht},\ \bibinfo {year}
  {1984})\ pp.\ \bibinfo {pages} {5--26}\BibitemShut {NoStop}%
\bibitem [{\citenamefont {{Yang}}\ \emph {et~al.}(2025)\citenamefont {{Yang}},
  \citenamefont {{Zhang}}, \citenamefont {{Zhu}}, \citenamefont {{Zhao}},\ and\
  \citenamefont {{Liu}}}]{2025JCAP...01..091Y}%
  \BibitemOpen
  \bibfield  {author} {\bibinfo {author} {\bibfnamefont {S.}~\bibnamefont
  {{Yang}}}, \bibinfo {author} {\bibfnamefont {Y.-P.}\ \bibnamefont {{Zhang}}},
  \bibinfo {author} {\bibfnamefont {T.}~\bibnamefont {{Zhu}}}, \bibinfo
  {author} {\bibfnamefont {L.}~\bibnamefont {{Zhao}}}, \ and\ \bibinfo {author}
  {\bibfnamefont {Y.-X.}\ \bibnamefont {{Liu}}},\ }\href {\doibase
  10.1088/1475-7516/2025/01/091} {\bibfield  {journal} {\bibinfo  {journal}
  {Journal of Cosmology and Astroparticle Physics}\ }\textbf {\bibinfo {volume}
  {2025}},\ \bibinfo {eid} {091} (\bibinfo {year} {2025})},\ \Eprint
  {http://arxiv.org/abs/2407.00283} {arXiv:2407.00283 [gr-qc]} \BibitemShut
  {NoStop}%
\bibitem [{\citenamefont {Shabbir}\ \emph
  {et~al.}(2025{\natexlab{b}})\citenamefont {Shabbir}, \citenamefont {Jamil},\
  and\ \citenamefont {Azreg-Aïnou}}]{SHABBIR2025101816}%
  \BibitemOpen
  \bibfield  {author} {\bibinfo {author} {\bibfnamefont {O.}~\bibnamefont
  {Shabbir}}, \bibinfo {author} {\bibfnamefont {M.}~\bibnamefont {Jamil}}, \
  and\ \bibinfo {author} {\bibfnamefont {M.}~\bibnamefont {Azreg-Aïnou}},\
  }\href {\doibase https://doi.org/10.1016/j.dark.2025.101816} {\bibfield
  {journal} {\bibinfo  {journal} {Physics of the Dark Universe}\ }\textbf
  {\bibinfo {volume} {47}},\ \bibinfo {pages} {101816} (\bibinfo {year}
  {2025}{\natexlab{b}})}\BibitemShut {NoStop}%
\bibitem [{\citenamefont {{Zhao}}\ \emph {et~al.}(2025)\citenamefont {{Zhao}},
  \citenamefont {{Tang}},\ and\ \citenamefont {{Xu}}}]{2025EPJC...85...36Z}%
  \BibitemOpen
  \bibfield  {author} {\bibinfo {author} {\bibfnamefont {L.}~\bibnamefont
  {{Zhao}}}, \bibinfo {author} {\bibfnamefont {M.}~\bibnamefont {{Tang}}}, \
  and\ \bibinfo {author} {\bibfnamefont {Z.}~\bibnamefont {{Xu}}},\ }\href
  {\doibase 10.1140/epjc/s10052-025-13767-0} {\bibfield  {journal} {\bibinfo
  {journal} {European Physical Journal C}\ }\textbf {\bibinfo {volume} {85}},\
  \bibinfo {eid} {36} (\bibinfo {year} {2025})},\ \Eprint
  {http://arxiv.org/abs/2411.01979} {arXiv:2411.01979 [gr-qc]} \BibitemShut
  {NoStop}%
\bibitem [{\citenamefont {Poisson}\ and\ \citenamefont
  {Will}(2014)}]{Poisson_Will_2014}%
  \BibitemOpen
  \bibfield  {author} {\bibinfo {author} {\bibfnamefont {E.}~\bibnamefont
  {Poisson}}\ and\ \bibinfo {author} {\bibfnamefont {C.~M.}\ \bibnamefont
  {Will}},\ }\href@noop {} {\emph {\bibinfo {title} {Gravity: Newtonian,
  Post-Newtonian, Relativistic}}}\ (\bibinfo  {publisher} {Cambridge University
  Press},\ \bibinfo {year} {2014})\BibitemShut {NoStop}%
\bibitem [{\citenamefont {{Meng}}\ \emph {et~al.}(2024)\citenamefont {{Meng}},
  \citenamefont {{Xu}},\ and\ \citenamefont {{Tang}}}]{2024arXiv241101858M}%
  \BibitemOpen
  \bibfield  {author} {\bibinfo {author} {\bibfnamefont {L.}~\bibnamefont
  {{Meng}}}, \bibinfo {author} {\bibfnamefont {Z.}~\bibnamefont {{Xu}}}, \ and\
  \bibinfo {author} {\bibfnamefont {M.}~\bibnamefont {{Tang}}},\ }\href
  {\doibase 10.48550/arXiv.2411.01858} {\bibfield  {journal} {\bibinfo
  {journal} {arXiv e-prints}\ ,\ \bibinfo {eid} {arXiv:2411.01858}} (\bibinfo
  {year} {2024})},\ \Eprint {http://arxiv.org/abs/2411.01858} {arXiv:2411.01858
  [gr-qc]} \BibitemShut {NoStop}%
\end{thebibliography}%

\end{document}